\documentclass[aps,twocolumn,superscriptaddress,pra,10pt,nofootinbib]{revtex4-2}
\usepackage{graphicx}
\usepackage{amsmath}
\usepackage{amssymb}
\usepackage{amsfonts}
\usepackage{color}
\usepackage{dsfont}
\usepackage{hyperref}
\usepackage{physics}
\usepackage{tikz}
\usetikzlibrary{positioning,shapes}

\usepackage{tikz,xcolor}
\definecolor{lime}{HTML}{A6CE39}
\DeclareRobustCommand{\orcidicon}{%
	\begin{tikzpicture}
	\draw[lime, fill=lime] (0,0) 
	circle [radius=0.16] 
	node[white] {{\fontfamily{qag}\selectfont \tiny ID}};
	\draw[white, fill=white] (-0.0625,0.095) 
	circle [radius=0.007];
	\end{tikzpicture}
	\hspace{-2mm}
}
\foreach \x in {A, ..., Z}{%
	\expandafter\xdef\csname orcid\x\endcsname{\noexpand\href{https://orcid.org/\csname orcidauthor\x\endcsname}{\noexpand\orcidicon}}
}

\begin{document}

\title[Bounding the rotating wave approximation for coupled harmonic oscillators]{Bounding the rotating wave approximation for coupled harmonic oscillators}

\date{\today}

\author{Tim Heib\orcidH{}}
\email{t.heib@fz-juelich.de}
\affiliation{Quantum Computing Analytics (PGI-12), Forschungszentrum J\"ulich, 52425 J\"ulich, Germany}
\affiliation{Theoretical Physics, Universit\"at des Saarlandes, 66123 Saarbr\"ucken, Germany}
\author{Paul Lageyre\orcidF{}}
\affiliation{Quantum Computing Analytics (PGI-12), Forschungszentrum J\"ulich, 52425 J\"ulich, Germany}
\author{Alessandro Ferreri\orcidA{}}
\affiliation{Quantum Computing Analytics (PGI-12), Forschungszentrum J\"ulich, 52425 J\"ulich, Germany}
\author{Frank K. Wilhelm\orcidC{}}
\affiliation{Quantum Computing Analytics (PGI-12), Forschungszentrum J\"ulich, 52425 J\"ulich, Germany}
\affiliation{Theoretical Physics, Universit\"at des Saarlandes, 66123 Saarbr\"ucken, Germany}
\author{G. S. Paraoanu\orcidE{}}
\affiliation{QTF Centre of Excellence,  Department of Applied Physics, Aalto University School of Science, FI-00076 AALTO, Finland}
\affiliation{InstituteQ, Aalto University, FI-00076 AALTO, Finland}
\author{Daniel Burgarth\orcidG{}}
\affiliation{Physics Department, Friedrich-Alexander Universit\"at of Erlangen-Nuremberg, Staudtstr. 7, 91058 Erlangen, Germany}
\author{Andreas Wolfgang Schell\orcidD{}}
\affiliation{Insitute of Solid-State and Semiconductor Physics, Johannes Kepler Universit\"at Linz, 4040 Linz, Austria}
\affiliation{Institut f\"ur Festk\"orperphysik, Leibniz Universit\"at Hannover, 30167 Hannover, Germany}
\affiliation{Physikalisch-Technische Bundesanstalt, 38116 Braunschweig, Germany}
\author{David Edward Bruschi\orcidB{}}
\email{david.edward.bruschi@posteo.net}
\email{d.e.bruschi@fz-juelich.de}
\affiliation{Quantum Computing Analytics (PGI-12), Forschungszentrum J\"ulich, 52425 J\"ulich, Germany}
\affiliation{Theoretical Physics, Universit\"at des Saarlandes, 66123 Saarbr\"ucken, Germany}

\begin{abstract}
In this work we study the validity of the rotating wave approximation of an ideal system composed of two harmonic oscillators evolving with a quadratic Hamiltonian and arbitrarily strong interaction. We prove its validity for arbitrary states by bounding the error introduced. We then restrict ourselves to the dynamics of Gaussian states and are able to fully quantify the deviation of arbitrary pure Gaussian states that evolve through different dynamics from a common quantum state. We show that this distance is fully determined by the first and second moments of the statistical distribution of the number of excitations created from the vacuum during an appropriate effective time-evolution. We use these results to completely control the dynamics for this class of states, therefore providing a toolbox to be used in quantum optics and quantum information. Applications and potential physical implementations are also discussed.
\end{abstract}

\maketitle

\par\noindent

	\section*{Introduction}
	Coupled quantum harmonic oscillators are the paramount tool for studying systems of interacting bosonic degrees of freedoms. Applications range across the most diverse fields of physics, from quantum optics \cite{Mandel:Wolf:1995} and quantum information \cite{quantuminfo}, to particle-creation phenomena within quantum field theory in curved spacetime \cite{Hawking:1974,Unruh:1976,Birrell:Davies:1984}. Dynamics of interacting harmonic oscillators allow for analytical solutions in the case of \textit{linear dynamics}, i.e., the dynamics induced by Hamiltonians that are quadratic in the quadrature operators \cite{Adesso:Ragy:2014,Serafini:2017}. While linear dynamics enjoys full understanding at the price of limited scope of applicability, such as the restriction to Gaussian states and the \textit{covariance matrix formalism} \cite{Adesso:Ragy:2014}, nonlinear dynamics brings a richer yet more elusive type of phenomena to the table \cite{Bose:Jacobs:1997,Bruschi:Xuereb:2018,Bruschi:2020,Schneiter:Qvarfort:2020}, which are currently being investigated \cite{Bruschi:Xuereb:2024}.
	
	In this work we aim at \textit{bounding the validity of the rotating wave approximation for coupled quantum harmonic oscillators}. The rotating wave approximation (RWA) is a powerful tool to approximate exact dynamics and obtain analytical results in the case of interacting light-matter systems \cite{Rivera:Kaminer:2020,Gutzler:Garg:2021,Guan:Park:2022} as well as systems of interacting harmonic oscillators \cite{Aspelmeyer:Kippenberg:2014}. The general idea behind this approximation is that, given the Hamiltonian of a quantum system, one can study the time evolution in the interaction picture using terms that are constant or oscillate slowly, and neglect the terms of the interaction Hamiltonian that oscillate faster.  Two crucial features are generally believed to be necessary for this approximation to be valid: small coupling as compared to the relevant frequency scales of the problem, and resonant frequencies. In its simplest form, the \textit{rotating wave approximation} is used in the well-known Jaynes-Cumming model \cite{Jaynes:Cummings:1963}, which has been extensively studied since its proposal~\cite{Larson:Mavrogordatos:2021} and for which a proof was recently provided \cite{Burgarth:Facchi:2023}. After many decades of development and applications, the study of the rotating wave approximation is still an active area of research since it can be used to shed light into complex but interesting dynamics \cite{Zeuch:Hassler:2020}. Limitations \cite{Schek:Sage:1979,Berlin:Aliaga:2004,Jing:Lu:2009,Fleming:Cummings:2010,Spiegelberg:Sjoqvist:2013,Cardenas:Teizeira:2017} as well as potential issues \cite{Ford:OConnell:1997,Fleming:Cummings:2010,Jorgensen:Wubs:2022,Rokaj:Mistakidis:2023} have been also pointed out and examined. Regardless, a full understanding is still outstanding.
	
In this work we provide an explicit error bound for the approximation as usually employed. We achieve this goal by using Fock states, which in turn proves the validity of the approximation for arbitrary quantum states due to the density of Fock states in the Hilbert space.
	We also specialize our analysis by restricting ourselves to initial Gaussian states and therefore remaining in the Gaussian-state domain due to the linearity of the dynamics considered. This allows us to employ the covariance matrix formalism \cite{Adesso:Ragy:2014}, which has specifically been designed to apply to such scenarios. We employ the fidelity of quantum states \cite{quantuminfo}, which is related directly to the Bures distance in the Hilbert space \cite{Bures:1969}, to measure to which degree the two final states differ as a function of time and the other parameters of the problem. We compute it explicitly for the set of Gaussian states and find that a difference is related to the squeezing present within the initial state, as well as that injected by the dynamics. Interestingly we show that the first and second moments of the statistical distribution of the average number of excitations created from the vacuum during an appropriate effective time evolution encode all information necessary to determine the distance. Furthermore, we also prove that the standard rotating wave approximation is recovered when the frequencies of the two oscillators coincide and the coupling-to-frequency ratio is small \cite{Burgarth:Facchi:2023}. 
	
	We apply our results to an initial two-mode single-mode-squeezed Gaussian state and to the vacuum state, for which we can obtain analytical closed expressions. We also discuss the usefulness of our techniques for analytical studies of superconducting circuits and nitrogen-vacancy (NV) center systems coupled to spin baths. This highlights the wide applicability of our results. 
	
	This work is organized as follows. In Section \ref{Sec:System:and:Tools} we introduce the necessary tools.  In Section \ref{Sec:Demonstrating:RWA} prove the validity of the rotating wave approximation for arbitrary states. In Section~\ref{Sec:Distinguishing:two:Gaussian:States} we specialize our work to Gaussian states. In Section \ref{Sec:Applications} we discuss potential physical implementations to which our work is relevant. We conclude our work with considerations in Section~\ref{Sec:Considerations}.

	\section{System setup and tools}\label{Sec:System:and:Tools}

	\subsection{Time evolution}
	Time evolution of physical systems is the core topic of study here. Given an initial state $\hat{\rho}(0)$ and a Hamiltonian $\hat{H}(t)$, the state at time $t$ is obtained via the von Neumann equation $\hat{\rho}(t)=\hat{U}(t)\hat{\rho}(0)\hat{U}^\dag(t)$, where the time-evolution operator $\hat{U}(t)$ induced by the Hamiltonian $\hat{H}(t)$ reads
	\begin{align*}
		\hat{U}(t)=\overset{\leftarrow}{\mathcal{T}}\exp\left[-\frac{i}{\hbar}\int_0^{t}\,\dd t'\,\hat{H}(t')\right].
	\end{align*}
	Here $\overset{\leftarrow}{\mathcal{T}}$ stands for the time-ordering operator.
	
	Obtaining the explicit expression of the state $\hat{\rho}(t)$, or even only the expression of the time-evolution operator $\hat{U}(t)$, at an arbitrary time and for arbitrary Hamiltonians is usually an impossible task.
	While this task can be achieved somewhat efficiently for finite dimensional systems \cite{Wei:Norman:1963,Wei:Norman:1964,Bose:Jacobs:1997,Ibarra-Sierra:Sandoval-Santana:2015}, and a growing body of work has tackled this problem with varying degrees of success for infinite-dimensional ones \cite{Bruschi:Xuereb:2024}, the quest for a general solution remains open.
	In the second part of this work we will restrict the analysis to Gaussian states, thereby allowing for a general solution of this problem. 
	
\subsection{System of interest}
	
	We focus on physical systems that can be effectively modeled as two quantum coupled harmonic oscillators with frequencies $\omega_\text{a}$ and $\omega_\text{b}$, and annihilation and creation operators $\hat{a},\hat{a}^\dag$ and $\hat{b},\hat{b}^\dag$ respectively, which satisfy the canonical commutation relations $[\hat{a},\hat{a}^\dag]=[\hat{b},\hat{b}^\dag]=1$ while all others vanish. 
	
	We consider a general Hamiltonian with arbitrary coupling constants of the form
	\begin{align}\label{basic:hamiltonian}
		\hat{H}=\hat{H}_0+\hbar\,g_\text{bs} \left(\hat{a}\,\hat{b}^\dag+\hat{a}^\dag\,\hat{b}\right)+\hbar\,g_\text{sq} \left(\hat{a}^\dag\,\hat{b}^\dag+\hat{a}\,\hat{b}\right),
	\end{align}
	where we have introduced the free Hamiltonian $\hat{H}_0=\hbar\,\omega_\text{a}\,\hat{a}^\dag\,\hat{a}+\hbar\,\omega_\text{b}\,\hat{b}^\dag\,\hat{b}$ and the constants $g_\text{bs}$ and $g_\text{sq}$ determine the coupling strength between the two oscillators. 
	
	The term in Equation~\eqref{basic:hamiltonian} driven by $\,g_\text{bs}$ is known in quantum optics as a beam-splitting operation which passively shifts excitations from one mode to another. Here passive means that the total number of excitations is conserved. On the contrary, the term driven by $\,g_\text{sq}$ is known as squeezing, specifically two-mode squeezing, which is an active transformation that changes the total number of particles of the system \cite{quantuminfo}. 
	
	An important aspect of our work is that it is not restricted to particular values of the couplings, as long as the decay rate $\kappa$ can be effectively set to zero (i.e., the dynamics are ideal, or unitary).
	This is an advantage with respect to many studies found in the literature, where there is an ongoing discussion about the various regimes of operation that depend on the relative magnitude of the couplings, the frequencies and the decay rates: one therefore has the weak, strong, ultra-strong, and deep-strong regimes
	\cite{Mancini:Tombesi:1994,Sudhir:Genoni:2012,Rodriguez:2016,Larson:Mavrogordatos:2021}.

	\subsection{Linear dynamics}
	Here we consider \textit{linear dynamics}, which are defined as those induced by Hamiltonians that are \textit{quadratic} in the creation and annihilation operators.\footnote{This use of the term ``linear'' has nothing to do with the fundamental linearity of quantum mechanics, and it will clarify below.} In such case, we can use the symplectic formalism to map the usually intractable problem of manipulating unitary operators to the much more tractable problem of multiplying finite-dimensional matrices. An extensive introduction can be found in the literature \cite{Adesso:Ragy:2014}. 
	
	We assume to have $N$ bosonic modes with frequencies $\omega_n$ and annihilation and creation operators $\hat{a}_n,\hat{a}^\dag_n$, which satisfy the canonical commutation relations $[\hat{a}_n,\hat{a}_{n'}^\dag]=\delta_{nn'}$ while all other commutators vanish. We collect the creation and the annihilation operators in the vector $\hat{\mathbb{X}}$ of operators defined as $\hat{\mathbb{X}}:=(\hat{a}_1,...,\hat{a}_N,\hat{a}_1^\dag,...,\hat{a}_N^\dag)^{\text{Tp}}$.
	Any linear unitary evolution of our two oscillators can be represented by a time-dependent $2N\times2N$ \textit{symplectic} matrix $\boldsymbol{S}(t)$ through the key equation 
	\begin{align}\label{time:evolution:operator:symplectic:representation}
		\hat{\mathbb{X}}(t)=\hat{U}(t)^\dag\,\hat{\mathbb{X}}\,\hat{U}(t)=\boldsymbol{S}(t)\,\hat{\mathbb{X}}.
	\end{align}
	The defining property of a symplectic matrix $\boldsymbol{s}$ is that it satisfies $\boldsymbol{s}\,\boldsymbol{\Omega}\,\boldsymbol{s}^\dag=\boldsymbol{s}^\dag\,\boldsymbol{\Omega}\,\boldsymbol{s}=\boldsymbol{\Omega}$, where $\boldsymbol{\Omega}$ is the \textit{symplectic form}. Given the choice of ordering of the operators in the vector $\hat{\mathbb{X}}$, we have that
	\begin{align*}
		\boldsymbol{s}
		=
		\begin{pmatrix}
			\boldsymbol{\alpha} & \boldsymbol{\beta}\\
			\boldsymbol{\beta}^* & \boldsymbol{\alpha}^*
		\end{pmatrix},
		\quad\quad
		\boldsymbol{\Omega}
		=
		-i\,\begin{pmatrix}
			\mathds{1}_N & 0\\
			0 & -\mathds{1}_N
		\end{pmatrix},
	\end{align*}
	where $\mathds{1}_N$ is the $N$-dimensional identity operator. Notice that the defining property of the symplectic matrix $\boldsymbol{s}$ is equivalent to the well-known Bogoliubov identities, which in matrix form read $\boldsymbol{\alpha}\,\boldsymbol{\alpha}^\dag-\boldsymbol{\beta}\,\boldsymbol{\beta}^\dag=\mathds{1}_N$ and $\boldsymbol{\alpha}\,\boldsymbol{\beta}^{\text{Tp}}-\boldsymbol{\beta}\,\boldsymbol{\alpha}^{\text{Tp}}=0$. Note here that we use the Bogoliubov matrices $\boldsymbol{\alpha},\boldsymbol{\beta}$ for time-independent symplectic matrices $\boldsymbol{s}$, while we use the Bogoliubov matrices $\boldsymbol{A}(t),\boldsymbol{B}(t)$ for time-dependent symplectic matrices $\boldsymbol{S}(t)$.
	
	Finally, any quadratic Hamiltonian $\hat{H}$ can be put in a matrix form $\boldsymbol{H}$ by the following
	\begin{align*}
		\hat{H}=\frac{\hbar}{2}\hat{\mathbb{X}}^\dag\,\boldsymbol{H}\,\hat{\mathbb{X}}, \quad\quad 
		\boldsymbol{H}
		=
		\begin{pmatrix}
			\boldsymbol{U} & \boldsymbol{V} \\
			\boldsymbol{V}^* & \boldsymbol{U}^*
		\end{pmatrix}.
	\end{align*}
	Here $\boldsymbol{U}=\boldsymbol{U}^\dag$ and $\boldsymbol{V}=\boldsymbol{V}^\text{Tp}$.
	
	In this language the time evolution operator $\hat{U}(t)$ enjoys the \textit{symplectic representation} $\boldsymbol{S}(t)$ of the form{\small
	\begin{align}\label{arbitrary:symplectic:matrix}
		\boldsymbol{S}(t)=\overset{\leftarrow}{\mathcal{T}}\,\exp\left[{\int_0^t\,\dd t'\,\boldsymbol{\Omega}\,\boldsymbol{H}(t')}\right]=
		\begin{pmatrix}
			\boldsymbol{A}(t) & \boldsymbol{B}(t) \\
			\boldsymbol{B}^*(t) & \boldsymbol{A}^*(t) 
		\end{pmatrix}.
	\end{align}}
	These tools are key to obtaining our results.
	
	\subsection{Covariance matrix formalism}
	Gaussian states are prominent across many areas of physics \cite{Mandel:Wolf:1995}. In conjunction with the techniques introduced above, they allow for a complete description and characterisation of the whole physical system using the covariance matrix formalism \cite{Adesso:Ragy:2014}. 
	
	A Gaussian state $\hat{\rho}_\text{G}$ of $N$ bosonic modes in the covariance matrix formalism is \textit{fully} characterised by the $2N$-dimensional vector of first moments $\underline{d}$ and the $2N\times 2N$ covariance matrix of second moments $\boldsymbol{\sigma}$ defined by their elements $d_n:=\text{Tr}( \hat{X}_n \hat{\rho}_\text{G})$ and $\sigma_{nm}:=\text{Tr}(\{\hat{X}_n,\hat{X}_m^\dag\}\hat{\rho}_\text{G})-2\,\text{Tr}(\hat{X}_n\hat{\rho}_\text{G})\text{Tr}(\hat{X}_m^\dag\hat{\rho}_\text{G})$. Here, $\{\cdot,\cdot\}$ is the anticommutator.
	
	If a unitary transformation $\hat{U}(t)$ induces linear dynamics of an initial Gaussian state $\hat{\rho}_\text{G}(0)$, then it is represented by a symplectic matrix $\boldsymbol{S}(t)$, and the von Neumann equation $\hat{\rho}_\text{G}(t)=\hat{U}(t)\,\hat{\rho}_\text{G}(0)\,\hat{U}^\dag(t)$ takes the form
	\begin{align}
		\boldsymbol{\sigma}(t) = &\boldsymbol{S}(t)\,\boldsymbol{\sigma}(0)\,\boldsymbol{S}^\dag(t).
	\end{align}
	This equation must be supplemented with the transformation $\underline{d}(t)=\boldsymbol{S}(t)\,\underline{d}(0)$ of the first moments.
	
	Williamson's theorem guarantees that any $2\,N\times2\,N$, positive-definite Hermitian (or symmetric) matrix can be put in diagonal form by means of a symplectic transformation \cite{Williamson:1923}. Concretely, if the matrix under consideration is a covariance matrix $\boldsymbol{\sigma}$, then $\boldsymbol{\sigma} = \boldsymbol{s}_0\,\boldsymbol{\nu}_\oplus\,\boldsymbol{s}_0^\dag$ where $\boldsymbol{s}_0$ is an appropriate symplectic matrix. The diagonal matrix $\boldsymbol{\nu}_\oplus$ is called the \textit{Williamson form} of the covariance matrix $\boldsymbol{\sigma}$ and has the expression $\boldsymbol{\nu}_\oplus=\text{diag}(\nu_1,...,\nu_N,\nu_1,...,\nu_N)$, where $\nu_n$ are called the \textit{symplectic eigenvalues} of $\boldsymbol{\sigma}$ and are computed as the (absolute value of the) eigenvalues of $i\,\boldsymbol{\Omega}\,\boldsymbol{\sigma}$. 
	Tracing over in this formalism requires merely the deletion of the columns and rows of the covariance matrix that correspond to the undesired subsystems. It is also possible to show that every one-mode reduced state of a Gaussian state is locally thermal, i.e., thermal up to local unitary transformations with expression $\boldsymbol{\sigma}^{(\text{red})}_n=\nu_n\mathds{1}_2$. This tells us that the general expression for the symplectic eigenvalues is $\nu_n=\coth\bigl(\frac{\hbar\,\omega_n}{2\,k_\text{B}\,T_n}\bigr)$, where $T_n$ is the local temperature of each subsystem. Clearly, when $T_n=0$ for all $n$ one has $\boldsymbol{\nu}_\oplus\equiv\mathds{1}_{2N}$ and the state is pure \cite{Adesso:Ragy:2014}.

	\subsection{Considerations on the magnitude of the coupling and its limitations}
	The Hamiltonian \eqref{basic:hamiltonian} models two coupled oscillators for arbitrary values of the coupling constants $g_\text{bs}$ and $g_\text{sq}$. We can now use the formalism explained above to quickly assess which are the limits of our model.
	The Hamiltonian matrix $\boldsymbol{H}$ reads
	\begin{align*}
		\boldsymbol{H}
		=
		\begin{pmatrix}
			\omega_\text{a} & g_\text{bs} & 0 & g_\text{sq}\\
			g_\text{bs} & \omega_\text{b} & g_\text{sq} & 0\\
			0 & g_\text{sq} & \omega_\text{a} & g_\text{bs}\\
			g_\text{ ssq} & 0 & g_\text{bs} & \omega_\text{b}
		\end{pmatrix}.
	\end{align*}
	This Hamiltonian can be used to model well-defined harmonic oscillators if it is bounded from below and the frequencies of the normal modes are real. We compute the symplectic eigenvalues $\kappa_\pm$ of the Hamiltonian matrix and find the expressions
	\begin{align}\label{normal:mode:frequencies:strong:coupling:general}
		\kappa_\pm^2=&\frac{1}{2}\,\left[(\omega_\text{a}^2+\omega_\text{b}^2)+2(g_\text{bs}^2-g_\text{sq}^2)\pm\Gamma\right],
	\end{align}
	where $\Gamma^2:=(\omega_\text{a}^2-\omega_\text{b}^2)^2+8\,\omega_\text{a}\,\omega_\text{b}\,(g_\text{bs}^2+g_\text{sq}^2)+4\,(\omega_\text{a}^2+\omega_\text{b}^2)\,(g_\text{bs}^2-g_\text{sq}^2)$.
	The full derivation is presented and discussed in the Appendix.
	
	The symplectic eigenvalues $\kappa_\pm$ of the Hamiltonian are nothing more than the eigenfrequencies of the normal modes \cite{Abdalla:1994,Mukhopadhyay:2018,Bruschi:Paraoanu:2021}. Therefore, we expect them to be real and positive if the system is to be used to model harmonic oscillators. This implies that the quantities that appear in the Hamiltonian must satisfy the constraint
	\begin{align}\label{Constraint:normal:modes}
		\omega_\text{a}^2\omega_\text{b}^2> 2\omega_\text{a}\omega_\text{b}(g_\text{bs}^2+g_\text{sq}^2)-(g_\text{bs}^2-g_\text{sq}^2)^2.
	\end{align}
	
	We now note that it is of particular interest to study the case where $g_\text{bs}=g_\text{sq}\equiv g$, where the coupling is of the position-position form $\hat{q}_{\text{a}}\hat{q}_{\text{b}}$. In this case, it is easy to show the constraint above becomes $g<g_\text{cr}=\sqrt{\omega_\text{a}\omega_\text{b}}/2 $, while when $g_\text{bs}=0$ it follows that the constraint becomes $g\equiv g_\text{sq}<g_\text{cr}=\sqrt{\omega_\text{a}\omega_\text{b}}$. 
	The latter constraint is the same in the reverse case when $g_\text{sq}=0$, as can be expected since \eqref{Constraint:normal:modes} is symmetric in the exchange of $g_\text{bs}$ and $g_\text{sq}$.

	\subsection{Comparison of quantum states}
\begin{figure}[t!]
		\includegraphics[width=1.0\linewidth]{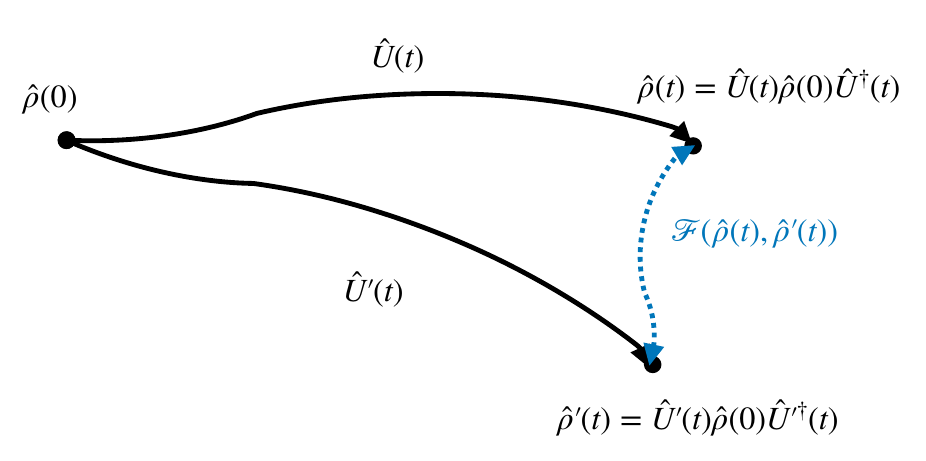}
		\caption{Schematic illustration of the comparison of two states obtained by unitary evolution of the \textit{same} initial state with two \textit{different} time-evolution operators.}\label{figure:two}
	\end{figure}

	We wish to compare two quantum states in a meaningful way. In particular, we are interested in comparing two states that are obtained by evolving the \textit{same} initial state via two time-evolution operators induced by two different Hamiltonians. 
	For an intuitive representation of the process of interest see Figure~\ref{figure:two}. We introduce here the quantities that will be used for our purposes.
	
	\subsubsection{Fidelity of arbitrary quantum states}
	A standard way of comparing two quantum states $\hat{\rho}$ and $\hat{\rho}'$ is to employ the \textit{fidelity} $\mathcal{F}$, which is defined as
	\begin{align}\label{quantum:state:fidelity}
		\mathcal{F}(\hat{\rho},\hat{\rho}'):=\left(\text{Tr}\left[\sqrt{\sqrt{\hat{\rho}}\,\hat{\rho}'\,\sqrt{\hat{\rho}}}\right]\right)^2
	\end{align}
	for arbitrary states \cite{quantuminfo}.
	For pure states $\hat{\rho}=|\psi\rangle\langle\psi|$ and $\hat{\rho}'=|\psi'\rangle\langle\psi'|$ the fidelity reduces to $\mathcal{F}(\hat{\rho},\hat{\rho}')=|\langle\psi|\psi'\rangle|^2$.

	\subsubsection{Fidelity of two-mode Gaussian states}
	In the case of a pair $\boldsymbol{\sigma}$ and $\boldsymbol{\sigma}'$ of two-mode Gaussian states \emph{with vanishing first moments}, which is of particular interest in this work, 
	explicit expressions for \eqref{quantum:state:fidelity} can been obtained \cite{Paraoanu:Scutaru:2000,Marian:Marian:2012}. The fidelity $\mathcal{F}(\boldsymbol{\sigma},\boldsymbol{\sigma}')$ for this case reads
	\begin{align*}
		\mathcal{F}(\boldsymbol{\sigma},\boldsymbol{\sigma}')=\frac{4}{\sqrt{\Lambda}+\sqrt{\Gamma}-\sqrt{(\sqrt{\Lambda}-\sqrt{\Gamma})^2-\Delta}},
	\end{align*}
	where we need the following quantities: $\Gamma:=\det(\mathds{1}_4-\boldsymbol{\Omega}\,\boldsymbol{\sigma}\,\boldsymbol{\Omega}\,\boldsymbol{\sigma}')$, $\Lambda:=\det(\mathds{1}_4+i\,\boldsymbol{\Omega}\,\boldsymbol{\sigma})\,\det(\mathds{1}_4+i\,\boldsymbol{\Omega}\,\boldsymbol{\sigma}')$ and $\Delta:=\det(\boldsymbol{\sigma}+\boldsymbol{\sigma}')$, see \cite{Marian:Marian:2012}.

	\section{Controlling the rotating wave approximation}\label{Sec:Demonstrating:RWA}
	The \textit{rotating wave approximation} is the claim that the dynamics induced by the Hamiltonian \eqref{basic:hamiltonian} for $g_\text{bs}=g_\text{sq}\equiv g$ is well approximated by the dynamics induced by the rotating wave Hamiltonian
	\begin{align}\label{RWA:hamiltonian}
		\hat{H}_\text{RWA}=\hat{H}_0+\hbar\,g \left(\hat{a}\,\hat{b}^\dag+\hat{a}^\dag\,\hat{b}\right),
	\end{align}
	at least for identical frequencies $\omega_{\text{a}}=\omega_{\text{b}}\equiv \omega$ and low coupling $g/\omega\ll1$, see \cite{Burgarth:Facchi:2023}. In other words, the Hamiltonians $\hat{H}$ and $\hat{H}_\text{RWA}$ induce indistinguishable (in the sense of expectation values of any observable) time evolutions for a certain amount of time. This statement relies on the fact that, in the interaction picture, the terms driven by $g_\text{sq}$ rotate with frequency $\Omega=\omega_\text{a}+\omega_\text{b}$, while the term driven by $g_\text{bs}$ rotate only with frequency $\Omega=\omega_\text{a}-\omega_\text{b}$. The rapid oscillations $\Omega=\omega_\text{a}+\omega_\text{b}$ are assumed to ``average out'' after a sufficiently long time, and therefore the contributions of the \textit{counterrotating} term $g \bigl(\hat{a}^\dag\,\hat{b}^\dag+\hat{a}\,\hat{b}\bigr)$ can be safely ignored. 
	
	We proceed to achieve our goal using the method developed in \cite{Burgarth:Facchi:2023}. This allows us
	to consider arbitrary states. The concrete task here is to compare the time evolution
	induced by the full Hamiltonian \eqref{basic:hamiltonian} with that induced by the rotating wave approximation Hamiltonian \eqref{RWA:hamiltonian}.
	
	Let us denote their unitary groups as $\hat{U}(t)=e^{-\frac{i}{\hbar}\hat{Ht}}$
	and $\hat{U}_{\text{RWA}}(t)=e^{-\frac{i}{\hbar}\hat{H}_{\text{RWA}}t}$ respectively. For simplicity we work on resonance, where $\omega_{\text{a}}=\omega_{\text{b}}\equiv \omega$, we assume equal coupling $g_\text{bs}=g_\text{sq}\equiv g$ for convenience, and
	transform the Hamiltonians into the rotating frame with respect to
	$H_{0}=\hbar\omega\left(b^{\dagger}b+a^{\dagger}a\right).$ This leads
	to the two operators
	\begin{align*}
		\hat{H}_{1}(t)&=\hbar g(\hat{a}\hat{b}^{\dagger}+\hat{a}^{\dagger}\hat{b})+\hbar g(e^{2i\omega t}\hat{a}^{\dagger}\hat{b}^{\dagger}+e^{-2i\omega t}\hat{a}\hat{b})\\
		\hat{H}_{2}&=\hbar g(\hat{a}\hat{b}^{\dagger}+\hat{a}^{\dagger}\hat{b}).
	\end{align*}
	Let us denote the induced time-evolution operators as $\hat{U}_{j}(t)$ with $j=1,2$. We note that, as also discussed in the literature \cite{Burgarth:Facchi:2023}, $\hat{H}_{2}$ is time-independent
	and, since it is a quadratic Hamiltonian, its dynamics is easily seen to leave invariant the Schwartz space of rapidly decreasing functions. Therefore most calculations remain essentially
	the same as those performed in previous work. We introduce the key quantity $\hat{S}_{21}(t) :=\int_{0}^{t}ds(\hat{H}_{2}-\hat{H}_{1}(s))$, which reads
	\begin{align*}
		\hat{S}_{21}(t) =-\frac{\hbar g\sin(\omega t)}{\omega}(e^{i\omega t}\hat{a}^{\dagger}\hat{b}^{\dagger}+e^{-i\omega t}\hat{a}\hat{b}).
	\end{align*}
	We continue with the analysis, and introduce the quantities $\Delta\hat{U}(t):=\hat{U}_{2}(t)-\hat{U}_{1}(t)$ and $\hat{K}(t):=[\hat{S}_{21}(t)\hat{H}_{2}-\hat{H}_{1}(t)\hat{S}_{21}(t)]\hat{U}_{2}(t)$ for convenience of presentation. Then, for all pure states $|\psi\rangle\in\mathcal{S}(\mathbb{R}^{2})$
	we have that
	{\small
		\begin{align*}
			i\hbar\Delta\hat{U}(t)|\psi\rangle= & \hat{S}_{21}(t)\hat{U}_{2}(t)|\psi\rangle+\frac{i}{\hbar}\hat{U}_{1}(t)\int_{0}^{t} \dd s\,\hat{U}_{1}^{\dagger}(s)\hat{K}(s) |\psi\rangle.
		\end{align*}
	}
	To obtain useful bounds, we focus on a Fock state $|n_{\text{a}},n_{\text{b}}\rangle$. The reason why restricting to Fock states only is enough for our purposes will become clear later. We
	can obtain an explicit bound by applying the triangle inequality to
	the above identity, and find{
	\begin{align*}
		\left\Vert \Delta\hat{U}(t)|n_\text{a},n_\text{b}\rangle\right\Vert \le&\;\frac{ g}{\omega} \left\Vert \hat{a}^{\dagger}\hat{b}^{\dagger}\,\hat{U}_{2}(t)|n_\text{a},n_\text{b}\rangle\right\Vert \\
        &\;+\frac{ g}{\omega}\left\Vert \hat{a}\hat{b}\,\hat{U}_{2}(t)|n_\text{a},n_\text{b}\rangle\right\Vert \\
        &\;+\frac{1}{\hbar^2}\int_{0}^{t}\dd s\,\left\Vert \hat{K}(s)|n_\text{a},n_\text{b}\rangle\right\Vert .
	\end{align*}}
	For the last term, we have by the triangle inequality
	\begin{align*}
		\left\Vert \hat{K}(s)|n_\text{a},n_\text{b}\rangle\right\Vert \le & \;\hbar\frac{ g}{\omega}\left\Vert \hat{a}^{\dagger}\hat{b}^{\dagger}\hat{H}_{2}\hat{U}_{2}(s)|n_\text{a},n_\text{b}\rangle\right\Vert\\
        &\;+\hbar\frac{ g}{\omega}\left\Vert \hat{a}\hat{b}\hat{H}_{2}\hat{U}_{2}(s)|n_\text{a},n_\text{b}\rangle\right\Vert \\
		&\;+\hbar\frac{ g}{\omega}\left\Vert \hat{H}_{1}(s)\hat{a}^{\dagger}\hat{b}^{\dagger}\hat{U}_{2}(s)|n_\text{a},n_\text{b}\rangle\right\Vert\\
        &\;+\hbar\frac{ g}{\omega}\left\Vert \hat{H}_{1}(s)\hat{a}\hat{b}\hat{U}_{2}(s)|n_\text{a},n_\text{b}\rangle\right\Vert .
	\end{align*}
	If we were now to proceed and use the explicit expression for the Hamiltonians $\hat{H}_{1}(t)$ and $\hat{H}_{2}$ in the bound for $||\hat{K}(s)|n_\text{a},n_\text{b}\rangle||$ presented above we would obtain fourth order polynomials in the
	annihilation and creation operators that act on the states $\hat{U}_{2}(s)|n_\text{a},n_\text{b}\rangle$. Since $\hat{H}_{2}$ conserves the total number of excitations $N_{\text{ab}}:=n_\text{a}+n_\text{b}$ present in the Fock state, we have that $\hat{U}_{2}(s)|n_\text{a},n_\text{b}\rangle=\sum_{p=0}^{N_{\text{ab}}} C_p(s)|N_{\text{ab}}-p,p\rangle$ is a superposition of $N_{\text{ab}}+1$ states with fixed total number $N_{\text{ab}}$ for appropriate coefficients $C_p(s)$, which are irrelevant for our purposes.
	For each term, we can loosely bound the action of the fourth order
	polynomials by a worst-case choice
	\begin{align*}
		\left\Vert \left(a^{\dagger}\right)^{4}|N_{\text{ab}},0\rangle\right\Vert \le\left(\sqrt{N_{\text{ab}}+4}\right)^{4}.
	\end{align*}
	We can therefore loosely bound
	\begin{align*}
		\left\Vert \hat{K}(t)|n_{a,}n_\text{b}\rangle\right\Vert \le 12\text{\ensuremath{\frac{\hbar^2g^2}{\omega}\left(N_{\text{ab}}+1\right)\left(N_{\text{ab}}+4\right)^{2}},}
	\end{align*}
	where the coefficient $12$ comes from the number of polynomials after
	expanding the Hamiltonians. Combining all together, we obtain
        \begin{align*}
		&\left\Vert \Delta\hat{U}(t)|n_\text{a},n_\text{b}\rangle\right\Vert \\
            &\quad\quad\le2\frac{ g}{\omega}\left(N_{\text{ab}}+4\right) \left(1+6 g\,t\left(N_{\text{ab}}+1\right)\left(N_{\text{ab}}+4\right)\right),
	\end{align*}
	where we applied the same technique on the second order polynomials.
	
	Defining $Z:=|| (\hat{U}(t)-\hat{U}_{\text{RWA}}(t))|n_\text{a},n_\text{b}\rangle||$ for convenience of presentation we finally have that
        \begin{align}
		Z\le2\frac{g}{\omega}\left(N_{\text{ab}}+4\right)\left(1+6\,g\,t\,\left(N_{\text{ab}}+1\right)\left(N_{\text{ab}}+4\right)\right).
	\end{align}
	As shown in the literature \cite{Burgarth:Facchi:2023}, the convergence
	speed depends explicitly on the excitation number and on the total evolution
	time. We expect that tighter bounds with lower order scaling in the
	excitation number can be found by a more careful analysis  \cite{Burgarth:Facchi:2023}.
	Lower bounds obtained in previous works show that the rotating wave approximation converges only strongly in the
	quantum Rabi model \cite{Rabi:1936,Xie:Zhong:2017}. Here, we expect the same. Regardless, using the density
	of Fock space in the full Hilbert space allows us to obtain, for any state $|\phi\rangle\in\mathcal{H}$,
	the following limit
	\begin{align}
		\lim_{\frac{g}{\omega}\rightarrow0}\left\Vert \left(\hat{U}(t)-\hat{U}_{\text{RWA}}(t)\right)|\phi\rangle\right\Vert =0.
	\end{align}
	This proves the rotating wave approximation for arbitrary states.

\section{Rotating wave approximation for Gaussian states via the fidelity}\label{Sec:Distinguishing:two:Gaussian:States}
	We have proven the rotating wave approximation for coupled quantum harmonic oscillators. Our proof relies on bounding the error and showing that the two evolutions converge.
	While this proof is rigorous, it does not provide full analytical control over the interplay between the two dynamics of interest.
	To overcome this problem, we restrict ourselves to Gaussian states. The price we pay by limiting the scope of applicability is greatly compensated by allowing for full analytical results.

	\subsection{Gaussian-state fidelity}
\begin{figure}[t!]
		\includegraphics[width=1.0\linewidth]{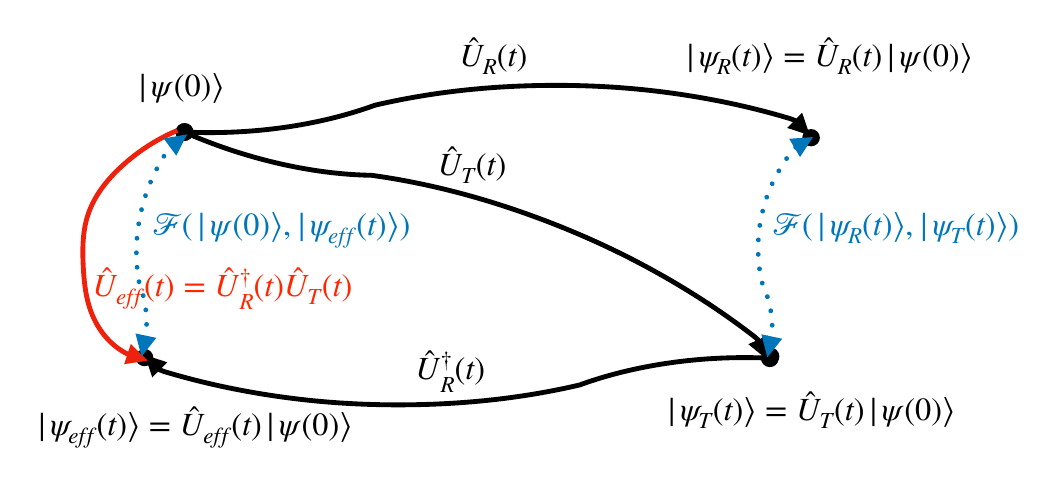}
		\caption{Schematic illustration of the state comparison via the fidelity. The two blue dotted lines stand to indicate that the fidelities coincide.}\label{figure:three}
	\end{figure}

	We now proceed to \textit{compute the distance between two (pure Gaussian) states obtained by evolving the same initial state with two different Hamitlonians}.
	To achieve our goal we first introduce a \textit{reference} state $|\psi_{\text{R}}(t)\rangle:=\hat{U}_{\text{R}}(t)\,|\psi(0)\rangle$ and a \textit{target} state $|\psi_{\text{T}}(t)\rangle:=\hat{U}_{\text{T}}(t)\,|\psi(0)\rangle$ that obtained evolving a common initial state $|\psi(0)\rangle$ via two different time-evolution operators $\hat{U}_{\text{R}}(t)$ and $\hat{U}_{\text{T}}(t)$. These evolution operators are induced by the \textit{reference} and \textit{target} Hamiltonians $\hat{H}_{\text{R}}(t)$ and $\hat{H}_{\text{T}}(t)$ respectively. We can then make use of the fidelity $\mathcal{F}(|\psi_{\text{R}}(t)\rangle,|\psi_{\text{T}}(t)\rangle)=|\langle\psi(0)|\hat{U}_{\text{R}}^{\dag}(t)\,\hat{U}_{\text{T}}(t)\,|\psi(0)\rangle|^2$ to quantify the distinguishability of the reference and target state. It is immediate to use the properties of the fidelity and verify that
	{\small
		\begin{align}\label{initial:fidelity:result}
			\mathcal{F}(|\psi_{\text{R}}(t)\rangle,|\psi_{\text{T}}(t)\rangle)=&\mathcal{F}(|\psi(0)\rangle,\hat{U}_{\text{R}}^{\dag}(t)\,\hat{U}_{\text{T}}(t)|\psi(0)\rangle).
		\end{align}
	}
	The crucial observation at this point is that the fidelity \eqref{initial:fidelity:result} is equivalent to the one obtained between the state $|\psi(0)\rangle$ and the state $|\psi_{\text{eff}}(t)\rangle:=\hat{U}_{\text{eff}}(t)|\psi(0)\rangle$, where we have introduced $\hat{U}_{\text{eff}}(t):=\hat{U}_{\text{R}}^{\dag}(t)\,\hat{U}_{\text{T}}(t)$. 
	A pictorial representation of the scheme can be found in Figure~\ref{figure:three}.
	
\begin{figure}[t!]
		\includegraphics[width=1.0\linewidth]{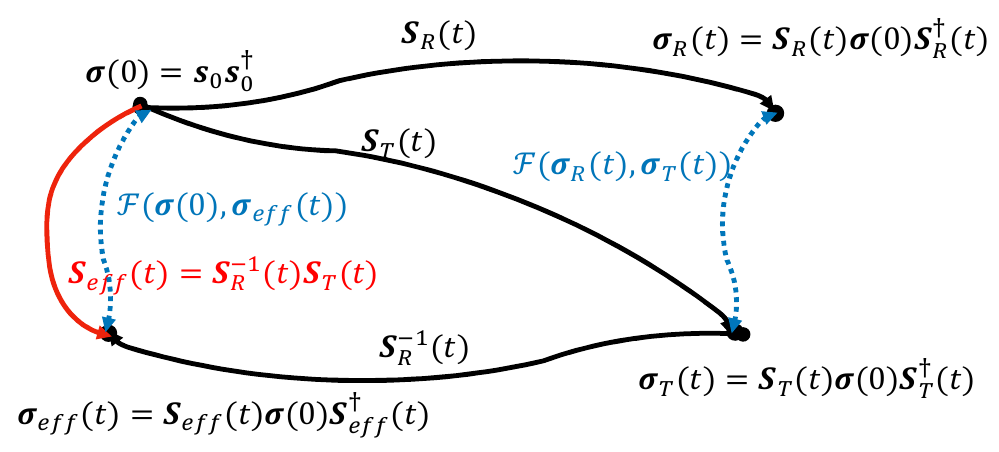}
		\caption{Schematic illustration of the equivalent situation depicted in Figure~\ref{figure:three} applied to the covariance matrix formalism.}\label{figure:four}
	\end{figure}

	Computing expression \eqref{initial:fidelity:result} is difficult for arbitrary initial states $|\psi(0)\rangle$. 
	As mentioned above, we restrict ourselves to the analysis of pure centred Gaussian states, i.e., pure Gaussian states with vanishing first moments. 
	This implies that the initial state with covariance matrix $\boldsymbol{\sigma}(0)$ can be decomposed as $\boldsymbol{\sigma}(0)=\boldsymbol{s}_0\,\boldsymbol{s}_0^\dag$ for an appropriate symplectic matrix $\boldsymbol{s}_0$. We follow the considerations just presented above and conclude that we need to compute the fidelity $\mathcal{F}_{\text{eff}}(t):=\mathcal{F}(\boldsymbol{\sigma}(0),\boldsymbol{\sigma}_{\text{eff}}(t))$ between the state $\boldsymbol{\sigma}(0)$ and the state $\boldsymbol{\sigma}_{\text{eff}}(t):=\boldsymbol{S}_\text{eff}(t)\boldsymbol{\sigma}(0)\,\boldsymbol{S}_\text{eff}^\dag(t)$, where $\boldsymbol{S}_\text{eff}(t):=\boldsymbol{S}^{-1}_\text{R}(t)\,\boldsymbol{S}_\text{T}(t)$. A pictorial representation of the the situation for the covariance matrix formalism can be found in Figure~\ref{figure:four}.
	
	When two states $\boldsymbol{\sigma}$ and $\boldsymbol{\sigma}'$ are pure, such as those that we consider, we are able to show that $\Lambda=0$ and $\Gamma=\Delta$. Explicit computations can be found in~\ref{Gaussian:fidelity:appendix}. Therefore, $\mathcal{F}(\boldsymbol{\sigma},\boldsymbol{\sigma}')$ reduces to
	\begin{align}\label{fidelity:pure:Gaussian:states}
		\mathcal{F}(\boldsymbol{\sigma},\boldsymbol{\sigma}')=4/\sqrt{\Gamma}.
	\end{align}
	We stress that this is a general result that depends only on the purity of the two states under consideration.
	
	This greatly simplified expression allows us to find in our particular case{\small
	\begin{align}\label{distance:fidelity:final:general}
		\mathcal{F}_{\text{eff}}(t)=\left|\det(\boldsymbol{A}_\text{f}(t))\right|^{-1}=(\textrm{det}(\mathds{1}_2+\boldsymbol{B}^\dag_{\text{f}}(t)\boldsymbol{B}_{\text{f}}(t)))^{-1/2},
	\end{align}}
	where we have introduced the \textit{final Bogoliubov matrix}
	{\small
		\begin{align}\label{B:f:Bogoliubov:coefficient:general}
			\boldsymbol{B}_{\text{f}}(t)=&\boldsymbol{\alpha}_0^\dag \boldsymbol{A}^\dag_\text{R}(t)\boldsymbol{A}_\text{T}(t)\boldsymbol{\beta}_0+\boldsymbol{\alpha}_0^\dag \boldsymbol{A}^\dag_\text{R}(t)\boldsymbol{B}_\text{T}(t)\boldsymbol{\alpha}^*_0\nonumber\\
			&-\boldsymbol{\beta}_0^\text{Tp} \boldsymbol{A}^\text{Tp}_\text{R}(t)\boldsymbol{B}^*_\text{T}(t)\boldsymbol{\beta}_0-\boldsymbol{\beta}_0^\text{Tp} \boldsymbol{A}^\text{Tp}_\text{R}(t)\boldsymbol{A}^*_\text{R}(t)\boldsymbol{\alpha}^*_0\nonumber\\
			&-\boldsymbol{\alpha}_0^\dag \boldsymbol{B}^\text{Tp}_\text{R}(t)\boldsymbol{B}^*_\text{T}(t)\boldsymbol{\beta}_0-\boldsymbol{\alpha}_0^\dag \boldsymbol{B}^\text{Tp}_\text{R}(t)\boldsymbol{A}^*_\text{T}(t)\boldsymbol{\alpha}^*_0\nonumber\\
			&+\boldsymbol{\beta}_0^\text{Tp} \boldsymbol{B}_\text{R}^\dag(t)\boldsymbol{A}_\text{T}(t)\boldsymbol{\beta}_0+\boldsymbol{\beta}_0^\text{Tp} \boldsymbol{B}^\dag_\text{R}(t)\boldsymbol{B}_\text{T}(t)\boldsymbol{\alpha}^*_0
		\end{align}
	}
	associated to the symplectic matrix $\boldsymbol{S}_\text{f}(t):=\boldsymbol{s}_0^{-1}\boldsymbol{S}_\text{eff}(t)\boldsymbol{s}_0=-\boldsymbol{\Omega}\boldsymbol{s}_0^\dag\boldsymbol{S}^\dag_\text{R}(t)\boldsymbol{\Omega}\boldsymbol{S}_\text{T}(t)\boldsymbol{s}_0$.
	The expressions \ref{distance:fidelity:final:general} and \ref{B:f:Bogoliubov:coefficient:general} constitute main results of this section.
	
	\textit{Gaussian-state fidelity: passive transformations}---We can apply this result to the case where $\boldsymbol{S}_\text{R}(t)$ represents a \textit{passive} transformation, i.e., one that does not alter the total particle content of the system. This is the case of the dynamics in the rotating wave approximation. In such case, one has that $\boldsymbol{B}_\text{R}(t)=0$ and that $\boldsymbol{S}_\text{R}(t)=\tilde{\boldsymbol{S}}_\text{R}(t)\oplus\tilde{\boldsymbol{S}}^*_\text{R}(t)$, where $\tilde{\boldsymbol{S}}_\text{R}(t)\tilde{\boldsymbol{S}}_\text{R}^\dag(t)=\mathds{1}_2$.
	
	We therefore obtain the same formal expression \eqref{distance:fidelity:final:general} for the fidelity of the two states, where now the Bogoliubov matrix $\boldsymbol{B}_{\text{f}}(t)$ reads
	{\small
		\begin{align}\label{B:f:Bogoliubov:coefficient}
			\boldsymbol{B}_{\text{f}}(t)&=\boldsymbol{\alpha}_0^\dag \tilde{\boldsymbol{S}}^\dag_\text{R}(t)\boldsymbol{A}(t)\boldsymbol{\beta}_0+\boldsymbol{\alpha}_0^\dag \tilde{\boldsymbol{S}}^\dag_\text{R}(t)\boldsymbol{B}(t)\boldsymbol{\alpha}^*_0\nonumber\\
    	   &\quad-\boldsymbol{\beta}_0^\text{Tp} \tilde{\boldsymbol{S}}^\text{Tp}_\text{R}(t)\boldsymbol{B}^*(t)\boldsymbol{\beta}_0-\boldsymbol{\beta}_0^\text{Tp} \tilde{\boldsymbol{S}}^\text{Tp}_\text{R}(t)\boldsymbol{A}^*(t)\boldsymbol{\alpha}^*_0.
		\end{align}
	}

	\subsection{Fidelity via the Bloch-Messiah decomposition}
	We now note that the Bloch-Messiah theorem guarantees that any two Bogoliubov matrices $\boldsymbol{A}(t)$ and $\boldsymbol{B}(t)$ can be decomposed as $\boldsymbol{A}(t)=\boldsymbol{U}(t)\,\boldsymbol{A}_\text{D}(t)\,\boldsymbol{V}^\dag(t)$ and $\boldsymbol{B}(t)=\boldsymbol{U}(t)\,\boldsymbol{B}_\text{D}(t)\,\boldsymbol{V}^\text{Tp}(t)$, where $\boldsymbol{U}(t)$ and $\boldsymbol{V}(t)$ are appropriate unitary matrices and $\boldsymbol{A}_\text{D}(t)$ and $\boldsymbol{B}_\text{D}(t)$ are nonnegative diagonal matrices satisfying $|\boldsymbol{A}_\text{D}(t)|^2-|\boldsymbol{B}_\text{D}(t)|^2=\mathds{1}_2$, see \cite{Braunstein:2005}. Here, the notation $|\boldsymbol{M}|$ refers to taking the absolute value for each entry of the matrix $\boldsymbol{M}$. This implies that $\boldsymbol{A}_\text{D}(t)=\text{diag}(\cosh r_+(t),\cosh r_-(t))$ and $\boldsymbol{B}_\text{D}(t)=\text{diag}(\sinh r_+(t),\sinh r_-(t))$ modulo irrelevant phases, where the time-dependent squeezing parameters $r_\pm(t)$ need to be determined. Therefore
	\begin{align}
		\mathcal{F}_{\text{eff}}(t)=(\cosh r_+(t)\,\cosh r_-(t))^{-1},
	\end{align}
	which is clearly smaller or equal to one.
	This decomposition can therefore be applied to the Bogoliubov coefficient $\boldsymbol{B}_\text{f}$ obtained in \eqref{B:f:Bogoliubov:coefficient:general}.

	\subsection{Controlling the rotating wave approximation for initial Gaussian-states}\label{Sec:Quantifying:The:Viability:od:the:RWA}
	The main result \eqref{distance:fidelity:final:general} quantifies the distance between the two states of interest for arbitrary couplings. Therefore, we wish to demonstrate that the rotating wave approximation is recovered in the limit considered above, and therefore holds for Gaussian states.
	
	To achieve our goal we consider the expression \eqref{distance:fidelity:final:general} supplemented by the matrix \eqref{B:f:Bogoliubov:coefficient}, where $\tilde{\boldsymbol{S}}_\text{R}(t)\equiv\tilde{\boldsymbol{S}}_\text{RWA}(t)$ is obtained from the full symplectic representation of the rotating wave dynamics $\boldsymbol{S}_\text{RWA}(t)=\tilde{\boldsymbol{S}}_\text{RWA}(t)\oplus\tilde{\boldsymbol{S}}_\text{RWA}^*(t)$, and $\tilde{\boldsymbol{S}}_\text{RWA}(t)$ is a $2\times2$ unitary matrix. An explicit expression for the dynamics induced by $\boldsymbol{S}_\text{RWA}(t)$, or equivalently $\tilde{\boldsymbol{S}}_\text{RWA}(t)$, has been given in~\ref{time:evolution:rwa:appendix}. One way to understand why this is the case is to recall that passive transformations can be immediately obtained from the general coupling case by setting $\boldsymbol{\beta}=0$, thus noting that $\boldsymbol{\alpha}$ must be a unitary (or orthogonal) matrix due to Bogoliubov identities. 
	In this case one has $\boldsymbol{B}(t)=0$ and $\boldsymbol{A}(t)\equiv\tilde{\boldsymbol{S}}_\text{RWA}(t)$.
	
	We now apply the following conditions: $\omega\equiv\omega_{\text{a}}=\omega_{\text{b}}$ and $\tilde{g}:=g/\omega\ll1$, and we have conveniently introduced the dimensionless time $\tau:=\omega t$. In the following we also assume that $\tilde{g}\tau=$ const. with $\tilde{g}^2\tau\ll1$. In~\ref{verifying:Validity:RWA:appendix} we show that, to zero order in $\tilde{g}$, we have $\boldsymbol{B}_{\text{f}}(\tau)=0$, which allows us to write $\boldsymbol{B}_{\text{f}}(\tau)=\boldsymbol{B}^{(1)}_{\text{f}}(\tau)\tilde{g}+\mathcal{O}(\tilde{g}^2)$. The fidelity, to lowest order, will consequently read
	\begin{align}\label{RWA:demonstration:expression:general}
		\mathcal{F}_{\text{eff}}(\tau)=1-\frac{1}{2}\Tr(\boldsymbol{B}^{(1)\dag}_{\text{f}}(\tau)\boldsymbol{B}^{(1)}_{\text{f}}(\tau))\tilde{g}^2.
	\end{align}
	This expression allows us to conclude that $\mathcal{F}_\text{eff}(\tau)\rightarrow1$ for $\tilde{g}\rightarrow0$, which definitely proves the validity of the rotating wave approximation as stated. This result applies independently of the specific expressions, however care needs to be taken depending on the choice of the initial state. We provide an explicit example for an initial squeezed state in Section~\ref{subsection:initial:single:mode:squeezed:state}.

	\subsection{Number expectation value}
	We now compute the expectation value of the (total) number of excitations present in the system. We will use this quantity to complement the study of our main result \eqref{distance:fidelity:final:general} and provide additional physical understanding of the rotating wave approximation.
	
	The total number of excitations is given by $\hat{N}_{}:=\hat{a}^\dag\hat{a}+\hat{b}^\dag\hat{b}$ in the Heisenberg picture. Therefore, returning for convenience to the Hilbert-space formulation, we see that the number expectation values read $N_{}(t):=\text{Tr}(\hat{N}_{}\hat{U}(t)\hat{\rho}(0)\hat{U}^\dag(t))$ and $N_\textrm{\text{RWA}}(t):=\text{Tr}(\hat{N}_{}\hat{U}_{\text{RWA}}(t)\hat{\rho}(0)\hat{U}_{\text{RWA}}^\dag(t))$ for the full evolution and the evolution in the rotating wave approximation respectively. Crucially, we note that the total number of excitations is conserved in the rotating wave approximation since $[\hat{N}_{},\hat{H}_{\text{RWA}}]=0$, and therefore $[\hat{N}_{},\hat{U}_{\text{RWA}}(t)]=0$. This reflects the fact that the beam-splitter operation is a \textit{passive operation} that conserves the total number of particles. We define $N_{}(0):=\text{Tr}(\hat{N}_{}\hat{\rho}(0))$ as the number of excitations in the initial state.
	
	We can then  use the number of excitations at time $t$ in the rotating wave approximation case to benchmark those obtained in the full case, as another mean to quantify the deviation of the latter case from the former. Thus, we introduce $\Delta N_{}(t):=N_{}(t)-N_\textrm{\text{RWA}}(t)$.
	We can then use the definition of the quantities above, employ the cyclic property of the trace and obtain
	\begin{align}
		\Delta N_{}(t)=&\text{Tr}(\hat{N}_{}\hat{U}_{\text{eff}}(t)\hat{\rho}(0)\hat{U}_{\text{eff}}^\dag(t))-N_{}(0),
	\end{align}
	where $\hat{U}_{\text{eff}}(t):=\hat{U}^\dag_{\text{RWA}}(t)\,\hat{U}(t)$ was introduced before. To obtain this result we have inserted twice the identity $\hat{U}^\dag_{\text{RWA}}(t)\hat{U}_{\text{RWA}}(t)=\mathds{1}$ and have used the fact that $[\hat{N}_{},\hat{U}_{\text{RWA}}(t)]=0$.
	
	We now return to the symplectic formalism. If $\boldsymbol{S}(t)$ is the symplectic matrix for the full evolution of a two-mode state, it is easy to see that 
	$\Delta N_{}(t)=\frac{1}{4}[\text{Tr}(\boldsymbol{S}(t)\boldsymbol{\sigma}(0)\boldsymbol{S}^\dag(t))-\text{Tr}(\boldsymbol{\sigma}(0))]
	=\frac{1}{4}[\text{Tr}(\boldsymbol{S}^\dag(t)\,\boldsymbol{S}(t)\,\boldsymbol{\sigma}(0))-\text{Tr}(\boldsymbol{\sigma}(0))]$. Following the discussion presented here, we employ the notation used in the previous subsections and we can equivalently write
	{\small
		\begin{align}
			\Delta N_{}(t)=&\frac{1}{4}\left[\text{Tr}\left(\boldsymbol{S}_\text{eff}^\dag(t)\boldsymbol{S}_\text{eff}(t)\boldsymbol{\sigma}(0)\right)-\text{Tr}(\boldsymbol{\sigma}(0))\right].
		\end{align}
	}
	This expression highlights the fact that the total average variation of excitations can, de facto, be obtained by computing the variation of excitations as if the initial state evolved only through the operator $\hat{U}_{\text{eff}}(t)$.
	
	We can provide a closed expression for the variation $\Delta N_{}(t)$ of the average number of particles in terms of Bogoliubov coefficients. We find 
	{\small
		\begin{align}\label{particle:number:variation}
			\Delta N_{}(t)=&\,\Tr\left(\boldsymbol{B}^\dagger(t)\boldsymbol{B}(t)\right)+2\Tr\left(\boldsymbol{B}^\text{Tp}(t)\boldsymbol{B}^*(t)\boldsymbol{\beta}_0\boldsymbol{\beta}_0^\dagger\right)\nonumber\\
			&-2\Re\left(\Tr\left(\boldsymbol{A}(t)\boldsymbol{B}^\textrm{Tp}(t)\boldsymbol{\alpha}_0\boldsymbol{\beta}_0^\textrm{Tp}\right)\right),
		\end{align}
	}
	where we have used the explicit expressions for the decomposition of the initial state $\boldsymbol{\sigma}(0)=\boldsymbol{s}_0\boldsymbol{s}_0^\dag$, and therefore $\boldsymbol{U}_0=\mathds{1}_2+2\boldsymbol{\beta}_0\boldsymbol{\beta}_0^\dag$ and $\boldsymbol{V}_0=2\boldsymbol{\alpha}_0\boldsymbol{\beta}_0^{\text{Tp}}$.

	\subsection{Relation between fidelity and the statistical distribution of the excitation number}
	Here we establish a relation between the fidelity and the statistical distribution of the number of created excitations.
	
\begin{figure}[t!]
		\includegraphics[width=1.0\linewidth]{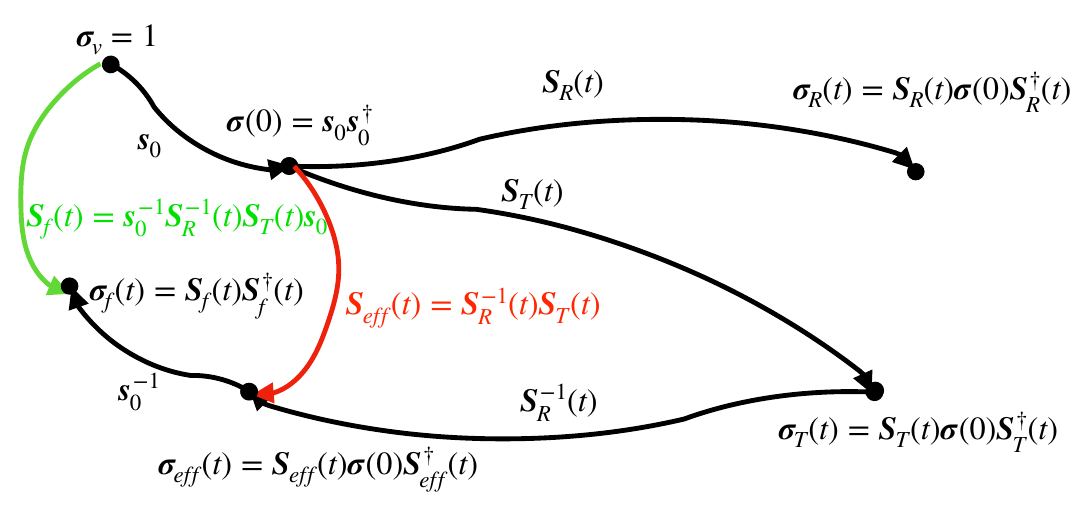}
		\caption{Schematic illustration of the equivalent time evolution due to $\boldsymbol{S}_\text{f}(t)$ starting from the vacuum state $\boldsymbol{\sigma}_v=\mathds{1}_N$.}\label{figure:five}
	\end{figure}

	In this work we have found that the fidelity has the expression $\mathcal{F}^{-2}(t)=\det\bigl(\mathds{1}_2+\boldsymbol{B}^\dag(t)\boldsymbol{B}(t)\bigr)$. This expression can be interpreted as follows: it is the fidelity obtained by comparing the vacuum state $\boldsymbol{\sigma}_0=\mathds{1}_4$ with the state $\boldsymbol{\sigma}(t)=\boldsymbol{S}(t)\boldsymbol{S}^\dag(t)$. For the sake of this subsection only, $\boldsymbol{S}(t)$ is an arbitrary symplectic matrix while $\boldsymbol{B}(t)$ is one of the blocks that defines it, see \eqref{arbitrary:symplectic:matrix}. We can compute the variation of the number expectation value $\Delta N(t)$ of excitations contained in $\boldsymbol{\sigma}(t)$, which reads $\Delta N(t)=\Tr\bigl(\boldsymbol{B}^\dag(t)\boldsymbol{B}(t)\bigr)$. Notice that this coincides with the average number of excitations since the vacuum state contains none. We can then use the expression of the determinant of a $2\times2$ matrix in terms of traces of appropriate matrices, and find
	{\small
		\begin{align}\label{Fidelity:initial:effective:final:state}
			\mathcal{F}^{-2}(t)=&1+\frac{3}{2}\Delta N(t)+\frac{1}{2}(\Delta N(t))^2-\frac{1}{4}\sigma^2_{\Delta N}(t).
		\end{align}
	}
	To arrive at this expression we have obtained the intermediate form $\mathcal{F}^{-2}(t)=1+\Delta N(t)+\frac{1}{2}(\Delta N(t))^2-\frac{1}{2}\Tr\bigl(\boldsymbol{B}^\dag(t)\boldsymbol{B}(t)\boldsymbol{B}^\dag(t)\boldsymbol{B}(t)\bigr)$, which we have re-written using the identity $\Tr\bigl(\boldsymbol{B}^\dag(t)\boldsymbol{B}(t)\boldsymbol{B}^\dag(t)\boldsymbol{B}(t)\bigr)=\frac{1}{2}\bigl[\Delta N^2(t)-2\Delta N(t)-(\Delta N(t))^2\bigr]$ demonstrated in the Appendix and shown in \eqref{useful:expression:quartic:betas:appendix}. The quantity $\Delta N^2(t)$ is the expectation value over the vacuum state of the square of the number operator, and $\sigma^2_{\Delta N}(t):=\Delta N^2(t)-(\Delta N(t))^2$ is the variance of the statistical quantity $\Delta N(t)$.
	A schematic representation of this scenario is given in Figure~\ref{figure:five}.
	
	\subsection{Bounding the fidelity}
	The expression \eqref{distance:fidelity:final:general} for the fidelity is in general difficult to compute analytically.
	We here provide upper and lower bounds that can be computed with relative ease as compared to the full expression.
	We use algebraic properties of the eigenvalues of $2\times2$ positive semidefinite matrices, and in~\ref{sec:bounding:fidelity:appendix} we obtain
	\begin{align}\label{fidelity:bounds}
		\frac{1}{1+\frac{1}{2}\Delta N_{\text{f}}(t)}\leq\mathcal{F}_\text{eff}(t)\leq\frac{1}{\sqrt{1+\Delta N_{\text{f}}(t)}},
	\end{align}
	where $\Delta N_{\text{f}}(t):=\text{Tr}\bigl(\boldsymbol{B}_\text{f}^\dagger(t)\boldsymbol{B}_\text{f}(t)\bigr)$ is the variation of the number of particles from the vacuum state associated to the time evolution matrix $\boldsymbol{S}_\text{f}(t)$.
	
	Computing $\text{Tr}\bigl(\boldsymbol{B}_\text{f}^\dagger(t)\boldsymbol{B}_\text{f}(t)\bigr)$ is simpler than computing $\text{det}\bigl(\mathds{1}+\boldsymbol{B}_\text{f}^\dagger(t)\boldsymbol{B}_\text{f}(t)\bigr)$. The reason for this can be seen by writing $\text{det}\bigl(\mathds{1}+\boldsymbol{B}_\text{f}^\dagger(t)\boldsymbol{B}_\text{f}(t)\bigr)=1+\text{Tr}\bigl(\boldsymbol{B}_\text{f}^\dagger(t)\boldsymbol{B}_\text{f}(t)\bigr)+\frac{1}{2}\text{Tr}^2\bigl(\boldsymbol{B}_\text{f}^\dagger(t)\boldsymbol{B}_\text{f}(t)\bigr)-\frac{1}{2}\text{Tr}\bigl(\boldsymbol{B}_\text{f}^\dagger(t)\boldsymbol{B}_\text{f}(t)\boldsymbol{B}_\text{f}^\dagger(t)\boldsymbol{B}_\text{f}(t)\bigr)$ in terms of traces of relevant matrices, which makes manifest the greater complexity in computing $\text{det}\bigl(\mathds{1}+\boldsymbol{B}_\text{f}^\dagger(t)\boldsymbol{B}_\text{f}(t)\bigr)$ due to the presence of the last term.
	We will see the advantage of having these bounds in the examples below.
	
	We also note that we can bound the fidelity using the average number of particles alone, thereby removing the necessity of computing (and therefore measuring) the variance $\sigma^2_{\Delta N}(t)=\Delta N^2(t)-(\Delta N(t))^2$ of the statistical quantity $\Delta N(t)$. This is a consequence of the fact that, as shown above, $\text{Tr}\bigl(\boldsymbol{B}_\text{f}^\dagger(t)\boldsymbol{B}_\text{f}(t)\boldsymbol{B}_\text{f}^\dagger(t)\boldsymbol{B}_\text{f}(t)\bigr)$ is directly related to  $\sigma^2_{\Delta N}(t)$.
	
	\subsection{Initial single-mode squeezed states}\label{subsection:initial:single:mode:squeezed:state}
	We now assume that we have an initial single mode squeezed state of both modes $\hat{a}$ and $\hat{b}$, which implies $\boldsymbol{U}_0=\cosh (2s)\,\mathds{1}_2$ and $\boldsymbol{V}_0=\sinh (2s)\,\mathds{1}_2$. 
	Here, $s$ is the squeezing parameter, which is the same for both modes, there is no phase in the squeezing direction. 
	This choice adds simplicity to the following calculations and can be done without loss of generality. Note that we can recover the initial vacuum state by simply setting $s=0$. We employ \eqref{particle:number:variation} and therefore obtain
	\begin{align}
		\Delta N_{}(t)&=\cosh(2s)\textrm{Tr}(\boldsymbol{B}^\dag(t) \boldsymbol{B}(t))\nonumber\\
            &\quad+\sinh(2s) \Re\left(\textrm{Tr}(\boldsymbol{A}^\dag(t) \boldsymbol{B}(t))\right).
	\end{align}
	Further lengthy algebra allows us to express $N_{}(t)$ in terms of the Bogoliubov coefficients, as well as the frequencies $\kappa_\pm$. We do not report the expression here to avoid overburdening the reader. It can be obtained directly when the need arises.
	
	We now turn to the demonstration that the rotating wave applies, as already discussed before. In particular, we wish to provide a more explicit expression for the general expression \eqref{RWA:demonstration:expression:general} for this particular case. As done above, we employ the expressions \eqref{distance:fidelity:final:general} and \eqref{B:f:Bogoliubov:coefficient} on resonance and in the limit of small coupling strength, i.e.,  $\omega_\textrm{b}:=\omega+\delta\omega$, $\tilde{\kappa}_\pm=\kappa_\pm/\omega$ $\tilde{g}:=g/\omega$, and $\tau:=\omega\,t$, and finally assuming that $\tilde{g}\ll1$.  
	
\begin{figure}
	\centering
	\includegraphics[width=1.0\linewidth]{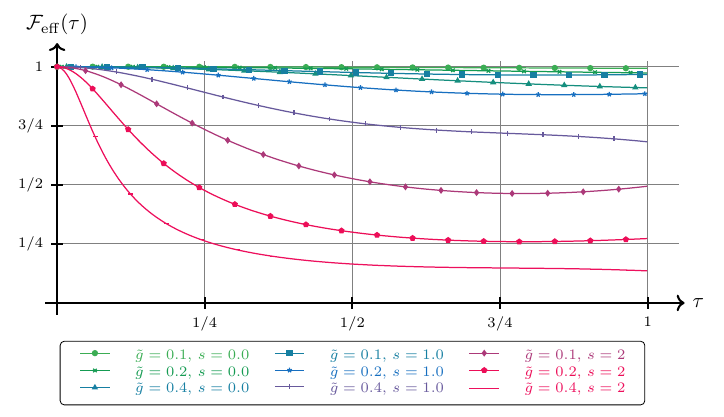}
	\caption{Plot of the fidelity $\mathcal{F}_\text{eff}(\tau)$ given by Equation \eqref{eqn:Fidelity:initial:single:mode:squeezed:state:main:text} in case of initial single-mode squeezed states for different combinations of the relative coupling strength $\tilde{g}=0.1,0.2,0.4$ and squeezing parameter $s=0,1,2$ as a function of $\tau\in[0,1]$. It is immediate to recognise the importance that the squeezing parameter $s$ has in reaching smaller values of the fidelity.}
	\label{fig:tba:01}
\end{figure}

	We have analyzed the perturbative regime and we have found that some care is necessary when attempting to obtain a meaningful expansion of the quantities of interest. Since they contain oscillatory functions, where the small parameters enter as variables multiplied by other degrees of freedom, a close inspection of the expressions requires us not only to assume $\tilde{g}\ll1$, but also to assume $\Tilde{g}\tau=\text{const.}$ as well as $\Tilde{g}^2\tau\equiv \frac{g}{\omega}gt\ll 1$, such that $\Tilde{g}^2\tau=\mathcal{O}(\Tilde{g})$. Furthermore, as can be seen from explicit expressions obtained in~\ref{verifying:Validity:RWA:appendix}, we also require the squeezing parameter $s$ to remain finite and small. In this regime, we can safely compute the perturbative expansions of interest and, specializing to the initial squeezed state as mentioned above, we obtain
	\begin{align}\label{eqn:Fidelity:initial:single:mode:squeezed:state:main:text}
		\mathcal{F}_{\text{eff}}^{-2}(\tau)=1+C_2(\tau,s)\Tilde{g}^2,
	\end{align}
	where $C_2(\tau,s)$ is a bounded function that can be found in~\ref{verifying:Validity:RWA:appendix}. To better appreciate the meaning of this result, in Figure~\ref{fig:tba:01} we provide a pictorial representation of (\ref{eqn:Fidelity:initial:single:mode:squeezed:state:main:text}) for different fixed values of $\tilde{g}$ and $s$.

	\textit{Initial vacuum state}---We can specialize our result even further and consider the simplest case possible of an initial vacuum state, for which we need to set $s=0$ from the start. The complicated expression \eqref{B:f:Bogoliubov:coefficient:general} drastically simplifies and expression \eqref{distance:fidelity:final:general} reduces to 
	\begin{align}
		\mathcal{F}^{-2}_{\text{eff}}(\tau)=\det\left(\mathds{1}_2+\boldsymbol{B}^\dag(\tau)\boldsymbol{B}(\tau)\right).\label{Fidelity:initial:vacuum:state}
	\end{align}
	This expression can be recast in terms of the first two moments of the distribution of the expectation value of the excitation number as obtained in \eqref{Fidelity:initial:effective:final:state}, where here $\Delta N_{}(\tau)=\Tr\bigl(\boldsymbol{B}^\dag(\tau)\boldsymbol{B}(\tau)\bigr)$. 
	
	The bounds on the fidelity \eqref{Fidelity:initial:vacuum:state} can be immediately computed using $\Delta N_{}(\tau)=\Tr\bigl(\boldsymbol{B}^\dag(\tau)\boldsymbol{B}(\tau)\bigr)$, where now the number of excitations created coincides with those created during the dynamics of interest. i.e., $\Delta N_\text{f}(\tau)\equiv\Delta N_{}(\tau)$. One concludes from the bounds above that, in the case of an initial vacuum states, the fidelity deteriorates as a monotonic function of the number of particles created in the system, vanishing in the limit $\Delta N_{}(\tau)\rightarrow\infty$ of infinite particles injected in the system.
	
\begin{figure}
	\centering
	\includegraphics[width=1.0\linewidth]{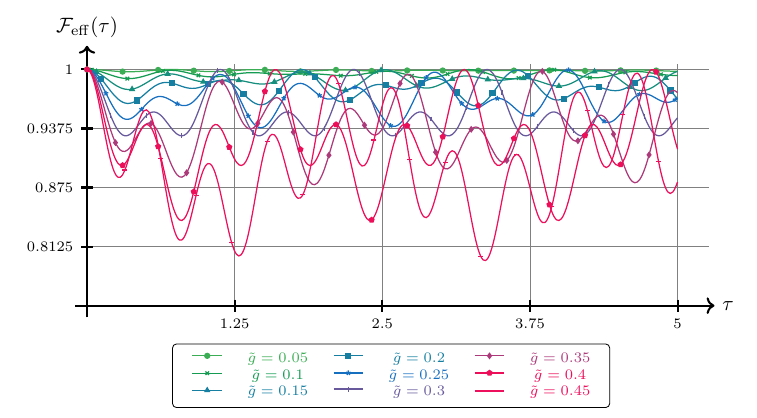}
	\caption{Pictorial depiction of the fidelity $\mathcal{F}_\text{eff}(\tau)$ from equation \eqref{eqn:fidelity:effective:initial:vacuum:state:main:text} for the initial vacuum state $\boldsymbol{\sigma}(0)=\mathds{1}_4$. In this plot, the fidelity is depicted for different values of $\tilde{g}=n/20$ for $n=1,...,9$ as a function of $\tau\in[0,22]$. The fidelity is bounded from below by $3/4$, since $\tilde{g}>1/2$ and there is no initial squeezing (see (\ref{eqn:fidelity:effective:initial:vacuum:state:main:text}). Thus, we have re-scaled the $y$-axis accordingly for easiness of presentation. Notice that increasing $\tilde{g}$ does not only reduce the achievable minimal values of the fidelity, but also leads the fidelity to have an increasingly complicated oscillatory pattern as a function of time $\tau$.}
	\label{fig:tba:02}
\end{figure}

	We can then study the validity of the rotating wave approximation for this case by imposing the resonance condition and performing a perturbative expansion for $\Tilde{g}\ll 1$. To lowest non-trivial order we obtain
	{
		\begin{align}\label{eqn:fidelity:effective:initial:vacuum:state:main:text}
			\mathcal{F}_{\text{eff}}(\tau)=1-\frac{\Tilde{g}^2}{2}\left(\sin^2(\Tilde{\kappa}_+\tau)+\sin^2(\Tilde{\kappa}_-\tau)\right),
		\end{align}
	}
	where one recalls the expression of the normal-mode frequencies $\Tilde{\kappa}_\pm=\sqrt{1\pm 2\Tilde{g}}$ for this particular case. Thus, the fidelity is bounded for the initial vacuum state for all times. We can use this expression for the fidelity to immediately obtain the Bures distance for this case, which reads $ D^2_\textrm{B}(\boldsymbol{\sigma}(0),\boldsymbol{\sigma}_\textrm{eff}(\tau))=\frac{1}{2}\left(\sin^2(\Tilde{\kappa}_+\tau)+\sin^2(\Tilde{\kappa}_-\tau)\right)\,\Tilde{g}^2$ to lowest nontrivial order. A pictoral visualization of equation \eqref{eqn:fidelity:effective:initial:vacuum:state:main:text} can be found in Figure~\ref{fig:tba:02}.
	
\section{Implementations}\label{Sec:Applications}
	We provide here potential implementations of the model used in this work to highlight the wide scope of applicability of our techniques. 
	
	\subsection{General modelling of the interaction}
 
\begin{figure}[h!]
		\includegraphics[width=\linewidth]{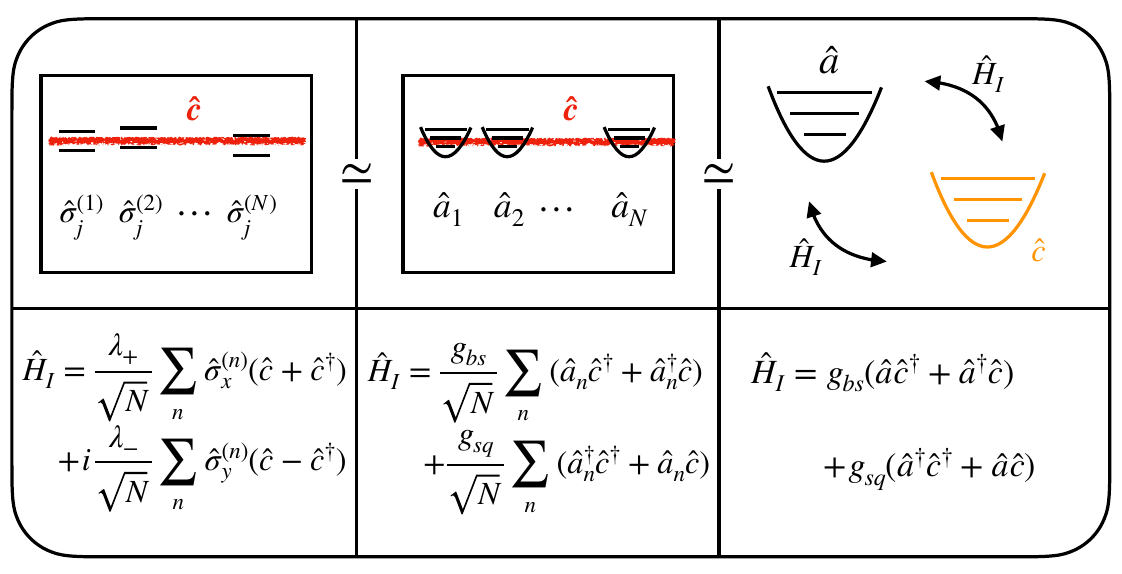}
		\caption{Here we provide a few systems that can be modelled by the same Hamiltonian \eqref{basic:hamiltonian}. The first is a (infinitely large) set of $N$ two-level systems with operators $\hat{\sigma}^{(n)}_j$ all coupled to the same cavity mode with operators $\hat{c},\hat{c}^\dag$. The couplings $\lambda_\pm$, once the Holstein-Primakoff transformation has been taken and all of the usual steps have been done to map the total spin $N/2$ to approximatively a harmonic oscillator \cite{Holstein:Primakoff:1940}, as per usual in the Dicke model \cite{Kirton:Roses:2018}, can be then used to directly determine $g_\text{bs}$ and $g_\text{sq}$. The second system is that of $N$ harmonic oscillators with operators $\hat{a}_n,\hat{a}_n^\dag$ all coupled to the same cavity mode with operators $\hat{c},\hat{c}^\dag$. A simple polaron-like transformation can map their combination into a polaron-like mode with operators $\hat{b},\hat{b}^\dag$, together with $N-1$ uncoupled oscillators that can be discarded as far as the dynamics of the two modes of interest are concerned. The final panel depicts the current Hamiltonian \eqref{basic:hamiltonian}.}\label{figure:zero}
	\end{figure}

    \vspace{-0.2cm}
	The Hamiltonian \eqref{basic:hamiltonian} does not apply only to systems that are composed of two harmonic oscillators. For example, we can also employ it to model a setup, commonly known as the Dicke-model \cite{Kirton:Roses:2018}, where a very large or infinite number of two-level systems are coupled to a common cavity field mode. In this case, a very large collection of finite-dimensional systems is approximated to an infinite dimensional one (i.e., a harmonic oscillator) by means of the Holstein-Primakoff transformation \cite{Holstein:Primakoff:1940}. We can also employ it to model a setup where a large number of vibrational degrees of freedom of molecules trapped in a cavity that interact with a common cavity mode \cite{delPino::Feist:2015}. 
	To see how this is possible, we take the latter case and observe that one might be given the following Hamiltonian
	$\hat{H}=\hat{H}_0+\sum_{n=1}^N\hbar\,\tilde{g}_\text{bs} \bigl(\hat{a}\,\hat{d}_n^\dag+\hat{a}^\dag\,\hat{d}_n\bigr)$
	where $\hat{H}_0:=\hbar\,\omega_\text{a}\,\hat{a}^\dag\,\hat{a}+\sum_{n=1}^N\hbar\,\omega_\text{d}\,\hat{d}_n^\dag\,\hat{d}_n$.  We then introduce the vectors of operators $\hat{X}:=(\hat{d}_1,\ldots,\hat{d}^\dag_N)^{\text{Tp}}$ and $\hat{Y}:=(\hat{b}_1,\ldots,\hat{b}^\dag_N)^{\text{Tp}}$ that are related by $\hat{Y}=\boldsymbol{R}\hat{X}$. Here we have that $[\hat{b}_n,\hat{b}_m]=\delta_{nm}$ while all others vanish, and that $\boldsymbol{R}$ is an orthogonal matrix, i.e., $\boldsymbol{R}\boldsymbol{R}^{\text{Tp}}=\mathds{1}_{2N}$.
	It is immediate to see that, if we impose
	$\hat{b}\equiv\hat{b}_1:=1/\sqrt{N}\sum_{n=1}^N\hat{d}_n$ and $g_\text{bs}:=\sqrt{N}\tilde{g}_\text{bs}$, then one has $[\hat{b},\hat{b}^\dag]=1$ and this Hamiltonian reduces to $\hat{H}=\hat{H}_0+\hbar\,g_\text{bs} \bigl(\hat{a}\,\hat{b}^\dag+\hat{a}^\dag\,\hat{b}\bigr)+\sum_{n=2}^N\hbar\,\omega_\text{d}\,\hat{b}_n^\dag\,\hat{b}_n$ with the new free-Hamiltonian $\hat{H}_0:=\hbar\,\omega_\text{a}\,\hat{a}^\dag\,\hat{a}+\hbar\omega_\text{d}\,\hat{b}^\dag\,\hat{b}$. We note that, in this form, $\hat{H}$ corresponds to the particular case of \eqref{basic:hamiltonian} for $g_\text{sq}=0$ once the uncoupled modes $n=1,...,N$ are discarded. It is immediate to extend this logic to design the initial multi-mode Hamiltonian and recover the full form of \eqref{basic:hamiltonian}. A pictorial representation of the equivalence of the different models is given in Figure~\ref{figure:zero}. 
	
	\subsection{Superconducting circuits}
	Circuit quantum electrodynamics offers a natural platform for studying effects beyond the rotating wave approximation \cite{Blais:Grimsmo:2021,Hillmann:Quijandria:2022}. This is enabled by the fact that the interaction between superconducting circuits leads 
	to much stronger coupling constants than what can be achieved for example in optical systems. Signatures of the counter-rotating terms have been observed as the Bloch-Siegert shift of spectral lines in strongly driven Cooper pair transitors \cite{Tuorila:2010} and of the cavity frequency in transmon qubits coupled to cavities \cite{Pietikainen:2017}. An outstanding problem is the extraction of the virtual photons predicted to exist in the ground state of such an arbitrarily strong interacting system \cite{Giannelli:Anfuso:2023,Giannelli:2024}. The interest here would be to obtain a controllable physical system described by the quantum Rabi model - the general fundamental model for the coupling of a qubit to a resonator. Another potential application in the field of quantum computing with superconducting qubits could be found in providing analytical bounds on fidelity of protocols that use the Single Excitation Subspace scheme, where states with a single excitation are exploited  \cite{Geller:Martinis:2015}.
	
	Rather than working directly with the system Hamiltonian and increasing the coupling, a different approach  is to realize a simulation of the arbitrary coupling regime. Such simulations have been realized experimentally in transmon-resonator systems \cite{Ustinov:2017,DiCarlo:2017}.
	This partly realizes the physics we are after here, since ultimately the transmon is an oscillator itself, albeit with a small anharmonicity. 
	Here we focus on a different approach, whereby the arbitrarily strong coupling is realized in a multiple rotating frame \cite{Huard:2018,Milman:2012}.
	
	Consider the following Hamiltonian, describing a system of two oscillators with frequencies $\epsilon_{\rm a}$ and $\epsilon_{\rm b}$. Their interaction, described by a term $(\hat{a} + \hat{a}^{\dag})(\hat{b}+\hat{b}^{\dag})$ is tunable and can be pumped at frequencies that are of the same order of magnitude as $\epsilon_{\rm a}$ and $\epsilon_{\rm b}$. We choose two pumps, one with amplitude $2 g_{\rm sq}$
	and frequency $\omega_{\rm sq}$, and the other with amplitude $2 g_{\rm bs}$ and frequency $\omega_{\rm bs}$. We can write the total Hamiltonian as $\hat{H}(t) = \hat{H}_0+\hat{H}_{\text{I}}(t)$, where $\hat{H}_0=\hbar \epsilon_{\rm a} \hat{a}^{\dag} \hat{a} +\hbar \epsilon_{\rm b} \hat{b}^{\dag} \hat{b}$ is the free Hamiltonian and 
	{\begin{align}
		\hat{H}_{\text{I}}(t) &=2 \hbar g_{\rm sq}\cos (\omega_{\rm sq}t) (\hat{a} + \hat{a}^{\dag})(\hat{b} + \hat{b}^{\dag})\nonumber\\
            &\quad+2 \hbar g_{\rm bs}\cos (\omega_{\rm bs}t) (\hat{a} + \hat{a}^{\dag})(\hat{b} + \hat{b}^{\dag}).
	\end{align}
 }
	Next, we move  to a doubly rotating frame defined by the unitary $\hat{U}(t) = \operatorname{exp}({i \hbar ((\epsilon_{\rm a}-\omega_{\rm a}) \hat{a}^{\dag}\hat{a} + (\epsilon_{\rm b}-\omega_{\rm b}) \hat{b}^{\dag}\hat{b}) t})$, obtaining a new Hamiltonian $\hat{H} \rightarrow \hat{U}\hat{H}\hat{U}^{\dag} + i \hbar \dot{\hat{U}}\hat{U}^{\dag}$. In this frame we neglect the oscillatory terms and obtain
	\begin{align*}
		\hat{H} = \hat{H}_0 + \hbar g_{\rm bs} (\hat{a}^{\dag} \hat{b} + \hat{a}\hat{b}^{\dag})  + \hbar g_{\rm sq} (\hat{a}^{\dag} \hat{b}^{\dag} + \hat{a}\hat{b}),
	\end{align*}
	where now $\hat{H}_0:=\hbar \omega_{\rm a}\hat{a}^{\dag} \hat{a} + \hbar \omega_{\rm b} \hat{b}^{\dag}\hat{b}$. This is identical to the Hamiltonian \eqref{basic:hamiltonian} considered in this work.
	
	This system has major experimental advantages, as the frequencies $\omega_{\rm a}$ and $\omega_{\rm b}$ are in fact
	detunings from $\epsilon_{\rm a}$ and $\epsilon_{\rm b}$ that can be adjusted in a wide range, and, similarly, the pump amplitudes can be switched on/off or tuned to exceed $\omega_{\rm a}$ and $\omega_{\rm b}$. We therefore have, in this rotating frame, a system that is manifestly in the regime of arbitrary coupling that has been analyzed in this work. All the tools presented here can be applied in a straightfoward fashion.
	We conclude by noting that values of $g_{\rm bs}$, $g_{\rm sq}$, $\omega_{\rm a}$, $\omega_{\rm b}$ of the order of $2\pi \times 10$ MHz have been already experimentally achieved \cite{Huard:2018}. 
	
	\subsection{Bath driving}
	We have argued that our main Hamiltonian \eqref{basic:hamiltonian} can model different concrete implementations. Among many systems of current interest one finds spin systems coupled to spin baths that interact with modes of light. A well-known example is a nitrogen-vacancy (NV) center in diamond coupled to a nuclear spin bath \cite{Farfurnik:Horowicz:2018,Balasubramanian:Osterkamp:2019,Zhou:Choi:2020,Joos:Bluvstein:2022}.
	Such systems are promising candidates for physical qubits to run a quantum computer \cite{Rembold:Oshnik:2020}. In general, the coupling that occurs is predominantly of the ZZ form -- i.e., a dipole-dipole interaction. Nevertheless, it might be possible under specific circumstances, e.g., in a specific frame, to obtain other types of interactions, such as XX, see \cite{Bermudez:Jelezko:2011,Xu:Yin:2019}. Regardless of the available capabilities of engineering such couplings within current setups, we are here mostly inspired by such types of interactions that appear in NV centers to study the induced dynamics in light of our main body of work.
	
	We envision a Hamiltonian of the form
	\begin{align}\label{nv:center:hamiltonian}
		\hat{H}=&\hat{H}_{\text{JC}}+\sum_n\hbar\xi_n'\hat{\sigma}_{\text{x}}\hat{\sigma}^{(n)}_{\text{x}}+\hbar g'\sum_n \left(\hat{a}+\hat{a}^\dag\right)\hat{\sigma}^{(n)}_{\text{x}}
	\end{align}
	for this case, where $\hat{H}_{\text{JC}}:=\hat{H}_0+\hbar\lambda\hat{\sigma}_{\text{x}}(\hat{a}+\hat{a}^\dag)$ and $\hat{H}_0=\hbar\,\omega_\text{a}\,\hat{a}^\dag\,\hat{a}+\hbar\omega_0\hat{\sigma}_{\text{z}}+\sum_n\hbar\omega_{\text{bth}}\hat{\sigma}^{(n)}_{\text{z}}$ for simplicity of notation.
	Here our core qubit has operators $\hat{\sigma}_j$ that satisfy the usual SU$(2)$ algebra $[\hat{\sigma}_j,\hat{\sigma}_k]=i\epsilon_{jk\ell}\hat{\sigma}_\ell$, the spin bath is described by $N$ operators $\hat{\sigma}^{(n)}_k$ that also satisfy the same algebra, and the mode of light has operators $\hat{a},\hat{a}^\dag$ that satisfy $[\hat{a},\hat{a}^\dag]=1$. The parameters $\lambda$, $\xi_n'$ and $g'$ are the coupling constants.
	
	We then use a Holstein-Primakoff-like transformation \cite{Holstein:Primakoff:1940}, which is implemented as follows: first, we introduce the two operators $\hat{b}\approx 1/\sqrt{N}\sum_{n=1}^N\hat{\sigma}^{(n)}_-$, $\hat{b}^\dag\approx 1/\sqrt{N}\sum_{n=1}^N\hat{\sigma}^{(n)}_+$. Then, we collect the spin bath operators in the vector $\underline{\hat{\sigma}}:=(\hat{\sigma}^{(1)}_{\text{x}},...,\hat{\sigma}^{(n)}_{\text{x}})^{\text{Tp}}$  of operators, and introduce the new vector $\underline{\hat{b}}:=\boldsymbol{R}\,\underline{\hat{\sigma}}$ of operators, where $\boldsymbol{R}$ is a $N\times N$ orthogonal matrix that satisfies $\boldsymbol{R}\boldsymbol{R}^{\text{Tp}}=\mathds{1}_N$ and we impose $\hat{b}\equiv\hat{b}_1=1/\sqrt{N}\sum_{n=1}^N\hat{\sigma}^{(n)}_-$. This constraint defines the first row of $\boldsymbol{R}$, and the others can be constructed by the standard Gram-Schmidt procedure. Since we will not be interested in their explicit expression, we do not present further details. 
	
	We can obtain the Hamiltonian \eqref{nv:center:hamiltonian} using these new quantities. We find
	\begin{align}\label{nv:center:hamiltonian:simplified}
		\hat{H}=&\hat{H}_0+\hbar\hat{\sigma}_{\text{x}}\left(\lambda(\hat{a}+\hat{a}^\dag)+\xi(\hat{b}+\hat{b}^\dag)\right)\nonumber\\
		&+ \hbar g(\hat{a}+\hat{a}^\dag)(\hat{b}+\hat{b}^\dag)+\sum_{n=2}^N\xi_n\hat{\sigma}_{\text{x}}(\hat{b}_n+\hat{b}^\dag_n),
	\end{align}
	where now $\hat{H}_0=\hbar\,\omega_\text{a}\,\hat{a}^\dag\,\hat{a}+\hbar\,\omega_{\text{bth}}\,\hat{b}^\dag\,\hat{b}+\hbar\omega_0\hat{\sigma}_{\text{z}}+\hbar \omega_{\text{bth}}\sum_{n=2}^N \hat{b}^\dag_n \hat{b}_n$. The couplings $\xi_n$ can be obtained as follows: we introduce the vectors $\underline{\xi}:=(\xi,\xi_2,...,\xi_N)^{\text{Tp}}$ and $\underline{\xi}':=(\xi_1',\xi_2',...,\xi_N')^{\text{Tp}}$, and have that $\underline{\xi}=\boldsymbol{R}\underline{\xi}'$. Notice that we also have $\hat{b}^\dag\hat{b}+\sum_{n=2}^N\hat{b}_n^\dag\hat{b}_n\approx\sum_{n=1}^N\hat{\sigma}_{\text{z}}^{(n)}$, which we recall applies for $N\gg1$ as discussed before. 
	
	We now note that we can re-write $\hat{H}=\hat{H}_{\text{sc}}+\hat{H}_{\text{qb}}$, where $\hat{H}_{\text{sc}}$ is the arbitrary-coupling Hamiltonian \eqref{basic:hamiltonian} considered in this work, while $\hat{H}_{\text{qb}}$ contains all other terms concerning the qubit and the spin bath. Therefore, the time evolution operator $\hat{U}(t)$ induced by the current Hamiltonian \eqref{nv:center:hamiltonian:simplified} can be split via 
	\begin{align}\label{spin:time:evolution:operator}
		\hat{U}(t)=\hat{U}_{\text{sc}}(t)\overset{\leftarrow}{\mathcal{T}}e^{-\frac{i}{\hbar}\int_0^t dt'\,\hat{H}_{\text{I}}(t')},
	\end{align}
	where $\hat{U}_{\text{sc}}(t):=\exp\bigl[-\frac{i}{\hbar} \hat{H}_{\text{sc}} t\bigr]$ and we have introduced the ``interaction Hamiltonian'' $\hat{H}_{\text{I}}(t):=\hat{U}^\dag_{\text{sc}}(t)[\hat{H}-\hat{H}_0-\hbar g(\hat{a}+\hat{a}^\dag)(\hat{b}+\hat{b}^\dag)] \hat{U}_{\text{sc}}(t)$.
	This explicitly reads
	\begin{align*}
		\hat{H}_{\text{I}}(t)&=\omega_0\hat{\sigma}_{\text{z}}+\hbar \omega_{\text{bth}}\sum_{n=2}^N \hat{b}^\dag_n \hat{b}_n+\sum_{n=2}^N\xi_n\hat{\sigma}_{\text{x}}(\hat{b}_n+\hat{b}^\dag_n)\\
        &\quad+\hbar\hat{\sigma}_{\text{x}}\left(\mu_{\text{a}}(t)\hat{a}+\mu_{\text{b}}(t)\hat{b}+\text{h.c.}\right),
	\end{align*}
	where the coefficients $\mu_{\text{a}}(t)$ and $\mu_{\text{b}}(t)$ can be obtained by inserting the expressions \eqref{time:evolution:operators:strong:coupling} for the evolution of the bosonic operators into $\lambda(\hat{a}+\hat{a}^\dag)+\xi(\hat{b}+\hat{b}^\dag)$. This implies that such coefficients are functions of $\lambda$ and $\mu$ as well.
	
	We can now use the techniques developed in this work to monitor the validity of the rotating wave approximation that can be applied to the evolution encoded in $\hat{U}_{\text{sc}}$. One possibility is that the coupling with the spin bath is weak, i.e., $|\xi|\ll1$ and $|\xi_n|\ll1$, and therefore the terms proportional to $\xi_n$ can be discarded, together with the free part term. However, in the last term it must not be that $|\mu_{\text{b}}(t)|\ll1$ since it depends on $\lambda$, $\xi$ and $g$.   
	
	It should be evident, at this point, that there is a crucial difference between applying the rotating wave approximation and maintaining the full arbitrary-coupling evolution. As can be seen from the explicit expressions \eqref{time:evolution:of:mode:operators} for the rotating wave approximation, if $|\xi|\ll1$ there is no ``amplification'' due to $g$, since the coefficients \eqref{time:evolution:of:mode:operators} are simple oscillating functions with order unit coefficients. On the contrary, if the rotating wave approximation does not apply, one expects that the full coefficients \eqref{time:evolution:operators:strong:coupling} will have (much) larger magnitudes as can be seen from the discussion pertaining the Bloch-Messiah decomposition. We conclude that, in the case of weak coupling to the spin bath but arbitrarily strong coupling of the bath to a mode of light, one can effectively write 
	\begin{align}
		\hat{H}_{\text{I}}(t)\approx\,\omega_0\hat{\sigma}_{\text{z}}+\hbar\hat{\sigma}_{\text{x}}\left(\mu_{\text{a}}(t)\hat{a}+\mu_{\text{b}}(t)\hat{b}+\text{h.c.}\right).
	\end{align}
	This means that the time evolution \eqref{spin:time:evolution:operator} is given by the product of two unitary operators: (i) the interaction of the light mode with the ensemble of bath spins (now acting as an overall bosonic mode with operators $\hat{b}$ and $\hat{b}^\dag$), and (ii) by the interaction of the target qubit with the light mode and the ensemble itself.
	We can therefore envisage that engineering the time-dependence of the coupling with light and the drive of the spin bath, as well as the strengths of such interactions, can provide a wider scope of manoeuvre to design single-qubit quantum gates \cite{Joos:Bluvstein:2022}, as well as to reduce the noise. An additional qubit can potentially be added to engineer two-qubit gates.
	Note that the applicability of the techniques developed here is restricted to the term $\hat{U}_{\text{sc}}(t)$ alone. The application of the rotating wave approximation to the qubit-field mode interaction is yet another available tool that can be used when necessity arises.

	\section{Final considerations}\label{Sec:Considerations}
	We continue here by presenting some final considerations regarding the scope and applicability of our work.
	
	\subsection{Applicability to Gaussian states}\label{Sec:Considerations:applicability:Gaussian:states}
	First, we note that our result \eqref{distance:fidelity:final:general} applies to Gaussian states only. In this sense, we have used the powerful techniques provided by the covariance matrix formalism to obtain an explicit dependence of the figure of merit chosen here, namely the fidelity of quantum states, on all relevant parameters. These include the defining parameters of the initial state, as well as the coupling strength of the interaction and time. While a final explicit expression as a function of all parameters can in principle be obtained by performing the matrix operations that appear in \eqref{B:f:Bogoliubov:coefficient:general}, it would be unnecessarily cumbersome and it would obscure the main message of this work. The main expression can be applied to specific scenarios when needed, and we have done so by employing an initial two single-mode squeezed state for the system.
	
	An important consequence of our result is that the distance of two arbitrary pure reference and target Gaussian states $\boldsymbol{\sigma}_\text{R}(t)=\boldsymbol{S}_\text{R}(t)\boldsymbol{\sigma}_0\boldsymbol{S}^\dag_\text{R}(t)$ and $\boldsymbol{\sigma}_\text{T}(t)=\boldsymbol{S}_\text{T}(t)\boldsymbol{\sigma}_0\boldsymbol{S}^\dag_\text{T}(t)$, as measured by the fidelity $\mathcal{F}(\boldsymbol{\sigma}_\text{R}(t),\boldsymbol{\sigma}_\text{T}(t))$, is uniquely determined by the first and second moments of the distribution of the average excitations created via an effective time evolution that transformed the vacuum state into the effective final pure state $\boldsymbol{\sigma}_\text{f}(t)=\boldsymbol{S}_\text{f}(t)\boldsymbol{S}^\dag_\text{f}(t)$, where $\boldsymbol{S}_\text{f}(t)=\boldsymbol{s}_0^{-1}\boldsymbol{S}_\text{R}^{-1}(t)\boldsymbol{S}_\text{T}(t)\boldsymbol{s}_0$. This can provide a useful tool for experiments where comparing two states is difficult but measuring the moments of the average particle number distribution is not.

	\subsection{Role of quantum correlations}\label{Sec:Considerations:quantum:correlations:role}
	The role of quantum correlations in determining the distance between quantum states has attracted minimal attention to date \cite{Buono:Nocerino:2016}. Given the importance that quantum resources have not only from a theoretical perspective but also for the development of concrete quantum technologies, it is crucial to shed light on the interplay of squeezing, in the context of continuous variables, and distances between states.
	
	In our work we first focussed on answering this question for the case of pure states, and we have found that the fidelity is maximal if and only if $\boldsymbol{B}^\dag_{\text{f}}(t)\boldsymbol{B}_{\text{f}}(t)=0$. Here, the key observation is that $\boldsymbol{B}_{\text{f}}(t)$ is a Bogoliubov matrix of the ``$\beta$-type'', that is, related to the creation of excitations. The condition $\boldsymbol{B}^\dag_{\text{f}}(t)\boldsymbol{B}_{\text{f}}(t)=0$ occurs at least in the cases where $\boldsymbol{\beta}_0\equiv\boldsymbol{B}(t)\equiv0$, which also correspond to the fact that $\boldsymbol{s}_0$ and $\boldsymbol{S}(t)$ are unitary and block diagonal. Intuition here leads us to observe that, in such circumstance, no squeezing is present in the initial state or in the time evolution. On the other hand, when $g\neq0$ we have that $\boldsymbol{B}_{\text{f}}(t)$ does not vanish and, in general, $\mathcal{F}_{\text{eff}}(t)<1$ for most times. This is a clear indication that squeezing is the feature directly responsible for the departure between the trajectories in the Hilbert space determined by the time evolution or, in other words, squeezing is a key resource to distinguish pure Gaussian states.
	
	We have seen that the fidelity reaches unity when $\boldsymbol{B}(t)=0$. Our main expressions \eqref{distance:fidelity:final:general} and \eqref{B:f:Bogoliubov:coefficient} imply that this can occur at finite times when the frequencies $\kappa_\pm$ of the normal modes satisfy the relation $\kappa_+/\kappa_-=q\in\mathbb{Q}$. This behaviour is common in systems of coupled harmonic oscillators, where beatings and interference lead to interesting constructive or destructive contributions when the ratios of the frequencies are rational numbers. Concretely, it is easy to see that, if $\kappa_+/\kappa_-=q$ then there are times $t_*$ for which $\exp[\pm i \boldsymbol{\kappa}t_*]=\mathds{1}_4$, where $\boldsymbol{\kappa}=\text{diag}(\kappa_+,\kappa_-,\kappa_+,\kappa_-)$, and thus \eqref{main:transformation:equation:main:text} guarantees that $\boldsymbol{B}(t_*)=0$, which in turn implies that $\boldsymbol{B}^\dag_{\text{f}}(t_*)=0$ (if no initial squeezing is present in the state). 
	
	\subsection{Validity of the rotating wave approximation}
	While we have obtained the fidelity to quantify the distance between the states evolved via the full Hamiltonian and the rotating wave approximation one, we have also used our analytical results to show that the rotating wave approximation holds. To do this, the standard approach is to assume that the frequencies of the oscillator are the same, i.e., $\omega\equiv\omega_{\text{a}}=\omega_{\text{b}}$, that the coupling-to-frequency ratio $\tilde{g}:=g/\omega$ is very small, and that the time is correspondingly large, i.e., $\tilde{g}(\omega t)$ is constant while  $\tilde{g}^2 (\omega t)\ll1$. In this limit, the rotating wave part $g[\hat{a}\hat{b}^\dag+\hat{a}^\dag\hat{b}]$ becomes fully time-independent within the interaction picture, while the counter-rotating or squeezing term $g[\hat{a}\hat{b}+\hat{a}^\dag\hat{b}^\dag]$ oscillates with a frequency of $2\omega$. We compute the fidelity in this limit and find that $\mathcal{F}\approx 1-\mathcal{O}((g/\omega)^2)$ in all cases with Gaussian states considered. We obtain the same result using bounds on the difference between various evolutions acting on Fock states. Convergence of the full evolution to the rotating wave approximation in the regime just described implies convergence on the whole Hilbert space, again proving the claim.
	Additionally, we note that if $\tilde{g}(\omega t)$ were not to be constant and thus $\tilde{g}^2 (\omega t)$ could be of the order of unity, it would follow that the matrix $\boldsymbol{B}_{\text{f}}(t)$ would also have zero order contributions, and therefore the fidelity would not in general approach unity, invalidating the claim. 
	
	Notice that the convergence described here occurs polynomially as a function of $\tilde{g}^2$. This is not surprising: the fidelity of two states that are closed to each other and are distinguished by an infinitesimal parameter $\dd\theta$ have distance of order $\dd\theta^2$. In our case, the fidelity is a quadratic function of the (Bures) distance for small deviations \cite{Bures:1969}, which proves this claim.

	\subsection{Future Scope}
 We take the opportunity here to briefly discuss some implications of our work that are important for current research beyond theoretical linear quantum optics and can therefore be addressed in future work.

In our work we consider only ideal systems, an approach that allows us to provide exact expressions for the case of lossless and noiseless systems with two modes, and thus gain predictive power. This is important because we can provide proof-of-principle understanding that can be used in all related studies. The downside of our approach is that we are restricted to very simple initial states, as well as ideal closed systems that are isolated form any environment. Therefore, in order to export the methods gained here to tackle the dynamics of realistic systems, where more harmonic oscillators interact and where noise, loss and decoherence are present, our work needs to be extended to the degree that the use of numerical simulations become inevitable. Nevertheless, the use of numerical methods has its own advantages: for example, we can try to answer the question of how the fidelity scales for complex systems as a function of the number of oscillators, or we can study how the results change when open systems are considered and steady states can be achieved. Since all of these scenarios are of practical importance for realistic implementations of coupled quantum harmonic oscillators, we believe that such questions should be addressed in the future.
 
Another aspect to be considered is that our study has been restricted to linear dynamics only. Nevertheless, nonlinearities are key in many areas of physics, such as condensed matter \cite{Bishop:Krumhansl:1980} and nonlinear quantum optics \cite{Chang:Vuletic:2014}, and they have become a key aspect of modern approaches aimed at fully controlling quantum dynamics \cite{Bruschi:Xuereb:2024}. While the rotating wave approximation for bosonic systems is formulated in its standard form in the way presented here, it is natural to ask what occurs when nonlinear terms are present. In such cases, one cannot take advantage of the powerful methods provided by the covariance matrix formalism, and now approaches must be developed. Nevertheless, in order to push the boundary of our understanding even before any new general development is put forward, one can add simple nonlinearities to the Hamiltonian, such as a \textit{Kerr nonlinearity}, and extend the methods developed here in the presence of these new dynamics. 
Any new results obtained in this way could provide further insights in this area of quantum physics and open new avenues for future research. We leave these developments to future work.

	\section{Conclusions}
	We have tackled the question of the validity of the rotating wave approximation for two coupled quantum harmonic oscillators. To achieve our goal we, bounded the error introduced by the dynamics in Fock space and showed convergence in the small coupling regime, thereby fully answering the question of the validity of the approximation  in the affirmative. 
	
	We have then proposed a way to tackle the question of the validity of the rotating wave approximation fully analytically for Gaussian states only, thereby taking advantage of the powerful tools provided by the covariance matrix formalism. The main idea here was to quantify the distance between the initial state evolved via dynamics with arbitrarily strong couplings and the same initial state evolved through the rotating wave Hamiltonian. Measuring the distance was realized via the fidelity of quantum states, which is directly related to the Bures distance of the Hilbert space. This allowed us to compute the fidelity analytically and we have found that it depends directly on the nonclassical resource known as squeezing, that can be present in the initial state or be induced by the full dynamics. If no squeezing is present, the rotating wave approximation coincides with the full dynamics. Our main expression can be applied regardless of the specific regime of operation, thus giving full control over the Gaussian-state scenario. An important corollary of our approach is a relation between the distance of two arbitrary pure Gaussian states and the first and second moments of the statistical distribution of the average number of excitations created from the vacuum state via an effective evolution that depends on the states of interest. This provides a useful tool for experiments where two arbitrary Gaussian states need to be compared.
	
	We have also proposed simple applications of our Gaussian-state results, thereby showcasing the utility of the analytical expressions obtained. Concrete implementations are presented and discussed for interacting quantum systems that are not composed solely of harmonic oscillators, but whose dynamics can be well approximated by those of coupled harmonic oscillators. These include superconducting circuits and nitrogen-vacancy (NV) center systems coupled to spin baths.
	
	We believe that our work provides a definitive answer to the question of the validity of the rotating wave approximation for two coupled quantum harmonic oscillators, as well as a non-perturbative quantification of the dynamics when initial Gaussian states are considered. This approach can therefore be used to further investigate the formal aspects of the theory.


\acknowledgments
We thank Pablo Tieben, Matthias M. M\"uller, and Daniel Zeuch for useful comments, discussions and suggestions. A.F., F.W.M. and D.E.B. acknowledge support from the joint project No. 13N15685 ``German Quantum Computer based on Superconducting Qubits (GeQCoS)'' sponsored by the German Federal Ministry of Education and Research (BMBF) under the \href{https://www.quantentechnologien.de/fileadmin/public/Redaktion/Dokumente/PDF/Publikationen/Federal-Government-Framework-Programme-Quantum-technologies-2018-bf-C1.pdf}{framework programme
		``Quantum technologies -- from basic research to the market''}. D.E.B. also acknowledges support from the German Federal Ministry of Education and Research via the \href{https://www.quantentechnologien.de/fileadmin/public/Redaktion/Dokumente/PDF/Publikationen/Federal-Government-Framework-Programme-Quantum-technologies-2018-bf-C1.pdf}{framework programme
		``Quantum technologies -- from basic research to the market''} under contract number 13N16210 ``SPINNING''. G.S.P. acknowledges support from the Research Council of Finland Centre of Excellence QTF (project 352925).

\bibliographystyle{apsrev4-2}
\bibliography{NarducciFidelityBib}


\appendix

\onecolumngrid

\newpage
\section{Time evolution of the mode operators}
	In this section we present the explicit expression of the time evolution induced by an arbitrary quadratic Hamiltonian. 
	The full derivation has already been treated in previous work \cite{Urzua:Ramos-Prieto:2019,Bruschi:Paraoanu:2021}. The cases where $\omega_\text{a}=\omega_\text{b}$ and $g_\text{bs}=g_\text{sq}$, as well as a more manageable expression for the time-evolution operator in the case of our Hamiltonian \eqref{basic:hamiltonian}, can be found in the literature \cite{Estes:Keil:1968}. Furthermore, one can also find examples of existing studies of the rotating wave approximation cases for which $g_\text{sq}=0$, see~\cite{Portes:Rodrigues:2008}.
	
	\subsection{Linear dynamics of $N$ interacting bosonic modes via the symplectic formalism}\label{time:evolution:strong:coupling:regime}
	In this work we consider $N$ bosonic modes with frequencies $\omega_n$ and annihilation and creation operators $\hat{a}_n,\hat{a}^\dag_n$, which satisfy the canonical commutation relations $[\hat{a}_n,\hat{a}_{n'}^\dag]=\delta_{nn'}$ while all other commutators vanish. We collect the creation and the annihilation operators in the vector $\hat{\mathbb{X}}$ of operators defined as $\hat{\mathbb{X}}:=(\hat{a}_1,...,\hat{a}_N,\hat{a}_1^\dag,...,\hat{a}_N^\dag)^{\text{Tp}}$.
	
	Any Hamiltonian $\hat{H}$ of bosonic systems that is quadratic in the creation and annihilation operators can be recast in the in the symplectic form via the relation $\hat{H}=\frac{\hbar}{2}\mathbb{X}^\dag\cdot\boldsymbol{H}\cdot\mathbb{X}$ as explained in the main text, thus obtaining the matrix $\boldsymbol{H}$ that has the generic expression
	\begin{align}
		\boldsymbol{H}
		=
		\begin{pmatrix}
			\boldsymbol{U} & \boldsymbol{V} \\
			\boldsymbol{V}^* & \boldsymbol{U}^*
		\end{pmatrix},
	\end{align}
	where $\boldsymbol{U}$ is a Hermitian and $\boldsymbol{V}$ a symmetric matrix. A time-dependent Hamiltonian $\hat{H}(t)$ induces time evolution in the symplectic formalism through the symplectic matrix
	\begin{align}\label{time:evolution:matrix:strong:coupling:appendix}
		\boldsymbol{S}(t)
		:=
		\overset{\leftarrow}{\mathcal{T}}\exp[\boldsymbol{\Omega}\,\int_0^t\,dt'\,\boldsymbol{H}(t')]
		\equiv
		\begin{pmatrix}
			\boldsymbol{A}(t) & \boldsymbol{B}(t) \\
			\boldsymbol{B}^*(t) & \boldsymbol{A}^*(t)
		\end{pmatrix},
	\end{align}
	where $\overset{\leftarrow}{\mathcal{T}}$ is the time-ordering operator.
	
	Recall that a matrix $\boldsymbol{S}$ is symplectic if it satisfies the constraints $\boldsymbol{S}\boldsymbol{\Omega}\boldsymbol{S}^\dag=\boldsymbol{\Omega}=\boldsymbol{S}^\dag\boldsymbol{\Omega}\boldsymbol{S}$, which allow for its inverse to obtained simply by $\boldsymbol{S}^{-1}=-\boldsymbol{\Omega}\boldsymbol{S}^\dag\boldsymbol{\Omega}$, see \cite{Bruschi:Paraoanu:2021}. The constraints, known as Bogoliubov identities are listed here explicitly for later convenience:
	\begin{align*}
		\left\{
		\begin{array}{lll}
			\mathds{1}_N&=&\boldsymbol{\alpha}\boldsymbol{\alpha}^\dag- \boldsymbol{\beta}\boldsymbol{\beta}^\dag\\
			0 &=&\boldsymbol{\alpha}\boldsymbol{\beta}^{\text{Tp}}- \boldsymbol{\beta}\boldsymbol{\alpha}^{\text{Tp}}\\
			0&=&\boldsymbol{\beta}^*\boldsymbol{\alpha}^\dag- \boldsymbol{\alpha}^*\boldsymbol{\beta}^\dag\\
			\mathds{1}_N&=&\boldsymbol{\alpha}^*\boldsymbol{\alpha}^{\text{Tp}}-\boldsymbol{\beta}^*\boldsymbol{\beta}^{\text{Tp}}
		\end{array}
		\right.,
		\quad\quad\quad
		\left\{
		\begin{array}{lll}
			\mathds{1}_N&=&\boldsymbol{\alpha}^\dag\boldsymbol{\alpha}- \boldsymbol{\beta}^{\text{Tp}}\boldsymbol{\beta}^*\\
			0 &=&\boldsymbol{\alpha}^\dag\boldsymbol{\beta}- \boldsymbol{\beta}^{\text{Tp}}\boldsymbol{\alpha}^*\\
			0&=&\boldsymbol{\beta}^\dag\boldsymbol{\alpha}- \boldsymbol{\alpha}^{\text{Tp}}\boldsymbol{\beta}^*\\
			\mathds{1}_N&=&\boldsymbol{\alpha}^{\text{Tp}}\boldsymbol{\alpha}^*-\boldsymbol{\beta}^\dag\boldsymbol{\beta}
		\end{array}
		\right..
	\end{align*}
	Throughout this work we use the Bogoliubov matrices $\boldsymbol{\alpha},\boldsymbol{\beta}$ for time-independent symplectic matrices, while we use the Bogoliubov matrices $\boldsymbol{A}(t),\boldsymbol{B}(t)$ for time-dependent ones.
	
	We note that Williamson's theorem \cite{Williamson:1923} guarantees that any positive definite Hamiltonian matrix $\boldsymbol{H}$ can be decomposed as $\boldsymbol{H}=\boldsymbol{s}^\dagger\,\boldsymbol{\kappa}\,\boldsymbol{s}$, where the symplectic matrix $\boldsymbol{s}$ is the one that diagonalises the Hamiltonian matrix $\boldsymbol{H}$ and $\boldsymbol{\kappa}=\text{diag}(\nu_1,...,\nu_N,\nu_1,...,\nu_N)=\tilde{\boldsymbol{\kappa}}\oplus\tilde{\boldsymbol{\kappa}}$, where $\tilde{\boldsymbol{\kappa}}:=\text{diag}(\nu_1,...,\nu_N)$ and $\nu_n$ are known as the symplectic eigenvalues and satisfy  $\nu_n\geq1$ for all $n$.
	
	Using the properties of $2N\times2N$ matrices, it can be shown that, for time-independent Hamiltonians, one has
	\begin{align}
		\boldsymbol{S}(t)=\boldsymbol{s}^{-1}\,\exp[\boldsymbol{\Omega}\,\boldsymbol{\kappa}\,t]\,\boldsymbol{s}=-\boldsymbol{\Omega}\,\boldsymbol{s}^\dag\,\boldsymbol{\Omega}\,\exp[\boldsymbol{\Omega}\,\boldsymbol{\kappa}\,t]\,\boldsymbol{s}.
	\end{align}
	Therefore, it is not difficult to obtain the general expression for the Bogoliubov matrices $\boldsymbol{A}(t)$ and $\boldsymbol{B}(t)$, which read
	\begin{align}\label{main:transformation:equation}
		\boldsymbol{A}(t)&=\boldsymbol{\alpha}^\dagger\,e^{-i\,\tilde{\boldsymbol{\kappa}}\,t}\,\boldsymbol{\alpha}-\boldsymbol{\beta}^\text{Tp}\,e^{i\,\tilde{\boldsymbol{\kappa}}\,t}\,\boldsymbol{\beta}^*,\quad\text{and}\quad
		\boldsymbol{B}(t)=\boldsymbol{\alpha}^\dagger\,e^{-i\,\tilde{\boldsymbol{\kappa}}\,t}\,\boldsymbol{\beta}-\boldsymbol{\beta}^\text{Tp}\,e^{i\,\tilde{\boldsymbol{\kappa}}\,t}\,\boldsymbol{\alpha}^*.
	\end{align}
	We can use these expression to write down the formal evolution of the mode operators. Defining $\hat{\mathbb{X}}_\text{red}:=(\hat{a}_1,...,\hat{a}_N)^\text{Tp}$, we have
	\begin{align}\label{time:evolution:operators:strong:coupling}
		\hat{\mathbb{X}}_\text{red}(t)=&\boldsymbol{A}(t)\hat{\mathbb{X}}_\text{red}+\boldsymbol{B}(t)\hat{\mathbb{X}}_\text{red}^*,\quad\text{and}\quad
		\hat{\mathbb{X}}_\text{red}^*(t)=\boldsymbol{B}^*(t)\hat{\mathbb{X}}_\text{red}+\boldsymbol{A}^*(t)\hat{\mathbb{X}}^*_\text{red}.
	\end{align}

	\subsection{Linear dynamics of 2 interacting bosonic modes via the symplectic formalism}\label{time:evolution:strong:coupling:regime:two:modes}
	Here we specify the previous general tools to 2 modes. The ambition is to obtain the Bogoliubov matrices $\boldsymbol{\alpha}$ and $\boldsymbol{\beta}$ as function of the known parameters contained in the Hamiltonian matrix $\boldsymbol{H}$. The computations given above have been obtained first in the literature \cite{Bruschi:Paraoanu:2021}.
	
	We start by collecting the creation and the annihilation operators in the vector $\hat{\mathbb{X}}$ of operators defined as $\hat{\mathbb{X}}:=(\hat{a},\hat{b},\hat{a}^\dag,\hat{b}^\dag)^{\text{Tp}}$. We then assume here that $g_\text{sq}=g_\text{bs}\equiv g$ for convenience. In this case we write
	\begin{align}
		\boldsymbol{\alpha}
		=
		\begin{pmatrix}
			\alpha_{11} & \alpha_{12} \\
			\alpha_{21} & \alpha_{22}
		\end{pmatrix},
		\quad\quad
		\boldsymbol{\beta}
		=
		\begin{pmatrix}
			\beta_{11} & \beta_{12} \\
			\beta_{21} & \beta_{22}
		\end{pmatrix},
	\end{align}
	where we note that $\boldsymbol{\alpha}=\boldsymbol{\alpha}^*$ and $\boldsymbol{\beta}=\boldsymbol{\beta}^*$. This, in turn, implies that $\boldsymbol{A}^{\text{Tp}}(t)=\boldsymbol{A}(t)$ and $\boldsymbol{B}^\dag(t)=-\boldsymbol{B}(t)$.
	The extension to the case where they are complex is straightforward.
	
	The time evolution matrix $\boldsymbol{S}(t)$ has the same formal expression given in \eqref{time:evolution:matrix:strong:coupling:appendix}, where the $2\times2$ matrices $\boldsymbol{A}(t)$ and $\boldsymbol{B}(t)$ read
	\begin{align}\label{main:transformation:equation:main:text}
		\boldsymbol{A}(t)&=\boldsymbol{\alpha}^\text{Tp}\,e^{-i\,\tilde{\boldsymbol{\kappa}}\,t}\,\boldsymbol{\alpha}-\boldsymbol{\beta}^\text{Tp}\,e^{i\,\tilde{\boldsymbol{\kappa}}\,t}\,\boldsymbol{\beta},\quad\quad
		\boldsymbol{B}(t)=\boldsymbol{\alpha}^\text{Tp}\,e^{-i\,\tilde{\boldsymbol{\kappa}}\,t}\,\boldsymbol{\beta}-\boldsymbol{\beta}^\text{Tp}\,e^{i\,\tilde{\boldsymbol{\kappa}}\,t}\,\boldsymbol{\alpha}.
	\end{align}
	Here $\tilde{\boldsymbol{\kappa}}=\text{diag}(\kappa_+,\kappa_-)$.
	These coefficients match those found in the literature for $\omega_\text{a}=\omega_\text{b}$, see \cite{Estes:Keil:1968}.
	In the case of arbitrary coupling constants $g_\text{sq}$ and $g_\text{bs}$, we have
	\begin{align}\label{normal:mode:frequencies:strong:coupling:general:appendix}
		\kappa_\pm^2=\frac{1}{2}\,\bigl[(\omega_\text{a}^2+\omega_\text{b}^2)+2\,(g_\text{bs}^2-g_\text{sq}^2)
		\pm\sqrt{(\omega_\text{a}^2-\omega_\text{b}^2)^2+8\,\omega_\text{a}\,\omega_\text{b}\,(g_\text{bs}^2+g_\text{sq}^2)+4\,(\omega_\text{a}^2+\omega_\text{b}^2)\,(g_\text{bs}^2-g_\text{sq}^2)}\bigr].
	\end{align}
	The operators evolve as
	\begin{align}\label{time:evolution:operators:strong:coupling:appendix}
		\begin{pmatrix}
			\hat{a}(t) \\
			\hat{b}(t) \\
		\end{pmatrix}
		=
		\boldsymbol{A}(t)
		\begin{pmatrix}
			\hat{a} \\
			\hat{b} 
		\end{pmatrix}
		+
		\boldsymbol{B}(t)
		\begin{pmatrix}
			\hat{a}^\dag \\
			\hat{b}^\dag
		\end{pmatrix}.
	\end{align}
	To complement the final expressions \eqref{time:evolution:operators:strong:coupling:appendix} above we need the following constraints
	\begin{align}\label{main:constraints:equation}
		\boldsymbol{U}&=\boldsymbol{\alpha}^\text{Tp}\,\boldsymbol{\kappa}\,\boldsymbol{\alpha}+\boldsymbol{\beta}^\text{Tp}\,\boldsymbol{\kappa}\,\boldsymbol{\beta},\nonumber\\
		\boldsymbol{V}&=\boldsymbol{\alpha}^\text{Tp}\,\boldsymbol{\kappa}\,\boldsymbol{\beta}+\boldsymbol{\beta}^\text{Tp}\,\boldsymbol{\kappa}\,\boldsymbol{\alpha},
	\end{align}
	which can be put in a more convenient form as we now proceed to do.
	
	Notice that we can multiply the first line of \eqref{main:constraints:equation} on the right by $\boldsymbol{\alpha}^\text{Tp}$ and the second line by $\boldsymbol{\beta}^\text{Tp}$ (assuming that $\boldsymbol{\beta}$ is a nonzero matrix), to obtain
	\begin{align}
		\boldsymbol{U}\,\boldsymbol{\alpha}^\text{Tp}&=\boldsymbol{\alpha}^\text{Tp}\,\boldsymbol{\kappa}\,\boldsymbol{\alpha}\,\boldsymbol{\alpha}^\text{Tp}+\boldsymbol{\beta}^\text{Tp}\,\boldsymbol{\kappa}\,\boldsymbol{\beta}\,\boldsymbol{\alpha}^\text{Tp},\nonumber\\
		\boldsymbol{V}\,\boldsymbol{\beta}^\text{Tp}&=\boldsymbol{\alpha}^\text{Tp}\,\boldsymbol{\kappa}\,\boldsymbol{\beta}\,\boldsymbol{\beta}^\text{Tp}+\boldsymbol{\beta}^\text{Tp}\,\boldsymbol{\kappa}\,\boldsymbol{\alpha}\,\boldsymbol{\beta}^\text{Tp}.
	\end{align}
	Subtracting and using the Bogoliubov identities we have $\boldsymbol{U}\,\boldsymbol{\alpha}^\text{Tp}-\boldsymbol{V}\,\boldsymbol{\beta}^\text{Tp}=\boldsymbol{\alpha}^\text{Tp}\,\boldsymbol{\kappa}$. Inverting the products and repeating we get $\boldsymbol{U}\,\boldsymbol{\beta}^\text{Tp}-\boldsymbol{V}\,\boldsymbol{\alpha}^\text{Tp}=-\boldsymbol{\beta}^\text{Tp}\,\boldsymbol{\kappa}$.
	Taking the transpose we obtain the equivalent sets of constraints
	\begin{align}\label{main:constraints:equation:alternative}
		\boldsymbol{\alpha}\,\boldsymbol{U}-\boldsymbol{\beta}\,\boldsymbol{V}&=\boldsymbol{\kappa}\,\boldsymbol{\alpha},\nonumber\\
		\boldsymbol{\beta}\,\boldsymbol{U}-\boldsymbol{\alpha}\,\boldsymbol{V}&=-\boldsymbol{\kappa}\,\boldsymbol{\beta}.
	\end{align}
	These can be then easily solved to obtain the coefficients of $\boldsymbol{\alpha}$ and $\boldsymbol{\beta}$. Notice that if $\boldsymbol{\beta}=0$ form the start, one has that $\boldsymbol{\alpha}$ is orthogonal (since $\boldsymbol{\alpha}$ is assumed to be real -- unitary otherwise) and one arrives at the same constraints \eqref{main:constraints:equation:alternative} by having set $\boldsymbol{\beta}=0$.
	
	We now start from \eqref{main:constraints:equation:alternative} which we write component-wise and have:
	\begin{align}
		\kappa_+\,\beta_{11}+\omega_\text{a}\,\beta_{11}+g\,\beta_{12}=&g\,\alpha_{12},\nonumber\\
		\kappa_+\,\alpha_{11}-\omega_\text{a}\,\alpha_{11}-g\,\alpha_{12}=&-g\,\beta_{12},\nonumber\\
		\kappa_+\,\beta_{12}+g\,\beta_{11}+\omega_\text{b}\,\beta_{12}=&g\,\alpha_{11},\nonumber\\
		\kappa_+\,\alpha_{12}-g\,\alpha_{11}-\omega_\text{b}\,\beta_{12}=&-g\,\beta_{11},\nonumber\\
		\kappa_-\,\beta_{21}+\omega_\text{a}\,\beta_{21}+g\,\beta_{22}=&g\,\alpha_{22},\nonumber\\
		\kappa_-\,\alpha_{21}-\omega_\text{a}\,\alpha_{21}-g\,\alpha_{22}=&-g\,\beta_{22},\nonumber\\
		\kappa_-\,\beta_{22}+g\,\beta_{21}+\omega_\text{b}\,\beta_{22}=&g\,\alpha_{21},\nonumber\\
		\kappa_-\,\alpha_{22}-g\,\alpha_{21}-\omega_\text{b}\,\beta_{22}=&-g\,\beta_{21}.
	\end{align}
	Now, combining lines $1$ and $2$, $3$ and $4$, $5$ and $6$, $7$ and $8$, we find 
	\begin{align}\label{intermediate:step:long:calculation}
		(\kappa_++\omega_\text{a})\,\beta_{11}=(\kappa_+-\omega_\text{a})\,\alpha_{11},\nonumber\\
		(\kappa_++\omega_\text{a})\,\beta_{11}+(\kappa_+-\omega_\text{a})\,\alpha_{11}=&2\,g\,(\alpha_{12}-\beta_{12}),\nonumber\\
		(\kappa_++\omega_\text{b})\,\beta_{12}=(\kappa_+-\omega_\text{b})\,\alpha_{12},\nonumber\\
		(\kappa_++\omega_\text{b})\,\beta_{12}+(\kappa_+-\omega_\text{b})\,\alpha_{12}=&2\,g\,(\alpha_{11}-\beta_{11}),\nonumber\\
		(\kappa_++\omega_\text{a})\,\beta_{21}=(\kappa_+-\omega_\text{a})\,\alpha_{21},\nonumber\\
		(\kappa_+-\omega_\text{a})\,\alpha_{21}+(\kappa_++\omega_\text{a})\,\beta_{21}=&2\,g\,(\alpha_{22}-\beta_{22}),\nonumber\\
		(\kappa_++\omega_\text{b})\,\beta_{22}=(\kappa_+-\omega_\text{b})\,\alpha_{22},\nonumber\\
		(\kappa_+-\omega_\text{b})\,\alpha_{22}+(\kappa_++\omega_\text{b})\,\beta_{22}=&2\,g\,(\alpha_{21}-\beta_{21}).\nonumber\\
	\end{align}
	We use the Bogoliubov identity $\boldsymbol{\alpha}\,\boldsymbol{\alpha}^{\textrm{Tp}}-\boldsymbol{\beta}\,\boldsymbol{\beta}^{\textrm{Tp}}=\mathds{1}_2$, which gives expressions of the form $\alpha_{11}^2+\alpha_{12}^2-\beta_{11}^2-\beta_{12}^2=1$, and the first line of \eqref{intermediate:step:long:calculation} to obtain the identity
	\begin{align}
		4\,\frac{\kappa_+\,\omega_\text{a}}{(\kappa_++\omega_\text{a})^2}\,\alpha_{11}^2+4\,\frac{\kappa_+\,\omega_\text{b}}{(\kappa_++\omega_\text{b})^2}\,\alpha_{12}^2=1,
	\end{align}
	which begs us to make the ansatz
	\begin{align}
		\alpha_{11}=\frac{\kappa_++\omega_\text{a}}{2\,\sqrt{\kappa_+\,\omega_\text{a}}}\cos\theta,\quad\quad\alpha_{12}=\frac{\kappa_++\omega_\text{b}}{2\,\sqrt{\kappa_+\,\omega_\text{b}}}\sin\theta.
	\end{align}
	Using the same reasoning for the other Bogoliubov coefficients, as well as the other Bogoliubov identities, lengthy algebra allows us to show that
	\begin{align}\label{bogoliubov:coefficients:appendix}
		\begin{matrix}
			\alpha_{11}=&\frac{\kappa_++\omega_\text{a}}{2\,\sqrt{\kappa_+\,\omega_\text{a}}}\cos\theta, & \beta_{11}=&\frac{\kappa_+-\omega_\text{a}}{2\,\sqrt{\kappa_+\,\omega_\text{a}}}\cos\theta, \\
			\alpha_{12}=&\frac{\kappa_++\omega_\text{b}}{2\,\sqrt{\kappa_+\,\omega_\text{b}}}\sin\theta, & \beta_{12}=&\frac{\kappa_+-\omega_\text{b}}{2\,\sqrt{\kappa_+\,\omega_\text{b}}}\sin\theta, \\
			\alpha_{21}=&\frac{\kappa_-+\omega_\text{a}}{2\,\sqrt{\kappa_-\,\omega_\text{a}}}\sin\theta, & \beta_{21}=&\frac{\kappa_--\omega_\text{a}}{2\,\sqrt{\kappa_-\,\omega_\text{a}}}\sin\theta, \\
			\alpha_{22}=&-\frac{\kappa_-+\omega_\text{b}}{2\,\sqrt{\kappa_-\,\omega_\text{b}}}\cos\theta, & \beta_{22}=&-\frac{\kappa_--\omega_\text{b}}{2\,\sqrt{\kappa_-\,\omega_\text{b}}}\cos\theta.
		\end{matrix}
	\end{align}
	together with the definition of $\theta$, which is given by the relation
	\begin{align}
		\tan(2\,\theta)=\frac{4\,g\,\sqrt{\omega_\text{a}\,\omega_\text{b}}}{(\omega_\text{a}^2-\omega_\text{b}^2)}.
	\end{align}
	The validity of these expressions requires the condition $0\leq \theta<\pi/2$. Finally, as an extra consistency verification we can see that for $\omega_\text{a}=\omega_\text{b}$ our coefficients reproduce the results existing in the literature  \cite{Estes:Keil:1968}. This implies that $\theta=\pi/4$.

	\subsection{Linear dynamics of 2 interacting bosonic modes in the rotating wave approximation via the symplectic formalism}\label{time:evolution:rwa:appendix}
	Here we briefly review the time evolution induced in the case of the rotating wave approximation.  This corresponds to setting $g_\text{sq}=0$ in the Hamiltonian \eqref{basic:hamiltonian}, which reduces to
	\begin{align}\label{RWA:hamiltonian:appendix}
		\hat{H}_\text{RWA}=\hbar\,\omega_\text{a}\,\hat{a}^\dag\,\hat{a}+\hbar\,\omega_\text{b}\,\hat{b}^\dag\,\hat{b}+\hbar\,g\,\left(\hat{a}\,\hat{b}^\dag+\hat{a}^\dag\,\hat{b}\right).
	\end{align}
	The rotating wave Hamiltonian $\hat{H}_\text{RWA}$ can be easily diagonalised and we can find the eigenfrequencies $\kappa_\pm$ of the normal modes, which we call $\omega_\pm\equiv\kappa_\pm$ for this regime. They read
	\begin{align}\label{normal:mode:frequencies:appendix}
		\omega_\pm=\frac{1}{2}\,\left[(\omega_\text{a}+\omega_\text{b})\pm\sqrt{(\omega_\text{a}-\omega_\text{b})^2+4\,g_\text{bs}^2}\right].
	\end{align}
	They match those found in the literature~\cite{Hertzog:Wang:2019}, as expected. Notice that, although one does not immediately achieve \eqref{normal:mode:frequencies:appendix} by setting $g_\text{sq}=0$ in \eqref{normal:mode:frequencies:strong:coupling:general}, it is immediate to check that $\omega_\pm^2=(\kappa_\pm|_{g_\text{sq}=0})^2$. 
	
	In the following it will prove convenient to introduce $\omega_\Delta:=\frac{1}{2}(\omega_\text{a}-\omega_\text{b})$ and $\omega_\Sigma:=\frac{1}{2}(\omega_\text{a}+\omega_\text{b})$, which allow us to rewrite \eqref{normal:mode:frequencies:appendix} as $\omega_\pm=\omega_\Sigma\pm\omega_\text{bs}$, with $\omega_\text{bs}:=\sqrt{\omega_\Delta^2+g_\text{bs}^2}$.

	We are now in the position of presenting the symplectic representation $\boldsymbol{S}_\text{RWA}(t)=\exp[\boldsymbol{\Omega}\,\boldsymbol{H}_\text{RWA}\,t]$ of the time evolution operator $\hat{U}_\text{RWA}(t)=\exp\bigl[-\frac{i}{\hbar}\,\hat{H}_\text{RWA}\,t\bigr]$. The Hamiltonian matrix $\boldsymbol{H}_\text{RWA}$ for this case reads
	\begin{align}\label{time:evolution:rotating:wave:approximation:appendix}
		\boldsymbol{H}_\text{RWA}
		=
		\begin{pmatrix}
			\omega_\text{a} & g & 0 & 0\\
			g & \omega_\text{b} & 0 & 0\\
			0 & 0 & \omega_\text{a} & g\\
			0 & 0 & g & \omega_\text{b}
		\end{pmatrix}.
	\end{align}
	This reads $\boldsymbol{S}_\text{RWA}(t)=\tilde{\boldsymbol{S}}_\text{RWA}(t)\oplus\tilde{\boldsymbol{S}}_\text{RWA}^*(t)$, where the $2\times2$ matrix $\tilde{\boldsymbol{S}}_\text{RWA}(t)$ is unitary and has the expression
	\begin{align}\label{time:evolution:of:mode:operators:matrix}
		\tilde{\boldsymbol{S}}_\text{RWA}(t) 
		=
		e^{-i\,\omega_\Sigma\,t}
		\begin{pmatrix}
			\chi_\varphi(t) & -i\,\xi_\varphi(t) \\
			-i\,\xi_\varphi(t) & \chi_\varphi^*(t) 
		\end{pmatrix},
	\end{align}
	and the coefficients in the matrix read $\chi_\varphi(t):=\cos(\omega_\text{bs}\,t)-i\,c_{2\varphi} \sin(\omega_\text{bs}\,t)$ and $\xi_\varphi(t):=s_{2\varphi} \sin(\omega_\text{bs}\,t)$. The angle $\varphi$ is defined through $c_{2\varphi}\equiv\cos(2\,\varphi):=\omega_\Delta/\omega_\text{bs}$ and $s_{2\varphi}\equiv\sin(2\,\varphi):=g_\text{bs}/\omega_\text{bs}$ and $\tan(2\,\varphi)=g_\text{bs}/\omega_\Delta$.
	Another convenient way to write $\tilde{\boldsymbol{S}}_\text{RWA}(t) $ is to employ the Pauli matrices $\boldsymbol{\sigma}_k$ and write
	\begin{align}
		\tilde{\boldsymbol{S}}_\text{RWA}(t) 
		=
		e^{-i\,\omega_\Sigma\,t}
		\left(\cos(\omega_\text{bs}\,t)\mathds{1}_2-i \vec{v}(t)\vec{\boldsymbol{\sigma}}\right),
	\end{align}
	where we have defined $\vec{v}(t):=(s_{2\varphi} \sin(\omega_\text{bs}\,t),0,c_{2\varphi} \sin(\omega_\text{bs}\,t))$, and $\vec{\boldsymbol{\sigma}}:=(\boldsymbol{\sigma}_{\text{x}},\boldsymbol{\sigma}_{\text{y}},\boldsymbol{\sigma}_{\text{z}})$ is a vector of Pauli matrices. Notice that $\cos^2(\omega_\text{bs}\,t)+|\vec{v}(t)|^2=1$ as expected.
	This expression allows us to also conveniently write $\boldsymbol{S}_\text{RWA}(t)$ as
	\begin{align}
		\boldsymbol{S}_\text{RWA}(t) 
		=
		\boldsymbol{S}_0(t)
		\left(
		\cos(\omega_\text{bs}\,t)
		\begin{pmatrix}
			\mathds{1}_2 & 0\\
			0 & \mathds{1}_2
		\end{pmatrix}
		-i 
		\begin{pmatrix}
			\vec{v}(t)\vec{\boldsymbol{\sigma}} & 0\\
			0 & -\vec{v}(t)\vec{\boldsymbol{\sigma}}
		\end{pmatrix}
		\right),
	\end{align}
	where $\boldsymbol{S}_0(t):=\text{diag}(e^{-i\,\omega_\Sigma\,t},e^{-i\,\omega_\Sigma\,t},e^{i\,\omega_\Sigma\,t},e^{i\,\omega_\Sigma\,t})$.
	
	This allows us to introduce the reduced operator vector $\hat{\mathbb{X}}_\text{R}:=(\hat{a},\hat{b})^\text{Tp}$ and therefore write $\hat{\mathbb{X}}_\text{R}(t)=\tilde{\boldsymbol{S}}_\text{RWA}(t) \hat{\mathbb{X}}_\text{R}$. Explicitly, we have
	\begin{align}\label{time:evolution:of:mode:operators}
		\hat{a}(t)=&
		e^{-i\,\omega_\Sigma\,t}\left(\chi_\varphi(t)\,\hat{a}-i\,\xi_\varphi(t)\,\hat{b}\right),\nonumber\\
		\,\hat{b}(t)=&
		e^{-i\,\omega_\Sigma\,t}\left(-i\,\xi_\varphi(t)\,\hat{a}+\chi_\varphi^*(t)\,\hat{b}\right).
	\end{align}
	These results have been obtained in the literature and they complement the remaining work. The are presented for the sake of the reader.

	\subsection{Comparing two dynamics: effective time evolution}\label{RWA:evolution:appendix}
	We proceed here by writing the expression \eqref{time:evolution:matrix:strong:coupling:appendix} as $\boldsymbol{S}(t)=\boldsymbol{S}_\text{RWA}(t)\,\boldsymbol{S}_\text{eff}(t)$, or equivalently $\boldsymbol{S}_\text{eff}(t)
	=\boldsymbol{S}_\text{RWA}^\dag(t)\,\boldsymbol{S}(t)$, where we have introduced 
	\begin{align}\label{general:s:matrix:ansatz:appendix}
		\boldsymbol{S}_\text{eff}(t)
		&=\overset{\leftarrow}{\mathcal{T}}\exp\left[g\,\boldsymbol{\Omega}\,\int_0^t\,\dd t'\,\boldsymbol{S}^\dag_\text{RWA}(t')\,\boldsymbol{P}\,\boldsymbol{S}_\text{RWA}(t')\right],
	\end{align}
	as well as the matrix
	\begin{align}
		\boldsymbol{P}:=&
		\begin{pmatrix}
			0 & \boldsymbol{\sigma}_\text{x} \\
			-\boldsymbol{\sigma}_\text{x} & 0 
		\end{pmatrix}
		=
		\begin{pmatrix}
			0 & 0 & 0 & 1 \\
			0 & 0 & 1 & 0 \\
			0 & -1 & 0 & 0 \\
			-1 & 0 & 0 & 0
		\end{pmatrix}.
	\end{align}
	Here, the term $\boldsymbol{S}_\text{eff}(t)$ encodes the ``correction'' that occurs by taking the rotating wave approximation in the Hamiltonian \eqref{basic:hamiltonian}.
	This can be also recast in the operator language as
	\begin{align}
		\hat{U}_\text{eff}(t)
		&=\overset{\leftarrow}{\mathcal{T}}\exp\left[-i\,g\,\int_0^t\,dt'\,\hat{U}^\dag_\text{RWA}(t')\,\left(\hat{a}^\dag\,\hat{b}^\dag+\hat{a}\,\hat{b}\right)\,\hat{U}_\text{RWA}(t')\right].
	\end{align}
	We now employ techniques presented in a different context in the literature \cite{Qvarfort:Serafini:2019} to compute \ref{general:s:matrix:ansatz:appendix}. We make the ansatz 
	\begin{align}\label{effective:s:matrix:ansatz:appendix}
		\boldsymbol{S}_\text{eff}(t)=\boldsymbol{L}(t)+g\,\boldsymbol{\Omega}\,\int_0^t dt'\,\boldsymbol{M}(t')\,\boldsymbol{L}(t'),
	\end{align}
	where $\boldsymbol{L}(t)$ is a block diagonal matrix and we have introduced $\boldsymbol{M}(t):=\boldsymbol{S}^\dag_\text{RWA}(t)\,\boldsymbol{P}\,\boldsymbol{S}_\text{RWA}(t)$ for notational convenience. It is immediate to verify that the following combinations of matrices are all identical to the identity: $\boldsymbol{M}^{-1}(t)=-\boldsymbol{M}(t)$ and $(\boldsymbol{\Omega}\boldsymbol{M}(t)\boldsymbol{\Omega})^2=\boldsymbol{M}(t)\boldsymbol{\Omega}\boldsymbol{M}(t)\boldsymbol{\Omega}=\boldsymbol{\Omega}\boldsymbol{M}(t)\boldsymbol{\Omega}\boldsymbol{M}(t)=-\mathds{1}_4$.
	
	We now take the time derivative of both sides of equations \eqref{general:s:matrix:ansatz:appendix} and \eqref{effective:s:matrix:ansatz:appendix}, equate and obtain
	\begin{align*}
		\dot{\boldsymbol{L}}(t)+\,g\,\boldsymbol{\Omega}\,\boldsymbol{M}(t)\,\boldsymbol{L}(t)&=g\,\boldsymbol{\Omega}\boldsymbol{M}(t)\boldsymbol{S}_\text{eff}(t)\\
		\dot{\boldsymbol{L}}(t)+\,g\,\boldsymbol{\Omega}\,\boldsymbol{M}(t)\,\boldsymbol{L}(t)&=g\,\boldsymbol{\Omega}\,\boldsymbol{M}(t)\left(\boldsymbol{L}(t)+g\,\boldsymbol{\Omega}\,\int_0^t dt'\,\boldsymbol{M}(t')\,\boldsymbol{L}(t')\right)\\
		\dot{\boldsymbol{L}}(t)&=g^2\,\boldsymbol{\Omega}\,\boldsymbol{M}(t)\,\boldsymbol{\Omega}\,\int_0^t dt'\,\boldsymbol{M}(t')\,\boldsymbol{L}(t')\\
		-\boldsymbol{\Omega}\,\boldsymbol{M}(t)\,\boldsymbol{\Omega}\,\dot{\boldsymbol{L}}(t)&=g^2\,\int_0^t dt'\,\boldsymbol{M}(t')\,\boldsymbol{L}(t').
	\end{align*}

	We now note that, since  $\boldsymbol{M}^{-1}(t)=-\boldsymbol{M}(t)$, it follows that $\boldsymbol{M}^2(t)=-\mathds{1}_4$, and therefore $\boldsymbol{M}(t)\dot{\boldsymbol{M}}(t)=-\dot{\boldsymbol{M}}(t)\boldsymbol{M}(t)$. 
	
	Taking the time derivative of both sides gives us
	\begin{align*}
		-\boldsymbol{\Omega}\,\dot{\boldsymbol{M}}(t)\,\boldsymbol{\Omega}\,\dot{\boldsymbol{L}}(t)-\boldsymbol{\Omega}\,\boldsymbol{M}(t)\,\boldsymbol{\Omega}\,\ddot{\boldsymbol{L}}(t)&=g^2\,\boldsymbol{M}(t)\,\boldsymbol{L}(t)\\
		-\boldsymbol{M}(t)\,\boldsymbol{\Omega}\,\dot{\boldsymbol{M}}(t)\,\boldsymbol{\Omega}\,\dot{\boldsymbol{L}}(t)+\ddot{\boldsymbol{L}}(t)&=-g^2\,\boldsymbol{L}(t),
	\end{align*}
	where one used $\boldsymbol{\Omega}\boldsymbol{M}(t)\boldsymbol{\Omega}=\boldsymbol{M}(t)$.
	
	This gives us the differential equation
	\begin{align*}
		\ddot{\boldsymbol{L}}(t)-\boldsymbol{M}(t)\,\boldsymbol{\Omega}\,\dot{\boldsymbol{M}}(t)\,\boldsymbol{\Omega}\,\dot{\boldsymbol{L}}(t)+g^2\,\boldsymbol{L}(t)=0.
	\end{align*}
	Finally, we use the identity $\boldsymbol{A}\boldsymbol{B}=\frac{1}{2}\{\boldsymbol{A},\boldsymbol{B}\}+\frac{1}{2}[\boldsymbol{A},\boldsymbol{B}]$, together with the fact that $(\boldsymbol{M}(t)\boldsymbol{\Omega})^2=-\mathds{1}_4$ to obtain
	\begin{align*}
		\ddot{\boldsymbol{L}}(t)-\frac{1}{2}[\boldsymbol{M}(t)\,\boldsymbol{\Omega},\dot{\boldsymbol{M}}(t)\,\boldsymbol{\Omega}]\dot{\boldsymbol{L}}(t)+g^2\,\boldsymbol{L}(t)=0.
	\end{align*}

		\section{Comparison via the fidelity of pure Gaussian states}\label{Gaussian:fidelity:appendix}
		Here we restrict ourselves to Gaussian states.
		We have already introduced Gaussian states and the covariance matrix formalism in the main text of the paper.
		Given two Gaussian states $\boldsymbol{\sigma}$ and $\boldsymbol{\sigma}'$ of two modes, the fidelity $\mathcal{F}(\boldsymbol{\sigma},\boldsymbol{\sigma}')$ between the two states that have \textit{vanishing initial first moments} has been already obtained in the literature \cite{Paraoanu:Scutaru:2000,Marian:Marian:2012}, and reads
		\begin{align}\label{fidelity:Gaussian:states:appendix}
			\mathcal{F}(\boldsymbol{\sigma},\boldsymbol{\sigma}'):=\frac{4}{\sqrt{\Lambda}+\sqrt{\Gamma}-\sqrt{(\sqrt{\Lambda}-\sqrt{\Gamma})^2-\Delta}},
		\end{align}
		where we have introduced
		\begin{align}\label{lambda:gamma:delta:appendix}
			\Gamma:=&\det(\mathds{1}_4-\boldsymbol{\Omega}\,\boldsymbol{\sigma}\,\boldsymbol{\Omega}\,\boldsymbol{\sigma}'),\nonumber\\
			\Lambda:=&\det(\mathds{1}_4+i\,\boldsymbol{\Omega}\,\boldsymbol{\sigma})\,\det(\mathds{1}_4+i\,\boldsymbol{\Omega}\,\boldsymbol{\sigma}'),\nonumber\\
			\Delta:=&\det(\boldsymbol{\sigma}+\boldsymbol{\sigma}').
		\end{align}
		Note that in our case we wish to study the fidelity $\mathcal{F}$ between an initial Gaussian state $\boldsymbol{\sigma}_0$ and the state $\boldsymbol{\sigma}(t):=\boldsymbol{S}_\text{eff}(t)\,\boldsymbol{\sigma}_0\,\boldsymbol{S}_\text{eff}^\dag(t)$. This observation will prove useful below.
		
		Crucially, we restrict ourselves to \textit{pure} Gaussian states $\boldsymbol{\sigma}$, i.e., $\det(\boldsymbol{\sigma})=1$. This also implies that $\boldsymbol{\sigma}=\boldsymbol{s}\,\boldsymbol{s}^\dag$ for an appropriate symplectic matrix $\boldsymbol{s}$, which is equivalent to saying that the Williamson form $\boldsymbol{\nu}_{\oplus}$ of the state $\boldsymbol{\sigma}$ is just the identity matrix \cite{Adesso:Ragy:2014}.
		
		Our initial state $\boldsymbol{\sigma}_0$ therefore reads $\boldsymbol{\sigma}_0:=\boldsymbol{s}_0\,\boldsymbol{s}^\dag_0$, while the state it will be compared to reads $\boldsymbol{\sigma}(t):=\boldsymbol{S}_\text{eff}(t)\,\boldsymbol{\sigma}_0\,\boldsymbol{S}_\text{eff}^\dag(t)$.
		In the following we separately compute the quantities \eqref{lambda:gamma:delta:appendix}.
		
		\begin{itemize}
			\item \textbf{Computing $\Lambda$}: It is immediate to show that, for pure states $\boldsymbol{\sigma},\boldsymbol{\sigma}'$, we have $\det(\mathds{1}_4+i\,\boldsymbol{\Omega}\,\boldsymbol{\sigma})=0$, since $\det(\mathds{1}_4+i\,\boldsymbol{\Omega}\,\boldsymbol{\sigma})=\prod_n(1-\nu^2_n)$, where $\nu_n$ are the symplectic eigenvalues of $\boldsymbol{\sigma}$ and for pure states we have $\nu_n\equiv1$. The same occurs for $\boldsymbol{\sigma}'$. Therefore, we conclude that $\Lambda=0$. 
			
			\item \textbf{Computing $\Delta$}: Recalling that the determinant of a symplectic matrix $\boldsymbol{S}$ is always unity, and using the fact that $\boldsymbol{S}^{-1}=-\boldsymbol{\Omega}\boldsymbol{S}^\dag\boldsymbol{\Omega}$ for any symplectic matrix $\boldsymbol{S}$, we have {
				\begin{align}
					\Delta:=&\,\det\left(\boldsymbol{\sigma}+\boldsymbol{\sigma}'\right)=\det\left(\boldsymbol{s}\,\boldsymbol{s}^\dag+\boldsymbol{\sigma}'\right)=\det\left(\mathds{1}_4+\boldsymbol{s}^{\dag-1}\,\boldsymbol{s}^{-1}\,\boldsymbol{\sigma}'\right)\nonumber\\
					=&\,\det\left(\mathds{1}_4-\boldsymbol{\Omega}\,\boldsymbol{s}\,\boldsymbol{s}^{\dag}\,\boldsymbol{\Omega}\,\boldsymbol{\sigma}'\right)=\det\left(\mathds{1}_4-\boldsymbol{\Omega}\,\boldsymbol{\sigma}\,\boldsymbol{\Omega}\,\boldsymbol{\sigma}'\right)=\Gamma.
			\end{align}}
			We have proven that, for pure states, it always occurs that $\Gamma=\Delta$ \textit{regardless} of the form of the states considered.
		\end{itemize}
		
		Here we have obtained an important result: the fidelity between two pure and centred Guassian states $\boldsymbol{\sigma},\boldsymbol{\sigma}'$ reads
		\begin{align}\label{fidelity:Gaussian:states:first}
			\mathcal{F}(\boldsymbol{\sigma}+\boldsymbol{\sigma}')=\frac{4}{\sqrt{\Gamma}}.
		\end{align}
		
		We continue with computing the key quantity $\Gamma$. 
		\begin{itemize}
			\item \textbf{Computing $\Gamma$}: We use the result just proven, namely that $\Gamma(t)=\Delta(t)$, and specialize to our case, where the two state considered are $\boldsymbol{\sigma}_0$ and $\boldsymbol{\sigma}(t)=\boldsymbol{S}_\text{eff}(t)\,\boldsymbol{\sigma}_0\,\boldsymbol{S}_\text{eff}^\dag(t)$. We then show that {\small
			\begin{align}
				\Gamma(t)&=\det\left(\boldsymbol{\sigma}_0+\boldsymbol{\sigma}(t)\right)=\det\left(\boldsymbol{\sigma}_0+\boldsymbol{S}_\text{eff}(t)\,\boldsymbol{\sigma}_0\,\boldsymbol{S}_\text{eff}^\dag(t)\right)=\det\left(\boldsymbol{\sigma}_0\,\boldsymbol{S}_\text{eff}^{\dag-1}(t)+\boldsymbol{S}_\text{eff}(t)\,\boldsymbol{\sigma}_0\right)\nonumber\\
				&=\det\left(-\boldsymbol{\sigma}_0\,\boldsymbol{\Omega}\,\boldsymbol{S}_\text{eff}(t)\,\boldsymbol{\Omega}+\boldsymbol{S}_\text{eff}(t)\,\boldsymbol{\sigma}_0\right)=\det\left(\boldsymbol{\sigma}_0\,\boldsymbol{\Omega}\,\boldsymbol{S}_\text{eff}(t)+\boldsymbol{S}_\text{eff}(t)\,\boldsymbol{\sigma}_0\,\boldsymbol{\Omega}\right),
			\end{align}}
			which gives us the result
			\begin{align}
				\Gamma(t)=\det(\boldsymbol{\sigma}_0\,\boldsymbol{\Omega}\,\boldsymbol{S}_\text{eff}(t)+\boldsymbol{S}_\text{eff}(t)\,\boldsymbol{\sigma}_0\,\boldsymbol{\Omega}).
			\end{align}
			We now use the expression of the initial state $\boldsymbol{\sigma}_0$ to obtain
			\begin{align}
				\Gamma(t)=\Delta(t)&=\det(\boldsymbol{\sigma}_0\,\boldsymbol{\Omega}\,\boldsymbol{S}_\text{eff}(t)+\boldsymbol{S}_\text{eff}(t)\,\boldsymbol{\sigma}_0\,\boldsymbol{\Omega})\nonumber\\
				&=\det(\boldsymbol{s}_0^{\dag}\,\boldsymbol{\Omega}\,\boldsymbol{S}_\text{eff}(t)+\boldsymbol{s}_0^{-1}\,\boldsymbol{S}_\text{eff}(t)\,\boldsymbol{s}_0\,\boldsymbol{s}_0^{\dag}\,\boldsymbol{\Omega})\nonumber\\
				&=\det(\boldsymbol{s}_0^{\dag}\,\boldsymbol{\Omega}\,\boldsymbol{s}_0\boldsymbol{s}_0^{-1}\boldsymbol{S}_\text{eff}(t)+\boldsymbol{s}_0^{-1}\,\boldsymbol{S}_\text{eff}(t)\,\boldsymbol{s}_0\,\boldsymbol{s}_0^{\dag}\,\boldsymbol{\Omega}\boldsymbol{s}_0\boldsymbol{s}_0^{-1})\nonumber\\
				&=\det(\boldsymbol{\Omega}\,\boldsymbol{s}_0^{-1}\boldsymbol{S}_\text{eff}(t)+\boldsymbol{s}_0^{-1}\,\boldsymbol{S}_\text{eff}(t)\,\boldsymbol{s}_0\,\boldsymbol{\Omega}\,\boldsymbol{s}_0^{-1})\nonumber\\
				&=\det(\boldsymbol{\Omega}\,\boldsymbol{s}_0^{-1}\boldsymbol{S}_\text{eff}(t)\boldsymbol{s}_0+\boldsymbol{s}_0^{-1}\,\boldsymbol{S}_\text{eff}(t)\,\boldsymbol{s}_0\,\boldsymbol{\Omega}),
			\end{align}
			which provides us with the expression
			\begin{align}
				\Gamma(t)=\Delta(t)=&\det(\boldsymbol{\Omega}\,\boldsymbol{S}_\text{f}(t)+\boldsymbol{S}_\text{f}(t)\,\boldsymbol{\Omega}).
			\end{align}
			Here we have introduced the symplectic matrix $\boldsymbol{S}_\text{f}(t):=\boldsymbol{s}_0^{-1}\boldsymbol{S}_\text{eff}(t)\boldsymbol{s}_0=\boldsymbol{s}_0^{-1}\boldsymbol{S}^{-1}_\text{R}(t)\boldsymbol{S}_\text{T}(t)\boldsymbol{s}_0$, with the standard symplectic matrix decomposition
			\begin{align}
				\boldsymbol{S}_\text{f}(t)
				=
				\begin{pmatrix}
					\boldsymbol{A}_\text{f}(t) & \boldsymbol{B}_\text{f}(t)\\
					\boldsymbol{B}_\text{f}^*(t) & \boldsymbol{A}_\text{f}^*(t)
				\end{pmatrix}.
			\end{align}
			We have the very compact expression
			\begin{align}
				\Gamma(t)=\Delta(t)=&16|\det(\boldsymbol{A}_\text{f}(t))|^2,
			\end{align}
			which in turn gives us the fidelity $\mathcal{F}_\text{eff}(t)\equiv\mathcal{F}(\boldsymbol{\sigma}_0,\boldsymbol{S}_\text{eff}(t)\,\boldsymbol{\sigma}_0\,\boldsymbol{S}_\text{eff}^\dag(t))$ as
			\begin{align}\label{fidelity:Gaussian:states:final:appendix}
				\mathcal{F}_\text{eff}(t)=\frac{1}{|\det(\boldsymbol{A}_\text{f}(t))|}=\frac{1}{\sqrt{\det(\mathds{1}_2+\boldsymbol{B}^\dag_\text{f}(t)\boldsymbol{B}_\text{f}(t))}},
			\end{align}
			where the last equality has been obtained using the identity $\boldsymbol{A}^\text{Tp}\boldsymbol{A}^*=\mathds{1}_N+\boldsymbol{B}^\dag\boldsymbol{B}$ valid for any submatrices of a symplectic matrix.
		\end{itemize}

		\section{Algebraic relations of expectation values}
		We are interested in computing the quantity of the form $\Tr\bigl(\boldsymbol{\beta}^\dag(t)\boldsymbol{\beta}(t)\boldsymbol{\beta}^\dag(t)\boldsymbol{\beta}(t)\bigr)$, which was obtained in the context of the relation between the fidelity and the moments of statistical distribution of the average number of excitations in a Gaussian state. The computations presented below are independent of the number of modes considered.
		
		To do this we note that $\hat{a}_k(t)=\sum_p(\alpha_{kp}(t)\hat{a}_p+\beta_{kp}(t)\hat{a}^\dag_p)$ according to our the evolution induced by quadratic Hamiltonians. 
		Let us define the total number operator $\hat{N}_{\text{tot}}(t)=\sum_k\hat{a}^\dag_k(t)\hat{a}_k(t)$, which coincides with the total variation $\Delta\hat{N}_{\text{tot}}(t):=\hat{N}_{\text{tot}}(t)-\hat{N}_{\text{tot}}(0)$ of excitations for our initial state since $\hat{N}_{\text{tot}}(0)|0\rangle=0$ and $\langle0|\hat{N}_{\text{tot}}(0)|0\rangle=0$. 
		Furthermore,
		\begin{align*}
			\langle0|(\Delta\hat{N}_{\text{tot}}(t))^2|0\rangle&=\langle0|(\hat{N}_{\text{tot}}(t)-\hat{N}_{\text{tot}}(0))(\hat{N}_{\text{tot}}(t)-\hat{N}_{\text{tot}}(0))|0\rangle\\
			&=\langle0|(\hat{N}_{\text{tot}}(t))^2|0\rangle-\langle0|\hat{N}_{\text{tot}}(0)\hat{N}_{\text{tot}}(t)|0\rangle-\langle0|\hat{N}_{\text{tot}}(t)\hat{N}_{\text{tot}}(0)|0\rangle+\langle0|(\hat{N}_{\text{tot}}(0))^2|0\rangle\\
			&=\langle0|(\hat{N}_{\text{tot}}(t))^2|0\rangle,
		\end{align*}
		which lets us conclude that also $\langle0|(\Delta\hat{N}_{\text{tot}}(t))^2|0\rangle=\langle0|(\hat{N}_{\text{tot}}(t))^2|0\rangle$. This will be necessary in the next step.
		
		We wish to compute the quantity $\Delta N^2_{\text{tot}}(t):=\langle0|(\Delta\hat{N}_{\text{tot}}(t))^2|0\rangle$, that is, the expectation value of the square of the total variation of excitations in the vacuum state. Below, we drop the explicit dependence on time in order to lighten the computations. 
		The first expression that we need is the one obtained before, namely
		\begin{align}
			\Delta N_{\text{tot}}{(t)}=\Tr(\boldsymbol{\beta}^\dag(t)\boldsymbol{\beta}(t)).
		\end{align}
		We then have {\small
		\begin{align*}
			\hfill\Delta N^2_{\text{tot}}{(t)}&=\langle0|(\Delta\hat{N}_{\text{tot}}(t))^2|0\rangle=\langle0|(\hat{N}_{\text{tot}}(t))^2|0\rangle\\
			&=\sum_{kk'}\langle0|\hat{a}^\dag_k(t)\hat{a}_k(t)\hat{a}^\dag_{k'}(t)\hat{a}_{k'}(t)|0\rangle\\
			&=\sum_{kk'}\sum_{pp'qq'}\langle0|(\alpha_{kp}(t)\hat{a}_p+\beta_{kp}(t)\hat{a}^\dag_p)^\dag(\alpha_{kp'}(t)\hat{a}_{p'}+\beta_{kp'}(t)\hat{a}^\dag_{p'})(\alpha_{k'q}(t)\hat{a}_q+\beta_{k'q}(t)\hat{a}^\dag_q)^\dag(\alpha_{k'q'}(t)\hat{a}_{q'}+\beta_{k'q'}(t)\hat{a}^\dag_{q'})|0\rangle\\
			&=\sum_{kk'}\sum_{pp'qq'}\bigl[\beta^*_{kp}(t)\alpha_{kp'}(t)\alpha_{k'q}^*(t)\beta_{k'q'}(t)\langle0|\hat{a}_p\hat{a}_{p'}\hat{a}^\dag_q\hat{a}^\dag_{q'}|0\rangle+\beta^*_{kp}(t)\beta_{kp'}(t)\beta_{k'q}^*(t)\beta_{k'q'}(t)\langle0|\hat{a}_p\hat{a}^\dag_{p'}\hat{a}_q\hat{a}_{q'}^\dag|0\rangle\bigr]\\
			&=\sum_{kk'}\sum_{pp'qq'}\bigl[\beta^*_{kp}(t)\alpha_{kp'}(t)\alpha_{k'q}^*(t)\beta_{k'q'}(t)\left(\delta_{p'q}\delta_{pq'}+\delta_{p'q'}\delta_{pq}\right)+\beta^*_{kp}(t)\beta_{kp'}(t)\beta_{k'q}^*(t)\beta_{k'q'}(t)\delta_{qq'}\delta_{pp'}\bigr]\\
			&=\sum_{kk'}\sum_{pq}\alpha_{kq}(t)\alpha^\dag_{qk'}(t)\beta_{k'p}(t)\beta^\dag_{pk}(t)
			+\sum_{kk'}\sum_{pp'}\beta_{k'p'}(t)\alpha^{\text{Tp}}_{p'k}(t)\beta^*_{kp}(t)\alpha^\dag_{pk'}(t)+\sum_{kk'}\sum_{pq}\beta_{kp}(t)\beta^\dag_{pk}(t)\beta_{k'q}(t)\beta^\dag_{qk'}(t).
		\end{align*}}
		We now proceed to cast the above expression into matrix form. We have
		\begin{align*}
			\Delta N^2_{\text{tot}}(t)&=\Tr(\boldsymbol{\alpha}(t)\boldsymbol{\alpha}^\dag(t)\boldsymbol{\beta}(t)\boldsymbol{\beta}^\dag(t))
			+\Tr(\boldsymbol{\beta}(t)\boldsymbol{\alpha}^{\text{Tp}}(t)\boldsymbol{\beta}^*(t)\boldsymbol{\alpha}^\dag(t))+\Tr^2(\boldsymbol{\beta}^\dag(t)\boldsymbol{\beta}(t))\\
			&=\Tr(\boldsymbol{\alpha}(t)\boldsymbol{\alpha}^\dag(t)\boldsymbol{\beta}(t)\boldsymbol{\beta}^\dag(t))
			+\Tr(\boldsymbol{\beta}(t)\boldsymbol{\alpha}^{\text{Tp}}(t)\boldsymbol{\alpha}^*(t)\boldsymbol{\beta}^\dag(t))+\Tr^2(\boldsymbol{\beta}^\dag(t)\boldsymbol{\beta}(t))\\
			&=\Tr(\boldsymbol{\alpha}(t)\boldsymbol{\alpha}^\dag(t)\boldsymbol{\beta}(t)\boldsymbol{\beta}^\dag(t))
			+\Tr(\boldsymbol{\beta}^\dag(t)\boldsymbol{\beta}(t)\boldsymbol{\alpha}^{\text{Tp}}(t)\boldsymbol{\alpha}^*(t))+\Tr^2(\boldsymbol{\beta}^\dag(t)\boldsymbol{\beta}(t))\\
			&=2\Tr(\boldsymbol{\beta}^\dag(t)\boldsymbol{\beta}(t))+2\Tr(\boldsymbol{\beta}^\dag(t)\boldsymbol{\beta}(t)\boldsymbol{\beta}^\dag(t)\boldsymbol{\beta}(t))
			+\Tr^2(\boldsymbol{\beta}^\dag(t)\boldsymbol{\beta}(t)).
		\end{align*}
		Therefore,
		\begin{align}\label{useful:expression:quartic:betas:appendix}
			\Tr(\boldsymbol{\beta}^\dag(t)\boldsymbol{\beta}(t)\boldsymbol{\beta}^\dag(t)\boldsymbol{\beta}(t))=\frac{1}{2}\left[\Delta N^2_{\text{tot}}(t)-2\Delta N_{\text{tot}}(t)-(\Delta N_{\text{tot}}(t))^2\right].
		\end{align}

		\section{Perturbative expansion of important quantities}
		Here we compute the perturbative expansion of important quantities when the assumption $\omega_{\text{a}}=\omega_{\text{b}}\equiv\omega$ of the rotating wave approximation applies. To proceed, we normalize all quantities by $\omega$. In particular, we have $\tilde{g}:=g/\omega$, $\tau:=\omega t$, $\tilde{\kappa}_\pm=\kappa_\pm/\omega$.
		In this case, it is immediate to see that the symplectic frequencies read $\tilde{\kappa}_\pm=\sqrt{1+2\tilde{g}}$.
		
		We start with the Bogoliubov coefficients $\boldsymbol{\alpha}$ and $\boldsymbol{\beta}$.
		Noting that here we have $\theta=\pi/4$, it is not difficult to see that
		\begin{align}
			\boldsymbol{\alpha}
			=\frac{1}{\sqrt{2}}
			\begin{pmatrix}
				1 & 1\\
				1 & -1
			\end{pmatrix}
			+\mathcal{O}(\tilde{g}^2),
			\quad\quad
			\boldsymbol{\beta}=\frac{1}{2\sqrt{2}}
			\begin{pmatrix}
				1 & 1\\
				-1 & 1
			\end{pmatrix}\tilde{g}
			+\mathcal{O}(\tilde{g}^2).
		\end{align}
		This then allows us to compute also
		$\boldsymbol{A}(\tau)$ and $\boldsymbol{B}(\tau)$. We thus have
		\begin{align}
			\boldsymbol{A}(\tau)
			=&\frac{1}{2}(e^{-i\tilde{\kappa}_+\tau}+e^{-i\tilde{\kappa}_-\tau})\mathds{1}_2
			+\frac{1}{2}(e^{-i\tilde{\kappa}_+\tau}-e^{-i\tilde{\kappa}_-\tau})\boldsymbol{\sigma}_{\text{x}}
			+\mathcal{O}(\tilde{g}^2),\nonumber\\
			\boldsymbol{B}(\tau)
			=&\frac{1}{4}(e^{-i\tilde{\kappa}_+\tau}-e^{-i\tilde{\kappa}_-\tau})\,\tilde{g}\,\mathds{1}_2
			+\frac{1}{4}(e^{-i\tilde{\kappa}_+\tau}+e^{-i\tilde{\kappa}_-\tau})\,\tilde{g}\,\boldsymbol{\sigma}_{\text{x}}
			-\text{c.c.}
			+\mathcal{O}(\tilde{g}^2).
		\end{align}
		Note that $\boldsymbol{B}(\tau)=\boldsymbol{\sigma}_{\text{x}}\boldsymbol{A}(\tau)\,\tilde{g}/2-\boldsymbol{\sigma}_{\text{x}}\boldsymbol{A}^*(\tau)\,\tilde{g}/2$.
		Furthermore, we compute also $\tilde{\boldsymbol{S}}_{\text{RWA}}(\tau)$. We have
		\begin{align}
			\tilde{\boldsymbol{S}}_{\text{RWA}}(\tau)
			=e^{-i\tau}\cos(\tilde{g}\tau)\mathds{1}_2-ie^{-i\tau}\sin(\tilde{g}\tau)\boldsymbol{\sigma}_{\text{x}}.
		\end{align}
		We then compute the perturbative expression of $\boldsymbol{B}_{\text{f}}(t)$ that we report here:{
		\begin{align}
			\boldsymbol{B}_{\text{f}}(t)=\boldsymbol{\alpha}_0^\dag \tilde{\boldsymbol{S}}^\dag_\text{RWA}(t)\boldsymbol{A}(t)\boldsymbol{\beta}_0+\boldsymbol{\alpha}_0^\dag \tilde{\boldsymbol{S}}^\dag_\text{RWA}(t)\boldsymbol{B}(t)\boldsymbol{\alpha}^*_0
			-\boldsymbol{\beta}_0^\text{Tp} \tilde{\boldsymbol{S}}^\text{Tp}_\text{RWA}(t)\boldsymbol{B}^*(t)\boldsymbol{\beta}_0-\boldsymbol{\beta}_0^\text{Tp} \tilde{\boldsymbol{S}}^\text{Tp}_\text{RWA}(t)\boldsymbol{A}^*(t)\boldsymbol{\alpha}^*_0.
		\end{align}}
		To lowest order we have
		\begin{align}
			\boldsymbol{B}_{\text{f}}(t)=\boldsymbol{\alpha}_0^\dag \tilde{\boldsymbol{S}}^\dag_\text{RWA}(t)\boldsymbol{A}(t)\boldsymbol{\beta}_0-\boldsymbol{\beta}_0^\text{Tp} \tilde{\boldsymbol{S}}^\text{Tp}_\text{RWA}(t)\boldsymbol{A}^*(t)\boldsymbol{\alpha}^*_0.
		\end{align}

		\section{Recovering the rotating wave approximation}\label{verifying:Validity:RWA:appendix}
		In~\ref{Gaussian:fidelity:appendix} we have shown that the fidelity has the expression
		\begin{align*}
			\mathcal{F}_\text{eff}(t)=\mathcal{F}(\boldsymbol{\sigma}_0,\boldsymbol{S}_{\text{eff}}(t)\boldsymbol{\sigma}_0\boldsymbol{S}_{\text{eff}}^\dag(t))=4/\sqrt{\Gamma(t)},
		\end{align*}
		where $\Gamma(t)=16|\det(\boldsymbol{A}_\textrm{f}(t))|^2=\det(\left\{\boldsymbol{\Omega},\boldsymbol{S}_\textrm{f}(t)\right\})$ 
		and $\boldsymbol{S}_\textrm{f}(t)=\boldsymbol{s}_0^{-1}\boldsymbol{S}_\textrm{RWA}^\dag(t)\boldsymbol{S}(t)\boldsymbol{s}_0$. The matrix $\boldsymbol{s}_0$ is a symplectic matrix since  it obeys the identity $\boldsymbol{s}_0\boldsymbol{\Omega}\boldsymbol{s}_0^\dag=\boldsymbol{s}_0^\dag\boldsymbol{\Omega}\boldsymbol{s}_0=\boldsymbol{\Omega}$. This implies that the inverse has the expression $\boldsymbol{s}_0^{-1}=-\boldsymbol{\Omega}\boldsymbol{s}_0^\dag\boldsymbol{\Omega}$. Hence:
		\begin{align}
			\boldsymbol{s}_0^{-1}&=-\begin{pmatrix}
				-i\mathds{1}_2&\boldsymbol{0}\\
				\boldsymbol{0}&i\mathds{1}_2
			\end{pmatrix}\begin{pmatrix}
				\boldsymbol{\alpha}_0&\boldsymbol{\beta}_0\\
				\boldsymbol{\beta}_0^*&\boldsymbol{\alpha}_0^*
			\end{pmatrix}^\dag\begin{pmatrix}
				-i\mathds{1}_2&\boldsymbol{0}\\
				\boldsymbol{0}&i\mathds{1}_2
			\end{pmatrix}\nonumber=\begin{pmatrix}
				\boldsymbol{\alpha}_0^\dag&-\boldsymbol{\beta}_0^\textrm{Tp}\\
				-\boldsymbol{\beta}_0^\dag&\boldsymbol{\alpha}_0^\textrm{Tp}
			\end{pmatrix}.\nonumber
		\end{align}
		We can use this expression to compute the symplectic matrix $\boldsymbol{S}_\textrm{f}(t)$, which reads:
            {\small
		\begin{align}
			\boldsymbol{S}_\textrm{f}&=\boldsymbol{s}_0^{-1}\boldsymbol{S}_\textrm{RWA}^\dag\boldsymbol{S}\boldsymbol{s}_0=\begin{pmatrix}
				\boldsymbol{\alpha}_0^\dag&-\boldsymbol{\beta}_0^\textrm{Tp}\\
				-\boldsymbol{\beta}_0^\dag&\boldsymbol{\alpha}_0^\textrm{Tp}
			\end{pmatrix}\begin{pmatrix}
				\Tilde{\boldsymbol{S}}_\textrm{RWA}&\boldsymbol{0}\\
				\boldsymbol{0}&\Tilde{\boldsymbol{S}}_\textrm{RWA}^*
			\end{pmatrix}^\dag\begin{pmatrix}
				\boldsymbol{A}&\boldsymbol{B}\\
				\boldsymbol{B}^*&\boldsymbol{A}^*
			\end{pmatrix}\begin{pmatrix}
				\boldsymbol{\alpha}_0&\boldsymbol{\beta}_0\\
				\boldsymbol{\beta}_0^*&\boldsymbol{\alpha}_0^*
			\end{pmatrix}\nonumber\\
			&=\begin{pmatrix}
				\boldsymbol{\alpha}_0^\dagger\Tilde{\boldsymbol{S}}_\textrm{RWA}^\dagger&-\boldsymbol{\beta}_0^\textrm{Tp}\Tilde{\boldsymbol{S}}_\textrm{RWA}^\textrm{Tp}\\
				-\boldsymbol{\beta}_0^\dagger\Tilde{\boldsymbol{S}}_\textrm{RWA}^\dagger&\boldsymbol{\alpha}_0^\textrm{Tp}\Tilde{\boldsymbol{S}}_\textrm{RWA}^\textrm{Tp}
			\end{pmatrix}\begin{pmatrix}
				\boldsymbol{A}\boldsymbol{\alpha}_0+\boldsymbol{B}\boldsymbol{\beta}_0^*&\boldsymbol{A}\boldsymbol{\beta}_0+\boldsymbol{B}\boldsymbol{\alpha}_0^*\\
				\boldsymbol{B}^*\boldsymbol{\alpha}_0+\boldsymbol{A}^*\boldsymbol{\beta}_0^*&\boldsymbol{B}^*\boldsymbol{\beta}_0+\boldsymbol{A}^*\boldsymbol{\alpha}_0^*
			\end{pmatrix}\nonumber\\
			&={\begin{pmatrix}
					\boldsymbol{\alpha}_0^\dagger\Tilde{\boldsymbol{S}}_\textrm{RWA}^\dagger\left(\boldsymbol{A}\boldsymbol{\alpha}_0+\boldsymbol{B}\boldsymbol{\beta}_0^*\right)-\boldsymbol{\beta}_0^\textrm{Tp}\Tilde{\boldsymbol{S}}_\textrm{RWA}^\textrm{Tp}\left(\boldsymbol{B}^*\boldsymbol{\alpha}_0+\boldsymbol{A}^*\boldsymbol{\beta}_0^*\right)&\boldsymbol{\alpha}_0^\dagger\Tilde{\boldsymbol{S}}_\textrm{RWA}^\dagger\left(\boldsymbol{A}\boldsymbol{\beta}_0+\boldsymbol{B}\boldsymbol{\alpha}_0^*\right)-\boldsymbol{\beta}_0^\textrm{Tp}\Tilde{\boldsymbol{S}}_\textrm{RWA}^\textrm{Tp}\left(\boldsymbol{B}^*\boldsymbol{\beta}_0+\boldsymbol{A}^*\boldsymbol{\alpha}_0^*\right)\\
					-\boldsymbol{\beta}_0^\dagger\Tilde{\boldsymbol{S}}_\textrm{RWA}^\dagger\left(\boldsymbol{A}\boldsymbol{\alpha}_0+\boldsymbol{B}\boldsymbol{\beta}_0^*\right)+\boldsymbol{\alpha}_0^\textrm{Tp}\Tilde{\boldsymbol{S}}_\textrm{RWA}^\textrm{Tp}\left(\boldsymbol{B}^*\boldsymbol{\alpha}_0+\boldsymbol{A}^*\boldsymbol{\beta}_0^*\right)&-\boldsymbol{\beta}_0^\dagger\Tilde{\boldsymbol{S}}_\textrm{RWA}^\dagger\left(\boldsymbol{A}\boldsymbol{\beta}_0+\boldsymbol{B}\boldsymbol{\alpha}_0^*\right)+\boldsymbol{\alpha}_0^\textrm{Tp}\Tilde{\boldsymbol{S}}_\textrm{RWA}^\textrm{Tp}\left(\boldsymbol{B}^*\boldsymbol{\beta}_0+\boldsymbol{A}^*\boldsymbol{\alpha}_0^*\right)
			\end{pmatrix}}\nonumber
		\end{align}}
		and thus we use the general expression
		\begin{align*}
			\boldsymbol{S}_\textrm{f} (t)=\begin{pmatrix}
				\boldsymbol{A}_\textrm{f}(t) & \boldsymbol{B}_\textrm{f}(t)\\
				\boldsymbol{B}_\textrm{f}^*(t) & \boldsymbol{A}_\textrm{f}^*(t)
			\end{pmatrix}.
		\end{align*}
		Here, the general Bogoliubov matrices $\boldsymbol{A}_\textrm{f}(t)$ and $\boldsymbol{B}_\textrm{f}(t)$  are given explicitly by
		\begin{align}
			\boldsymbol{A}_\textrm{f}(t)=\boldsymbol{\alpha}_0^\dagger\Tilde{\boldsymbol{S}}_\textrm{RWA}^\dagger(t)\boldsymbol{A}(t)\boldsymbol{\alpha}_0+\boldsymbol{\alpha}_0^\dagger\Tilde{\boldsymbol{S}}_\textrm{RWA}^\dagger(t)\boldsymbol{B}(t)\boldsymbol{\beta}_0^*
			-\boldsymbol{\beta}_0^\textrm{Tp}\Tilde{\boldsymbol{S}}_\textrm{RWA}^\textrm{Tp}(t)\boldsymbol{B}^*(t)\boldsymbol{\alpha}_0-\boldsymbol{\beta}_0^\textrm{Tp}\Tilde{\boldsymbol{S}}_\textrm{RWA}^\textrm{Tp}(t)\boldsymbol{A}^*(t)\boldsymbol{\beta}_0^*,\nonumber\\
			\boldsymbol{B}_\textrm{f}(t)=\boldsymbol{\alpha}_0^\dagger\Tilde{\boldsymbol{S}}_\textrm{RWA}^\dagger(t)\boldsymbol{A}(t)\boldsymbol{\beta}_0+\boldsymbol{\alpha}_0^\dagger\Tilde{\boldsymbol{S}}_\textrm{RWA}^\dagger(t)\boldsymbol{B}(t)\boldsymbol{\alpha}_0^*
			-\boldsymbol{\beta}_0^\textrm{Tp}\Tilde{\boldsymbol{S}}_\textrm{RWA}^\textrm{Tp}(t)\boldsymbol{B}^*(t)\boldsymbol{\beta}_0-\boldsymbol{\beta}_0^\textrm{Tp}\Tilde{\boldsymbol{S}}_\textrm{RWA}^\textrm{Tp}(t)\boldsymbol{A}^*(t)\boldsymbol{\alpha}_0^*.
		\end{align}

		\subsection{Recovering the rotating wave approximation: initial passive state}
		We want to consider now only the case where the initial state contains no squeezing, i.e., $\boldsymbol{s}_0$ is a unitary matrix. This implies that $\boldsymbol{\beta}_0=\boldsymbol{0}$ and also $\boldsymbol{\alpha}_0$ is unitary as well. Computations drastically simplify as now $\boldsymbol{A}_\text{f}(t)=\boldsymbol{\alpha}_0^\dagger\Tilde{\boldsymbol{S}}_{\textrm{RWA}}(t)\boldsymbol{A}(t)\boldsymbol{\alpha}_0$
		and therefore $|\text{det}(\boldsymbol{A}_\textrm{f}(t))|=|\text{det}(\boldsymbol{\alpha}_0^\dagger\Tilde{\boldsymbol{S}}_{\textrm{RWA}}(t)\boldsymbol{A}(t)\boldsymbol{\alpha}_0)|=|\det(\boldsymbol{A}(t))|$. For simplicity we will only consider the case ${g}\geq 0$. The computation for $\leq 0$ can be done analogously.
		
		Here we have the expression $\boldsymbol{A}(t)=\boldsymbol{\alpha}^\textrm{Tp}e^{-i\boldsymbol{\kappa}t}\boldsymbol{\alpha}-\boldsymbol{\beta}^\textrm{Tp}e^{i\boldsymbol{\kappa}t}\boldsymbol{\beta}$ that can be computed easily. Some algebra gives us the elements of $\boldsymbol{A}(t)$, which explicitly read
		\begin{align}
			 A_{11}(t)&=\left(\cos^2(\theta)\cos(\kappa_+t)+\sin^2(\theta)\cos(\kappa_-t)\right)-\frac{1}{2}i\left(\frac{\kappa_+^2+\omega_\textrm{a}^2}{\kappa_+\omega_\textrm{a}}\cos^2(\theta)\sin(\kappa_+t)+\frac{\kappa_-^2+\omega_\textrm{a}^2}{\kappa_-\omega_\textrm{a}}\sin^2(\theta)\sin(\kappa_-t)\right),\nonumber\\
			 A_{12}(t)&=\frac{\omega_\textrm{a}+\omega_\textrm{b}}{4\sqrt{\omega_\textrm{a}\omega_\textrm{b}}}\sin(2\theta)\left(\cos(\kappa_+t)-\cos(\kappa_-t)\right)-\frac{i}{4}\frac{\sin(2\theta)}{\sqrt{\omega_\textrm{a}\omega_\textrm{b}}}\left(\frac{\kappa_+^2+\omega_\textrm{a}\omega_\textrm{b}}{\kappa_+}\sin(\kappa_+t)-\frac{\kappa_-^2+\omega_\textrm{a}\omega_\textrm{b}}{\kappa_-}\sin(\kappa_-t)\right),\nonumber\\
			 A_{21}(t)&=A_{12}(t),\nonumber\\
			 A_{22}(t)&=\left(\sin^2(\theta)\cos(\kappa_+t)+\cos^2(\theta)\cos(\kappa_-t)\right)-\frac{1}{2}i\left(\frac{\kappa_+^2+\omega_\textrm{b}^2}{\kappa_+\omega_\textrm{b}}\sin^2(\theta)\sin(\kappa_+t)+\frac{\kappa_-^2+\omega_\textrm{b}^2}{\kappa_-\omega_\textrm{b}}\cos^2(\theta)\sin(\kappa_-t)\right).
		\end{align}
		This agrees with the results found in the literature \cite{Bruschi:Paraoanu:2021}.
		Hencefourth, we can compute the determinant of $\boldsymbol{A}(t)$ and find
		\begin{align}
			\det(\boldsymbol{A}(t))&=\left(\alpha_{11}^2e^{-i\kappa_+t}+\alpha_{21}^2e^{-i\kappa_-t}-\beta_{11}^2e^{i\kappa_+ t}-\beta_{21}^2e^{i\kappa_-t}\right)\left(\alpha_{12}^2e^{-i\kappa_+t}+\alpha_{22}^2e^{-i\kappa_-t}-\beta_{12}^2e^{i\kappa_+ t}-\beta_{22}^2e^{i\kappa_-t}\right)\nonumber\\
			&\quad-\left(\alpha_{11}\alpha_{12}e^{-i\kappa_+t}+\alpha_{21}\alpha_{22}e^{-i\kappa_-t}-\beta_{11}\beta_{12}e^{i\kappa_+t}-\beta_{21}\beta_{22}e^{i\kappa_-t}\right)^2\nonumber\\
			&=e^{-i(\kappa_++\kappa_-)t}(\alpha_{11}\alpha_{22}-\alpha_{12}\alpha_{21})^2+e^{i(\kappa_++\kappa_-)t}(\beta_{11}\beta_{22}-\beta_{12}\beta_{21})^2\nonumber\\
			&\quad-e^{-i(\kappa_+-\kappa_-)t}(\alpha_{11}\beta_{22}-\alpha_{12}\beta_{21})^2-e^{i(\kappa_+-\kappa_-)t}(\alpha_{22}\beta_{11}-\alpha_{21}\beta_{12})^2\nonumber\\
			&\quad-(\alpha_{11}\beta_{12}-\alpha_{12}\beta_{11})^2-(\alpha_{21}\beta_{22}-\alpha_{22}\beta_{21})^2.\nonumber
		\end{align}
		The Bogoliubov matrices $\boldsymbol{\alpha}$ and $\boldsymbol{\beta}$ satisfy the Bogoliubov identities  $\boldsymbol{\alpha}^\textrm{Tp}\boldsymbol{\alpha}-\boldsymbol{\beta}^\textrm{Tp}\boldsymbol{\beta}=\mathds{1}_2$ and $\boldsymbol{\alpha}^\textrm{Tp}\boldsymbol{\beta}-\boldsymbol{\beta}^\textrm{Tp}\boldsymbol{\alpha}=\boldsymbol{0}$. This implies that $\det(\boldsymbol{A}(t=0))=1$, and we can therefore simplify the previous expression to obtain{
		\begin{align*}
			\det(\boldsymbol{A}(t))&=1+\left(\det(\boldsymbol{\alpha})\right)^2\left(e^{-i(\kappa_++\kappa_-)t}-1\right)+\left(\det(\boldsymbol{\beta})\right)^2\left(e^{i(\kappa_++\kappa_-)t}-1\right)\nonumber\\
			&\quad+(\alpha_{11}\beta_{22}-\alpha_{12}\beta_{21})^2\left(1-e^{-i(\kappa_+-\kappa_-)t}\right)+(\alpha_{22}\beta_{11}-\alpha_{21}\beta_{12})^2\left(1-e^{i(\kappa_+-\kappa_-)t}\right)\nonumber\\
			&=1+\left(\cos((\kappa_++\kappa_-)t)-1\right)\left(\left(\det(\boldsymbol{\alpha})\right)^2+\left(\det(\boldsymbol{\beta})\right)^2\right)\nonumber\\
			&\quad\quad+\left(1-\cos((\kappa_+-\kappa_-)t)\right)\left((\alpha_{11}\beta_{22}-\alpha_{12}\beta_{21})^2+(\alpha_{22}\beta_{11}-\alpha_{21}\beta_{12})^2\right)\nonumber\\
			&\quad+i\Bigl(\sin((\kappa_++\kappa_-)t)\left(-\left(\det(\boldsymbol{\alpha})\right)^2+\left(\det(\boldsymbol{\beta})\right)^2\right)\nonumber\\
			&\quad\quad+\sin((\kappa_+-\kappa_-)t)\left((\alpha_{11}\beta_{22}-\alpha_{12}\beta_{21})^2-(\alpha_{22}\beta_{11}-\alpha_{21}\beta_{12})^2\right)\Bigl)\nonumber\\
			&=1+(\alpha_{11}\beta_{22}-\alpha_{12}\beta_{21})^2+(\alpha_{22}\beta_{11}-\alpha_{21}\beta_{12})^2-\left(\det(\boldsymbol{\alpha})\right)^2-\left(\det(\boldsymbol{\beta})\right)^2\nonumber\\
			&\quad\quad+\cos(\kappa_+t)\cos(\kappa_-t)\left(\left(\det(\boldsymbol{\alpha})\right)^2+\left(\det(\boldsymbol{\beta})\right)^2-(\alpha_{11}\beta_{22}-\alpha_{12}\beta_{21})^2-(\alpha_{22}\beta_{11}-\alpha_{21}\beta_{12})^2\right)\nonumber\\
			&\quad\quad+\sin(\kappa_+t)\sin(\kappa_-t)\left(-(\alpha_{11}\beta_{22}-\alpha_{12}\beta_{21})^2-(\alpha_{22}\beta_{11}-\alpha_{21}\beta_{12})^2-\left(\det(\boldsymbol{\alpha})\right)^2-\left(\det(\boldsymbol{\beta})\right)^2\right)\nonumber\\
			&\quad+i\Bigl(\sin(\kappa_+t)\cos(\kappa_-t)\left(-\left(\det(\boldsymbol{\alpha})\right)^2+\left(\det(\boldsymbol{\beta})\right)^2+(\alpha_{11}\beta_{22}-\alpha_{12}\beta_{21})^2-(\alpha_{22}\beta_{11}-\alpha_{21}\beta_{12})^2\right)\nonumber\\
			&\quad\quad+\cos(\kappa_+t)\sin(\kappa_-t)\left(-\left(\det(\boldsymbol{\alpha})\right)^2+\left(\det(\boldsymbol{\beta})\right)^2-(\alpha_{11}\beta_{22}-\alpha_{12}\beta_{21})^2+(\alpha_{22}\beta_{11}-\alpha_{21}\beta_{12})^2\right)\Bigl)\nonumber\\
			&=1+\left(\cos(\kappa_+t)\cos(\kappa_-t)-1\right)\left(\left(\det(\boldsymbol{\alpha})\right)^2+\left(\det(\boldsymbol{\beta})\right)^2-(\alpha_{11}\beta_{22}-\alpha_{12}\beta_{21})^2-(\alpha_{22}\beta_{11}-\alpha_{21}\beta_{12})^2\right)\nonumber\\
			&\quad\quad-\sin(\kappa_+t)\sin(\kappa_-t)\left(\left(\det(\boldsymbol{\alpha})\right)^2+\left(\det(\boldsymbol{\beta})\right)^2+(\alpha_{11}\beta_{22}-\alpha_{12}\beta_{21})^2+(\alpha_{22}\beta_{11}-\alpha_{21}\beta_{12})^2\right)\nonumber\\
			&\quad+i\Bigl(\sin(\kappa_+t)\cos(\kappa_-t)\left(-\left(\det(\boldsymbol{\alpha})\right)^2+\left(\det(\boldsymbol{\beta})\right)^2+(\alpha_{11}\beta_{22}-\alpha_{12}\beta_{21})^2-(\alpha_{22}\beta_{11}-\alpha_{21}\beta_{12})^2\right)\nonumber\\
			&\quad\quad+\cos(\kappa_+t)\sin(\kappa_-t)\left(-\left(\det(\boldsymbol{\alpha})\right)^2+\left(\det(\boldsymbol{\beta})\right)^2-(\alpha_{11}\beta_{22}-\alpha_{12}\beta_{21})^2+(\alpha_{22}\beta_{11}-\alpha_{21}\beta_{12})^2\right)\Bigl).
		\end{align*}}
		This expression is cumbersome, and therefore we introduce for latter convenience the following four quantities:
		\begin{align}
			 q_1:=& \left(\det(\boldsymbol{\alpha})\right)^2+\left(\det(\boldsymbol{\beta})\right)^2-(\alpha_{11}\beta_{22}-\alpha_{12}\beta_{21})^2-(\alpha_{22}\beta_{11}-\alpha_{21}\beta_{12})^2,\nonumber\\
			 q_2:=&\left(\det(\boldsymbol{\alpha})\right)^2+\left(\det(\boldsymbol{\beta})\right)^2+(\alpha_{11}\beta_{22}-\alpha_{12}\beta_{21})^2+(\alpha_{22}\beta_{11}-\alpha_{21}\beta_{12})^2,\nonumber\\
			 q_3:=&-\left(\det(\boldsymbol{\alpha})\right)^2+\left(\det(\boldsymbol{\beta})\right)^2+(\alpha_{11}\beta_{22}-\alpha_{12}\beta_{21})^2-(\alpha_{22}\beta_{11}-\alpha_{21}\beta_{12})^2,\nonumber\\
			 q_4:=&-\left(\det(\boldsymbol{\alpha})\right)^2+\left(\det(\boldsymbol{\beta})\right)^2-(\alpha_{11}\beta_{22}-\alpha_{12}\beta_{21})^2+(\alpha_{22}\beta_{11}-\alpha_{21}\beta_{12})^2.
		\end{align}
		Recall that we have set $g_\textrm{bs}=g_\textrm{sq}=g$. Thus we can use equations \eqref{bogoliubov:coefficients:appendix} to compute
		\begin{align}
			\det(\boldsymbol{\alpha})&=-\frac{(\kappa_++\omega_\textrm{a})(\kappa_-+\omega_\textrm{b})}{4\sqrt{\kappa_+\kappa_-\omega_\textrm{a}\omega_\textrm{b}}}\cos^2(\theta)-\frac{(\kappa_++\omega_\textrm{b})(\kappa_-+\omega_\textrm{a})}{4\sqrt{\kappa_+\kappa_-\omega_\textrm{a}\omega_\textrm{b}}}\sin^2(\theta),\nonumber\\
			\det(\boldsymbol{\beta})&=-\frac{(\kappa_+-\omega_\textrm{a})(\kappa_--\omega_\textrm{b})}{4\sqrt{\kappa_+\kappa_-\omega_\textrm{a}\omega_\textrm{b}}}\cos^2(\theta)-\frac{(\kappa_+-\omega_\textrm{b})(\kappa_--\omega_\textrm{a})}{4\sqrt{\kappa_+\kappa_-\omega_\textrm{a}\omega_\textrm{b}}}\sin^2(\theta),\nonumber\\
			\alpha_{11}\beta_{22}-\alpha_{12}\beta_{21}&=-\frac{(\kappa_++\omega_\textrm{a})(\kappa_--\omega_\textrm{b})}{4\sqrt{\kappa_+\kappa_-\omega_\textrm{a}\omega_\textrm{b}}}\cos^2(\theta)-\frac{(\kappa_++\omega_\textrm{b})(\kappa_--\omega_\textrm{a})}{4\sqrt{\kappa_+\kappa_-\omega_\textrm{a}\omega_\textrm{b}}}\sin^2(\theta),\nonumber\\
			\alpha_{22}\beta_{11}-\alpha_{21}\beta_{12}&=-\frac{(\kappa_+-\omega_\textrm{a})(\kappa_-+\omega_\textrm{b})}{4\sqrt{\kappa_+\kappa_-\omega_\textrm{a}\omega_\textrm{b}}}\cos^2(\theta)-\frac{(\kappa_+-\omega_\textrm{b})(\kappa_-+\omega_\textrm{a})}{4\sqrt{\kappa_+\kappa_-\omega_\textrm{a}\omega_\textrm{b}}}\sin^2(\theta).
		\end{align}
		Recall that the rotating wave approximation is best tackled when setting $\omega_\textrm{b}=\omega_\textrm{a}\equiv\omega$ and taking the limit for $g/\omega\ll1$.
		To achieve our goal we introduce 
		$\omega_\textrm{a}\equiv\omega$, $\omega_\textrm{b}=\omega+\delta\omega$ and the dimensionless time $\tau:=\omega\,t$. We then introduce the following quantities $\varepsilon:=\delta\omega/\omega$, $\Tilde{g}:=g/\omega$, and $\Tilde{\kappa}_\pm:=\kappa_\pm/\omega$, and we are interested in the regime for which $|\varepsilon|\ll\Tilde{g}$. The overall idea is to compute the perturbative expansion of the four quantities $q_{1}$, $q_2$, $q_3$, and $q_4$. These expressions allow us to re-write the normal frequencies as
		\begin{align*}
			\Tilde{\kappa}_\pm^2&=\frac{1}{2}\left(1+(1+\varepsilon)^2\pm\sqrt{(1-(1+\varepsilon)^2)^2+16(1+\varepsilon)\Tilde{g}^2}\right).
		\end{align*}
		We also recall that the functions $\cos(2\theta)$ and $\sin(2\theta)$ are defined via the constraint
		$\tan(2\theta)=\frac{4g\sqrt{\omega_\text{a}\omega_\text{b}}}{\omega_\text{a}^2-\omega_\text{b}^2}$. We have:
		\begin{align}
			\sin(2\theta)&=\frac{4g\sqrt{\omega_\text{a}\omega_\text{b}}}{\sqrt{16g^2\omega_\text{a}\omega_\text{b}+(\omega_\text{a}^2-\omega_\text{b}^2)^2}},\nonumber\\
			\cos(2\theta)&=\frac{\omega_\text{a}^2-\omega_\text{b}^2}{\sqrt{16g^2\omega_\text{a}\omega_\text{b}+(\omega_\text{a}^2-\omega_\text{b}^2)^2}}.
		\end{align}
		In terms of our new variables we have
		\begin{align*}
			\cos(2\theta)&=\frac{1-(1+\varepsilon)^2}{\sqrt{16\Tilde{g}^2(1+\varepsilon)+(1-(1+\varepsilon)^2)^2}}=-\frac{\varepsilon}{2{\Tilde{g}}}+\mathcal{O}\left((\varepsilon/{\Tilde{g}})^3\right),\\
			\sin(2\theta)&=\frac{4\Tilde{g}\sqrt{1+\varepsilon}}{\sqrt{16\Tilde{g}^2(1+\varepsilon)+(1-(1+\varepsilon)^2)^2}}=1-\frac{\varepsilon^2}{8\Tilde{g}^2}+\mathcal{O}\left(\left(\varepsilon/\Tilde{g}\right)^4\right).
		\end{align*}
		We are working in the regime $|\varepsilon|\ll\Tilde{g}$, which implies that $|\cos(2\theta)|<1/\sqrt{2}$. This, in turn, implies $2\theta\in(\pi/4,3\pi/4)$, which furthermore tells us that $\cos(\theta)>0$. Thus
		\begin{align*}
			\cos(\theta)=&|\cos(\theta)|=\sqrt{\frac{1}{2}2\cos^2(\theta)}=\sqrt{\frac{1}{2}\left(1+\cos^2(\theta)-\sin^2(\theta)\right)}=\sqrt{\frac{1+\cos(2\theta)}{2}},\\
			\sin(\theta)=&\frac{\sin(2\theta)}{2\cos(\theta)}.
		\end{align*}
		Finally, we can write
		\begin{align}
			\cos(\theta)&=\frac{1}{\sqrt{2}}\left(1-\frac{1}{4}\frac{\varepsilon}{\Tilde{g}}\right)+\mathcal{O}\left(\frac{\varepsilon^2}{\Tilde{g}^2}\right),\nonumber\\
			\sin(\theta)&=\frac{1}{\sqrt{2}}\left(1+\frac{1}{4}\frac{\varepsilon}{\Tilde{g}}\right)+\mathcal{O}\left(\frac{\varepsilon^2}{\Tilde{g}^2}\right).
		\end{align}
		One can easily check that the fundamental relation $\cos^2(\theta)+\sin^2(\theta)=1$ is observed to first non-trivial perturbative order.
		We are ready to compute the quantities $q_1$, $q_2$, $q_3$, and $q_4$. The computations are divided per each $q_j$.
		\begin{itemize}
			\item \textbf{Computing} $\boldsymbol{q_1}$: We start with the first coefficient and lengthy algebra gives us 
			\begin{align*}
				q_1&=\frac{1}{16\omega_\textrm{a}\omega_\textrm{b}}\left(\omega_\textrm{a}^2+14\omega_\textrm{a}\omega_\textrm{b}+\omega_\textrm{b}^2-\left(\omega_\textrm{a}-\omega_\textrm{b}\right)^2\left(\cos^2(2\theta)-\sin^2(2\theta)\right)\right),
			\end{align*}
			which requires us to also evaluate the following:
			\begin{align}
				\cos^2(2\theta)-\sin^2(2\theta)&=\frac{-16\Tilde{g}^2(1+\varepsilon)+(1-(1+\varepsilon)^2)^2}{16\Tilde{g}^2(1+\varepsilon)+(1-(1+\varepsilon)^2)^2}\nonumber\\
				&=-1+\frac{\varepsilon^2}{2\Tilde{g}^2}+\mathcal{O}(\varepsilon^4/\Tilde{g}^4)+\mathcal{O}(\varepsilon^4/\Tilde{g}^2).\nonumber
			\end{align}
			Putting all together, we obtain
			\begin{align}
				 q_1&=\frac{1}{16}\left(1+14(1+\varepsilon)+(1+\varepsilon)^2-\varepsilon^2 \left(\cos^2(2\theta)-\sin^2(2\theta)\right)\right)\nonumber\\
                 &=1+\frac{\varepsilon^2}{8}+\mathcal{O}(\varepsilon^3/\tilde{g}^3).
			\end{align}
			\item  \textbf{Computing} $\boldsymbol{q_2}$: We continue with the second coefficient, and lengthy algebra gives us{
			\begin{align}
				 q_2 &=\frac{4\left(\omega_\textrm{a}^2+\kappa_-^2\right)\left(\omega_\textrm{b}^2+\kappa_+^2\right)\sin^4(\theta)+4\left(\omega_\textrm{a}^2+\kappa_+^2\right)\left(\omega_\textrm{b}^2+\kappa_-^2\right)\cos^4(\theta)+2\sin^2(2\theta)\left(\omega_\textrm{a}\omega_\textrm{b}+\kappa_+^2\right)\left(\omega_\textrm{a}\omega_\textrm{b}+\kappa_-^2\right)}{16\kappa_+\kappa_-\omega_\textrm{a}\omega_\textrm{b}}\nonumber\\
				&=\frac{4\left(\omega_\textrm{a}^2\omega_\textrm{b}^2+\kappa_+^2\kappa_-^2\right)+\kappa_+^2\left((\omega_\textrm{b}-\omega_\textrm{a})\cos(2\theta)+\omega_\textrm{a}+\omega_\textrm{b}\right)^2+\kappa_-^2\left((\omega_\textrm{a}-\omega_\textrm{b})\cos(2\theta)+\omega_\textrm{a}+\omega_\textrm{b}\right)^2}{16\kappa_+\kappa_-\omega_\textrm{a}\omega_\textrm{b}}.\nonumber
			\end{align}}
			We now need the following pertubrative expressions
			\begin{align*}
				\frac{\omega_\textrm{a}\omega_\textrm{b}}{\kappa_+\kappa_-}
				&=\frac{1}{\sqrt{1-4\Tilde{g}^2}}\left(1-2\Tilde{g}^2\frac{\varepsilon}{1-4\Tilde{g}^2}\right)+\mathcal{O}(\varepsilon^2),\\
				\frac{\kappa_+\kappa_-}{\omega_\textrm{a}\omega_\textrm{b}}&=
				\sqrt{1-4\Tilde{g}^2}\left(1+2\Tilde{g}^2\frac{\varepsilon}{1-4\Tilde{g}^2}\right)+\mathcal{O}(\varepsilon^2),\\
				\frac{\omega_\textrm{a}^2\omega_\textrm{b}^2+\kappa_+^2\kappa_-^2}{4\kappa_+\kappa_-\omega_\textrm{a}\omega_\textrm{b}}&=
				\frac{1}{2}\frac{1}{\sqrt{1-4\Tilde{g}^2}}\left(1-2\Tilde{g}^2-\frac{4\Tilde{g}^4\varepsilon}{1-4\Tilde{g}^2}\right)+\mathcal{O}(\varepsilon^2).
			\end{align*}
			One also computes{\small
			\begin{align}
				&\quad\frac{\kappa_+^2\left((\omega_\textrm{b}-\omega_\textrm{a})\cos(2\theta)+\omega_\textrm{a}+\omega_\textrm{b}\right)^2+\kappa_-^2\left((\omega_\textrm{a}-\omega_\textrm{b})\cos(2\theta)+\omega_\textrm{a}+\omega_\textrm{b}\right)^2}{16\kappa_+\kappa_-\omega_\textrm{a}\omega_\textrm{b}}\nonumber\\
				&=\frac{\kappa_+^2\left(\varepsilon\cos(2\theta)+2+\varepsilon\right)^2+\kappa_-^2\left(-\varepsilon\cos(2\theta)+2+\varepsilon\right)^2}{16\Tilde{\kappa}_+\Tilde{\kappa}_-(1+\varepsilon)}\nonumber\\
				&=\frac{\left(1+(1+\varepsilon)^2\right)\left((2+\varepsilon)^2+\varepsilon^2\cos^2(2\theta)\right)+2\sqrt{(1-(1+\varepsilon)^2)^2+16(1+\varepsilon)\Tilde{g}^2}(2+\varepsilon)\varepsilon\cos(2\theta)}{16\sqrt{(1+\varepsilon)^3(1+\varepsilon-4\Tilde{g}^2)}}.\nonumber
			\end{align}}
			These terms can expanded separately for computational ease. One has
			\begin{align}
				\frac{\left(1+(1+\varepsilon)^2\right)\left((2+\varepsilon)^2+\varepsilon^2\cos^2(2\theta)\right)}{16\sqrt{(1+\varepsilon)^3(1+\varepsilon-4\Tilde{g}^2)}}&=\frac{\left(1+(1+\varepsilon)^2\right)\left((2+\varepsilon)^2+\frac{\varepsilon^4}{4\Tilde{g}^2}\right)}{16\sqrt{(1+\varepsilon)^3(1+\varepsilon-4\Tilde{g}^2)}}\nonumber\\
				&=\frac{1}{\sqrt{1-4\Tilde{g}^2}}\left(\frac{1}{2}-\frac{\Tilde{g}^2}{1-4\Tilde{g}^2}\varepsilon\right)+\mathcal{O}(\varepsilon^2),\nonumber\\
				\frac{2\sqrt{(1-(1+\varepsilon)^2)^2+16(1+\varepsilon)\Tilde{g}^2}(2+\varepsilon)\varepsilon\cos(2\theta)}{16\sqrt{(1+\varepsilon)^3(1+\varepsilon-4\Tilde{g}^2)}}&
                =-\frac{2\sqrt{(1-(1+\varepsilon)^2)^2+16(1+\varepsilon)\Tilde{g}^2}(2+\varepsilon)\frac{\varepsilon^2}{\sqrt{4\Tilde{g}^2}}}{16\sqrt{(1+\varepsilon)^3(1+\varepsilon-4\Tilde{g}^2)}}
				=\mathcal{O}(\varepsilon^2)\nonumber.
			\end{align}
			We finally conclude that
			\begin{align}
				q_2&=\frac{1}{\sqrt{1-4\Tilde{g}^2}}\left(1-\Tilde{g}^2-\Tilde{g}^2\frac{1+2\Tilde{g}^2}{1-4\Tilde{g}^2}\varepsilon\right)+\mathcal{O}(\varepsilon^2).
			\end{align}
			Note that this expansion only requires the assumption $|\varepsilon|\ll 1$ to be valid.
			\item  \textbf{Computing} $\boldsymbol{q_3}$: We continue with the third coefficient, and lengthy algebra gives us
			\begin{align}
				 q_3&=\frac{1}{16\kappa_+\kappa_-\omega_\textrm{a}\omega_\textrm{b}}\left(-8\omega_\textrm{b}\kappa_-\left(\omega_\textrm{a}^2+\kappa_+^2\right)\cos^2(\theta)-8\omega_\textrm{a}\kappa_-\left(\omega_\textrm{b}^2+\kappa_+^2\right)\sin^2(\theta)\right)\nonumber\\
				&=-\frac{1}{2\Tilde{\kappa}_+(1+\varepsilon)}\left((1+\varepsilon)\left(1+\Tilde{\kappa}_+^2\right)\cos^2(\theta)+\left((1+\varepsilon)^2+\Tilde{\kappa}_+^2\right)\sin^2(\theta)\right)\nonumber\\
				&=-\frac{1}{2\Tilde{\kappa}_+}\left(1+\varepsilon\sin^2(\theta)+\Tilde{\kappa}_+^2\frac{1+\varepsilon\cos^2(\theta)}{1+\varepsilon}\right).\nonumber
			\end{align}
			One then requires the following perturbative expressions:
			\begin{align}
				\varepsilon\sin^2(\theta)&=\varepsilon\left(\frac{1}{\sqrt{2}}\left(1+\frac{\varepsilon}{4\Tilde{g}}+\mathcal{O}(\varepsilon^2/\Tilde{g}^2)\right)\right)^2=\frac{\varepsilon}{2}+\mathcal{O}\left(\varepsilon^2/\tilde{g}\right),\nonumber
            \end{align}{\small
            \begin{align}
				\Tilde{\kappa}_+^2\frac{1+\varepsilon\cos^2(\theta)}{1+\varepsilon}&=\frac{1}{2}\left(1+(1+\varepsilon)^2+\sqrt{(1-(1+\varepsilon)^2)^2+16(1+\varepsilon)\Tilde{g}^2}\right)\frac{1+\frac{\varepsilon}{2}\left(1-\frac{\varepsilon}{\sqrt{4\Tilde{g}^2}}\right)}{1+\varepsilon}\nonumber\\
				&=\left(1+2\Tilde{g}\right)+\frac{\varepsilon}{2}+\mathcal{O}(\varepsilon^2).\nonumber
			\end{align}}
			This finally leads to
			\begin{align}
				q_3&=-\frac{1+\Tilde{g}+\varepsilon/2}{\sqrt{\frac{1}{2}\left(1+(1+\varepsilon)^2+\sqrt{(1-(1+\varepsilon)^2)^2+16(1+\varepsilon)\Tilde{g}^2}\right)}}\nonumber\\
                &=-\frac{1}{\sqrt{1+2\Tilde{g}}}\left(1+\Tilde{g}-\frac{\Tilde{g}^2\varepsilon}{2+4\Tilde{g}}\right)+\mathcal{O}(\varepsilon^2/\tilde{g}).
			\end{align}
			\item \textbf{Computing} $\boldsymbol{q_4}$: We conclude with the fourth and final coefficient, and lengthy algebra gives us
			\begin{align}
				 q_4&=\frac{1}{16\kappa_+\kappa_-\omega_\textrm{a}\omega_\textrm{b}}\left(-8\omega_\textrm{a}\kappa_+\left(\omega_\textrm{b}^2+\kappa_-^2\right)\cos^2(\theta)-8\omega_\textrm{b}\kappa_+\left(\omega_\textrm{a}^2+\kappa_-^2\right)\sin^2(\theta)\right)\nonumber\\
				&=-\frac{1}{2\Tilde{\kappa}_-(1+\varepsilon)}\left(\left(1+2\varepsilon+\varepsilon^2+\Tilde{\kappa}_-^2\right)\cos^2(\theta)+\left(1+\varepsilon+\Tilde{\kappa}_-^2+\varepsilon\Tilde{\kappa}_-^2\right)\sin^2(\theta)\right)\nonumber\\
				&=-\frac{1}{2\Tilde{\kappa}_-}\left(1+\varepsilon\cos^2(\theta)+\Tilde{\kappa}_-^2\frac{1+\varepsilon\sin^2(\theta)}{1+\varepsilon}\right)\nonumber.
			\end{align}
			We require the following perturbative expansions
			\begin{align}
				\varepsilon\cos^2(\theta)&={\varepsilon}\left(\frac{1}{\sqrt{2}}\left(1-\frac{\varepsilon}{4\Tilde{g}}\right)\right)^2=\frac{\varepsilon}{2}+\mathcal{O}(\varepsilon^2/\tilde{g}),\nonumber
            \end{align}{\small
            \begin{align}
				\Tilde{\kappa}_-^2\frac{1+\varepsilon\sin^2(\theta)}{1+\varepsilon}&=\frac{1}{2}\left(1+(1+\varepsilon)^2-\sqrt{(1-(1+\varepsilon)^2)^2+16(1+\varepsilon)\Tilde{g}^2}\right)\frac{1+\frac{\varepsilon}{2}\left(1+\frac{\varepsilon}{\sqrt{4\Tilde{g}^2}}\right)}{1+\varepsilon}\nonumber\\
				&=1-2\Tilde{g}+\frac{\varepsilon}{2}+\mathcal{O}(\varepsilon^2)\nonumber.
			\end{align}}
			Finally, we obtain
			\begin{align}
				q_4&=-\frac{1-\Tilde{g}+\varepsilon/2}{\sqrt{\frac{1}{2}\left(1+(1+\varepsilon)^2-\sqrt{(1-(1+\varepsilon)^2)^2+16(1+\varepsilon)\Tilde{g}^2}\right)}}\nonumber\\
                &=-\frac{1}{\sqrt{1-2\Tilde{g}}}\left(1-\Tilde{g}-\frac{\Tilde{g}^2\varepsilon}{2-4\Tilde{g}}\right)+\mathcal{O}(\varepsilon^2/\tilde{g}).
			\end{align}
		\end{itemize}
		
		We list here for completeness the perturbative expansions of the coefficients in the regime $|\varepsilon|\ll\Tilde{g}$. We have:
		\begin{align}
			q_1&=1+\frac{\varepsilon^2}{8}+\mathcal{O}(\varepsilon^3/\tilde{g}^3),\nonumber\\
			q_2&=\frac{1}{\sqrt{1-4\Tilde{g}^2}}\left(1-\Tilde{g}^2-\Tilde{g}^2\frac{1+2\Tilde{g}^2}{1-4\Tilde{g}^2}\varepsilon\right)+\mathcal{O}(\varepsilon^2),\nonumber\\
			q_3&=-\frac{1}{\sqrt{1+2\Tilde{g}}}\left(1+\Tilde{g}-\frac{\Tilde{g}^2\varepsilon}{2+4\Tilde{g}}\right)+\mathcal{O}(\varepsilon^2/\tilde{g}),\nonumber\\
			q_4&=-\frac{1}{\sqrt{1-2\Tilde{g}}}\left(1-\Tilde{g}-\frac{\Tilde{g}^2\varepsilon}{2-4\Tilde{g}}\right)+\mathcal{O}(\varepsilon^2/\tilde{g}).\label{Appendix:Calculations:Heib:q1234}
		\end{align}
		We now proceed to impose the resonance condition $\varepsilon\equiv 0$, and we also assume small relative coupling strength $\Tilde{g}\ll 1$.  In this case one finds $\Tilde{\kappa}_\pm^2=1\pm 2\Tilde{g}$ and $\cos(2\theta)=0$, $\sin(2\theta)=1$. This implies $2\theta=\pi/2$, which in turn implies $\sin(\theta)=1/\sqrt{2}$ and $\cos(\theta)=1/\sqrt{2}$. Thus, we only need to expand equations \eqref{Appendix:Calculations:Heib:q1234} and obtain
		\begin{align}
			q_1&=1,\nonumber\\
			q_2&=\frac{1-\Tilde{g}^2}{\sqrt{1-4\Tilde{g}^2}}=1+\Tilde{g}^2+\mathcal{O}(\Tilde{g}^4),\nonumber\\
			q_3&=-\frac{1+\Tilde{g}}{\sqrt{1+2\Tilde{g}}}=-1-\frac{1}{2}\Tilde{g}^2+\mathcal{O}(\Tilde{g}^3),\nonumber\\
			q_4&=-\frac{1-\Tilde{g}}{\sqrt{1-2\Tilde{g}}}=-1-\frac{1}{2}\Tilde{g}^2+\mathcal{O}(\Tilde{g}^3).
		\end{align}
		We recall that we wish to compute $\mathcal{F}_\text{eff}(t)\equiv\mathcal{F}(\boldsymbol{\sigma}_0,\boldsymbol{S}_{\text{eff}}(t)\boldsymbol{\sigma}_0\boldsymbol{S}_{\text{eff}}^\dag(t))=1/|\det(\boldsymbol{A}_\textrm{f}(t))|$, which depends on the $q_j$ coefficients via the expression 
		\begin{align*}
			|\det(\boldsymbol{A}_\textrm{f}(t))|^2=\left(1+\left(\cos(\kappa_+t)\cos(\kappa_-t)-1\right)q_1-\sin(\kappa_+t)\sin(\kappa_-t)q_2\right)^2
            +\left(\sin(\kappa_+t)\cos(\kappa_-t)q_3+\cos(\kappa_+t)\sin(\kappa_-t)q_4\right)^2.
		\end{align*}
		Using the perturbative expressions presented above, after some algebra we obtain
		\begin{align}
			|\det(\boldsymbol{A}_\textrm{f}(\tau))|^2&=1+\Tilde{g}^2\left(\sin^2(\Tilde{\kappa}_+\tau)+\sin^2(\Tilde{\kappa}_-\tau)\right)+\mathcal{O}(\Tilde{g}^3)\nonumber
		\end{align}
		Thus, we obtain the final perturbative expression for the fidelity, which reads
		\begin{align}
			\mathcal{F}_\text{eff}(t)\equiv\mathcal{F}(\boldsymbol{\sigma}_0,\boldsymbol{S}_{\text{eff}}(\tau)\boldsymbol{\sigma}_0\boldsymbol{S}_{\text{eff}}^\dag(\tau))&=1-\frac{\Tilde{g}^2}{2}\left(\sin^2(\Tilde{\kappa}_+\tau)+\sin^2(\Tilde{\kappa}_-\tau)\right)+\mathcal{O}(\Tilde{g}^3).\label{Eqn:Eff:Fidelity:Vacuum:State:result:Appendix}
		\end{align}
		Note that $\mathcal{F}_\text{eff}(t)\leq1$ at all times, as expected.

		\subsection{Recovering the rotating wave approximation: general case}
		In this section, we consider general initial pure  Gaussian states with vanishing first moments. We start by expanding the matrices $\boldsymbol{A}(\tau)$ and $\boldsymbol{B}(\tau)$ on resonance (i.e., $\omega_{\text{a}}=\omega_{\text{b}}=\omega$) in powers of $\Tilde{g}\ll 1$. We will consider for simplicity only the case $\Tilde{g}\geq 0$, the case $\Tilde{g}<0$ can be obtained by repeating the same calculations. We have
		\begin{align}
			 A_{11}(t)&=\left(\cos^2(\theta)\cos(\kappa_+t)+\sin^2(\theta)\cos(\kappa_-t)\right)-\frac{1}{2}i\left(\frac{\kappa_+^2+\omega_\textrm{a}^2}{\kappa_+\omega_\textrm{a}}\cos^2(\theta)\sin(\kappa_+t)+\frac{\kappa_-^2+\omega_\textrm{a}^2}{\kappa_-\omega_\textrm{a}}\sin^2(\theta)\sin(\kappa_-t)\right)\nonumber\\
			&=\frac{1}{2}\left(\cos(\Tilde{\kappa}_+\tau)+\cos(\Tilde{\kappa}_-\tau)\right)-\frac{i}{4}\left(\frac{2+2\Tilde{g}}{\sqrt{1+2\Tilde{g}}}\sin(\Tilde{\kappa}_+\tau)+\frac{2-2\Tilde{g}}{\sqrt{1-2\Tilde{g}}}\sin(\Tilde{\kappa}_-\tau)\right)\nonumber\\
			&=\frac{1}{2}\left(\cos(\Tilde{\kappa}_+\tau)+\cos(\Tilde{\kappa}_-\tau)\right)-\frac{2+\Tilde{g}^2}{4}i\left(\sin(\Tilde{\kappa}_+\tau)+\sin(\Tilde{\kappa}_-\tau)\right)\nonumber+\mathcal{O}(\Tilde{g}^3),
		\end{align}
		\begin{align}
			 A_{12}(t)&=\frac{\omega_\textrm{a}+\omega_\textrm{b}}{4\sqrt{\omega_\textrm{a}\omega_\textrm{b}}}\sin(2\theta)\left(\cos(\kappa_+t)-\cos(\kappa_-t)\right)-\frac{i}{4}\frac{\sin(2\theta)}{\sqrt{\omega_\textrm{a}\omega_\textrm{b}}}\left(\frac{\kappa_+^2+\omega_\textrm{a}\omega_\textrm{b}}{\kappa_+}\sin(\kappa_+t)-\frac{\kappa_-^2+\omega_\textrm{a}\omega_\textrm{b}}{\kappa_-}\sin(\kappa_-t)\right)\nonumber\\
			&=\frac{1}{2}\left(\cos(\Tilde{\kappa}_+\tau)-\cos(\Tilde{\kappa}_-\tau)\right)-\frac{i}{4}\left(\frac{2+2\Tilde{g}}{\sqrt{1+2\Tilde{g}}}\sin(\Tilde{\kappa}_+\tau)-\frac{2-2\Tilde{g}}{\sqrt{1-2\Tilde{g}}}\sin(\Tilde{\kappa}_-\tau)\right)\nonumber\\
			&=\frac{1}{2}\left(\cos(\Tilde{\kappa}_+\tau)-\cos(\Tilde{\kappa}_-\tau)\right)-\frac{2+\Tilde{g}^2}{4}i\left(\sin(\Tilde{\kappa}_+\tau)-\sin(\Tilde{\kappa}_-\tau)\right)\nonumber+\mathcal{O}(\Tilde{g}^3).
		\end{align}
		We note that $\omega_\textrm{a}=\omega_\textrm{b}$ implies $A_{21}(\tau)=A_{12}(\tau)$ and $A_{22}(\tau)=A_{11}(\tau)$. By computing the following perturbative expansions
		\begin{align}
			\cos(\Tilde{\kappa}_\pm\tau)-i(1+\Tilde{g}^2/2)\sin(\Tilde{\kappa}_\pm\tau)=&e^{-i\Tilde{\kappa}_\pm\tau}+\mathcal{O}(\Tilde{g}^2).\nonumber
		\end{align}
		We are then immediately able to find
		\begin{align}
			\boldsymbol{A}(\tau)&=
			\frac{1}{2}\left(e^{-i\Tilde{\kappa}_+\tau}+e^{-i\Tilde{\kappa}_-\tau}\right)\mathds{1}_2+
			\frac{1}{2}\left(e^{-i\Tilde{\kappa}_+\tau}-e^{-i\Tilde{\kappa}_-\tau}\right)\boldsymbol{\sigma}_{\text{x}}
			+\mathcal{O}(\Tilde{g}^2)=h_+(\tau)\mathds{1}_2+h_-(\tau)\boldsymbol{\sigma}_\text{x}+\mathcal{O}(\Tilde{g}^2)\nonumber,
		\end{align}
		where we introduced the functions $h_\pm(\tau)=(\exp(-i\Tilde{\kappa}_+\tau)\pm\exp(-i\Tilde{\kappa}_-\tau))/2$.
		One can now quickly derive the result from the previous section, by computing $\det(\boldsymbol{A}(\tau))$ using the expressions above. Here we can use the results provided in the literature to compute the perturbative expansion of matrix $\boldsymbol{B}(\tau)$, see~\cite{Bruschi:Paraoanu:2021}. We also have
		\begin{align}
			 B_{11}(t)&=-\frac{i}{2}\left(\frac{\kappa_+^2-\omega_\textrm{a}^2}{\kappa_+\omega_\textrm{a}}\cos^2(\theta)\sin(\kappa_+ t)+\frac{\kappa_-^2-\omega_\textrm{a}^2}{\kappa_-\omega_\textrm{a}}\sin^2(\theta)\sin(\kappa_-t)\right)\nonumber\\
			&=-\frac{i}{4}\left(\frac{2\Tilde{g}}{\sqrt{1+2\Tilde{g}}}\sin(\Tilde{\kappa}_+\tau)+\frac{-2\Tilde{g}}{\sqrt{1-2\Tilde{g}}}\sin(\Tilde{\kappa}_-\tau)\right)\nonumber\\
			 &=-\frac{i}{2}\left(\Tilde{g}\left(\sin(\Tilde{\kappa}_+\tau)-\sin(\Tilde{\kappa}_-\tau)\right)-\Tilde{g}^2\left(\sin(\Tilde{\kappa}_+\tau)+\sin(\Tilde{\kappa}_-\tau)\right)\right)+\mathcal{O}(\Tilde{g}^3),\nonumber\\
			 B_{12}(t)&=\frac{\omega_\textrm{a}-\omega_\textrm{b}}{4\sqrt{\omega_\textrm{a}\omega_\textrm{b}}}\sin(2\theta)\left(\cos(\kappa_+t)-\cos(\kappa_-t)\right)-i\frac{\sin(2\theta)}{4\sqrt{\omega_\textrm{a}\omega_\textrm{b}}}\left(\frac{\kappa_+^2-\omega_\textrm{a}\omega_\textrm{b}}{\kappa_+}\sin(\kappa_+t)-\frac{\kappa_-^2-\omega_\textrm{a}\omega_\textrm{b}}{\kappa_-}\sin(\kappa_-t)\right)\nonumber\\
			&=-\frac{i}{4}\left(\frac{2\Tilde{g}}{\sqrt{1+2\Tilde{g}}}\sin(\Tilde{\kappa}_+\tau)-\frac{-2\Tilde{g}}{\sqrt{1-2\Tilde{g}}}\sin(\Tilde{\kappa}_-\tau)\right)\nonumber\\
			&=-\frac{i}{2}\left(\Tilde{g}\left(\sin(\Tilde{\kappa}_+\tau)+\sin(\Tilde{\kappa}_-\tau)\right)-\Tilde{g}^2\left(\sin(\Tilde{\kappa}_+\tau)-\sin(\Tilde{\kappa}_-\tau)\right)\right)+\mathcal{O}(\Tilde{g}^3),\nonumber
		\end{align}
		together with the properties $B_{21}(\tau)=-B_{12}^*(\tau)=B_{12}(\tau)$ and $B_{22}(\tau)=B_{11}(\tau)$, see \cite{Bruschi:Paraoanu:2021}. Once more, noting that $\Tilde{g}\left(1\mp\Tilde{g}\right)\sin(\Tilde{\kappa}_\pm\tau)=\Tilde{g}\sin(\Tilde{\kappa}_\pm\tau)+\mathcal{O}(\Tilde{g}^2)$, and then defining $p_\pm(\tau):=-(\sin(\Tilde{\kappa}_+\tau)\pm\sin(\Tilde{\kappa}_-\tau))/2$, we have
		\begin{align}
			\boldsymbol{B}(\tau)
			&=\frac{\Tilde{g}}{4}\left(e^{-i\tilde{\kappa}_+\tau}-e^{-i\tilde{\kappa}_-\tau}\right)\mathds{1}_2
			+\frac{\Tilde{g}}{4}\left(e^{-i\tilde{\kappa}_+\tau}+e^{-i\tilde{\kappa}_-\tau}\right)\boldsymbol{\sigma}_{\text{x}}-\text{c.c.} +\mathcal{O}(\Tilde{g}^2).\nonumber
		\end{align}
		Note that one can write to lowest order $\boldsymbol{B}(\tau)=\Tilde{g}\boldsymbol{\sigma}_\text{x}\boldsymbol{A}(\tau)/2-\text{c.c.}$ To complement these expressions, we note that also have
		\begin{align}
			\Tilde{\boldsymbol{S}}_\textrm{RWA}(\tau)&
			=e^{-i\tau}\cos(\Tilde{g}\tau)\mathds{1}_2-ie^{-i\tau}\sin(\Tilde{g}\tau)\boldsymbol{\sigma}_{\text{x}}\nonumber.
		\end{align}
		For simplicity, we restrict the following computations to first order in $\Tilde{g}$ and omit all higher order terms. We note that all important matrices $\boldsymbol{A}$, $\boldsymbol{B}$, and $\Tilde{\boldsymbol{S}}_\textrm{RWA}$, in the regime where $\omega_\textrm{a}=\omega_\textrm{b}\equiv \omega$, have the generic form $\boldsymbol{M}= \mu\,\mathds{1}_2+\xi\,\boldsymbol{\sigma}_{\text{x}}$ and commute therefore with each other.
		
		We will compute now $\boldsymbol{B}_\text{f}^\dagger(\tau)\boldsymbol{B}_\text{f}(\tau)$ by order in $\Tilde{g}$. We start by computing $\boldsymbol{B}_\text{f}(\tau)$ in lowest order. Here one finds:{\small
		\begin{align}
			\boldsymbol{B}_\text{f}(\tau)&=\boldsymbol{\alpha}_0^\dagger\Tilde{\boldsymbol{S}}_\text{RWA}^\dagger(\tau)(h_+(\tau)\mathds{1}_2+h_-(\tau)\boldsymbol{\sigma}_\text{x})\boldsymbol{\beta}_0-\boldsymbol{\beta}_0^\textrm{Tp}\Tilde{\boldsymbol{S}}_\text{RWA}^\textrm{Tp}(\tau)(h_+^*(\tau)\mathds{1}_2+h_-^*(\tau)\boldsymbol{\sigma}_\text{x})\boldsymbol{\alpha}_0^*+\mathcal{O}(\Tilde{g}).\nonumber
		\end{align}}
		Next, we expand $\Tilde{\kappa}_\pm$ around $\Tilde{g}=0$ and find $\Tilde{\kappa}_\pm=1\pm\Tilde{g}-\Tilde{g}^2/2\pm\Tilde{g}^3/2+\mathcal{O}(\Tilde{g}^4)$. This allows us to write:
		\begin{align}
			h_\pm(\tau)&=\frac{1}{2}e^{-i\left(1-\frac{\Tilde{g}^2}{2}\right)\tau}\left(e^{-i\Tilde{g}\tau}\pm e^{i\Tilde{g}\tau}\right)+\mathcal{O}(\Tilde{g}^3\tau)\nonumber
		\end{align}
		and consequently:
		\begin{align}
			h_+(\tau)\mathds{1}_2+h_-(\tau)\boldsymbol{\sigma}_\text{x}&=e^{-i\left(1-\frac{\Tilde{g}^2}{2}\right)\tau}\left(\cos(\Tilde{g}\tau)\mathds{1}_2-i\sin(\Tilde{g}\tau)\boldsymbol{\sigma}_\text{x}\right)+\mathcal{O}(\Tilde{g}^3\tau).\nonumber
		\end{align}
		This implies:
		\begin{align}
			\Tilde{\boldsymbol{S}}_\text{RWA}^\dagger(\tau)(h_+(\tau)\mathds{1}_2+h_-(\tau)\boldsymbol{\sigma}_\text{x})&=e^{\frac{1}{2}i\Tilde{g}^2\tau}\mathds{1}_2+\mathcal{O}(\Tilde{g}^3\tau).\nonumber
		\end{align}
		We can recover the RWA if we set $\Tilde{g}\tau=\operatorname{const.}$, which implies in the limit $\Tilde{g}\ll 1$ and consequently $\Tilde{g}^2\tau\ll 1$ where we further require $\Tilde{g}^2\tau=\mathcal{O}(\Tilde{g})$. Using the Bogoliubov identities we find in lowest order in $\Tilde{g}$:
		\begin{align}
			\boldsymbol{B}_\textrm{f}(\tau)&=\boldsymbol{\alpha}_0^\dagger e^{\frac{i}{2}\Tilde{g}^2\tau}\mathds{1}_2\boldsymbol{\beta}_0-\boldsymbol{\beta}_0^\textrm{Tp}e^{-\frac{i}{2}\Tilde{g}^2\tau}\mathds{1}_2\boldsymbol{\alpha}_0^*+\mathcal{O}(\Tilde{g})+\mathcal{O}(\Tilde{g}^3\tau)=2i\sin\left(\frac{1}{2}\Tilde{g}^2\tau\right)\boldsymbol{\alpha}_0^\dag\boldsymbol{\beta}_0+\mathcal{O}(\Tilde{g}).\nonumber
		\end{align}
		This yields:
		\begin{align}
			\boldsymbol{B}_\textrm{f}^\dag(\tau)\boldsymbol{B}_\textrm{f}(\tau)=4\sin^2\left(\frac{1}{2}\Tilde{g}^2\tau\right)\boldsymbol{\beta}_0^\dag\boldsymbol{\alpha}_0\boldsymbol{\alpha}_0^\dag\boldsymbol{\beta}_0+\mathcal{O}(\Tilde{g})+\mathcal{O}(\Tilde{g}^4\tau),\nonumber
		\end{align}
		which in turn implies
		\begin{align}
			\det(\mathds{1}_2+\boldsymbol{B}_\textrm{f}^\dag(\tau)\boldsymbol{B}_\textrm{f}(\tau))=1+\mathcal{O}(\Tilde{g})+\mathcal{O}(\Tilde{g}^2\tau).\nonumber
		\end{align}
		Consequently, we have that $\mathcal{F}_\textrm{eff}(\tau)\to 1$ in the limit $\Tilde{g}\to 0$ and $\Tilde{g}^2\tau\to 0$, and we recover the RWA as claimed.
		
		Now we proceed to derive the exact expression for the lowest order term  in $\Tilde{g}$ in the expansion of $\text{det}(\mathds{1}_2+\boldsymbol{B}_\textrm{f}^\dag(\tau)\boldsymbol{B}_\textrm{f}(\tau))$. To achieve this task we must also consider higher-order contributions of $\boldsymbol{B}_\textrm{f}(\tau)$. It is convenient to introduce the matrices:{
		\begin{align}
			\boldsymbol{h}(\tau):=h_+(\tau)\mathds{1}_2+h_-(\tau)\boldsymbol{\sigma}_\text{x},\;\;\;\boldsymbol{p}_1(\tau):=p_-(\tau)\mathds{1}_2+p_+(\tau)\boldsymbol{\sigma}_\text{x},\;\;\;\boldsymbol{p}_2(\tau):=p_+(\tau)\mathds{1}_2+p_-(\tau)\boldsymbol{\sigma}_\text{x},\nonumber
		\end{align}}
		such that one has:
		\begin{align}
			\boldsymbol{A}(\tau)&=\boldsymbol{h}(\tau)+i\boldsymbol{p}_2(\tau)\Tilde{g}^2/2+\mathcal{O}(\Tilde{g}^3),\quad\text{and}\quad\boldsymbol{B}(\tau)=i\boldsymbol{p}_1(\tau)\Tilde{g}-i\boldsymbol{p}_2(\tau)\Tilde{g}^2+\mathcal{O}(\Tilde{g}^3).\nonumber
		\end{align}
		Next, we define the three matrices
		\begin{align}
			\boldsymbol{f}_0(\tau):=\Tilde{\boldsymbol{S}}_\textrm{RWA}^\dagger(\tau)\boldsymbol{h}(\tau),\quad\boldsymbol{f}_1(\tau):=\Tilde{\boldsymbol{S}}_\textrm{RWA}^\dagger(\tau)\boldsymbol{p}_1(\tau),\quad\text{and}\quad\boldsymbol{f}_2(\tau):=\Tilde{\boldsymbol{S}}_\textrm{RWA}^\dag(\tau)\boldsymbol{p}_2(\tau),\nonumber
		\end{align}
		which allows us to write
		\begin{align}
			\boldsymbol{B}_\textrm{f}(\tau)&=\boldsymbol{B}_\textrm{f}^{(0)}(\tau)+\boldsymbol{B}_\textrm{f}^{(1)}(\tau)\Tilde{g}+\boldsymbol{B}_\textrm{f}^{(2)}(\tau)\Tilde{g}^2+\mathcal{O}(\Tilde{g}^3).\nonumber
		\end{align}
		It is important to notice that the matrices $\boldsymbol{f}_0(\tau)$, $\boldsymbol{f}_1(\tau)$, and $\boldsymbol{f}_2(\tau)$ are all symmetric and commute with each other. One notices furthermore that $\boldsymbol{f}_0(\tau)$ is unitary.
		
		Here we list the explicit expressions for the matrices that appear above:
		\begin{align}
			\boldsymbol{B}_\textrm{f}^{(0)}(\tau)&:=\boldsymbol{\alpha}_0^\dag\boldsymbol{f}_0(\tau)\boldsymbol{\beta}_0-\boldsymbol{\beta}_0^\textrm{Tp}\boldsymbol{f}_0^\dag(\tau)\boldsymbol{\alpha}_0^*,\nonumber\\
			\boldsymbol{B}_\textrm{f}^{(1)}(\tau)&:=i\boldsymbol{\alpha}_0^\dag\boldsymbol{f}_1(\tau)\boldsymbol{\alpha}_0^*+i\boldsymbol{\beta}_0^\textrm{Tp}\boldsymbol{f}_1^\dag(\tau)\boldsymbol{\beta}_0,\nonumber\\
			\boldsymbol{B}_\textrm{f}^{(2)}(\tau)&:=\frac{i}{2}\left(\boldsymbol{\alpha}_0^\dag\boldsymbol{f}_2(\tau)\boldsymbol{\beta}_0+\boldsymbol{\beta}_0^\textrm{Tp}\boldsymbol{f}_2^\dagger(\tau)\boldsymbol{\alpha}_0^*\right)-i\left(\boldsymbol{\alpha}_0^\dag\boldsymbol{f}_2(\tau)\boldsymbol{\alpha}_0^*+\boldsymbol{\beta}_0^\textrm{Tp}\boldsymbol{f}_2^\dag(\tau)\boldsymbol{\beta}_0\right).\nonumber
		\end{align}
		Let us now tackle the expansion of the matrices above.
		\begin{itemize}
			\item \textbf{Expanding $\boldsymbol{B}_\textrm{f}^{(0)}(\tau)$}. We start with the lowest order coefficient and have
			\begin{align}
				\boldsymbol{f}_0(\tau)&=e^{\frac{i}{2}\Tilde{g}^2\tau}\left(\mathds{1}_2-\frac{i}{2}\Tilde{g}^3\tau\boldsymbol{\sigma}_\text{x}\right)+\mathcal{O}(\Tilde{g}^3).\nonumber
			\end{align}
			We can use the Bogoliubov identities and obtain
			\begin{align}
				\boldsymbol{B}_\textrm{f}^{(0)}(\tau)&=2i\sin\left(\frac{1}{2}\Tilde{g}^2\tau\right)\boldsymbol{\alpha}_0^\dag\boldsymbol{\beta}_0-\frac{i}{2}\Tilde{g}^3\tau\left(e^{\frac{i}{2}\Tilde{g}^2\tau}\boldsymbol{\alpha}_0^\dag\boldsymbol{\sigma}_\text{x}\boldsymbol{\beta}_0+e^{-\frac{i}{2}\Tilde{g}^2\tau}\boldsymbol{\beta}_0^\textrm{Tp}\boldsymbol{\sigma}_\text{x}\boldsymbol{\alpha}_0^*\right)+\mathcal{O}(\Tilde{g}^3)\nonumber\\
				&=2i\sin\left(\frac{1}{2}\Tilde{g}^2\tau\right)\boldsymbol{\alpha}_0^\dag\boldsymbol{\beta}_0+\mathcal{O}(\Tilde{g}^2)=\mathcal{O}(\Tilde{g})\nonumber.
			\end{align}
			\item \textbf{Expanding $\boldsymbol{B}_\textrm{f}^{(1)}(\tau)$}.
			Next, we need to expand the quantities $p_\pm(\tau)$, and we find:
			\begin{align}
				p_+(\tau)&=-\sin\left(\tau-\frac{1}{2}\Tilde{g}^2\tau\right)\cos(\Tilde{g}\tau)+\mathcal{O}(\Tilde{g}^2), \quad\text{and}\quad
                p_-(\tau)=-\cos\left(\tau-\frac{1}{2}\Tilde{g}^2\tau\right)\sin(\Tilde{g}\tau)+\mathcal{O}(\Tilde{g}^2).\nonumber
            \end{align}
			This implies:{
			\begin{align}
				\boldsymbol{f}_1(\tau)
				=-e^{i\tau}\biggl(\cos(\Tilde{g}\tau)\sin(\Tilde{g}\tau)e^{i\left(\tau-\frac{1}{2}\Tilde{g}^2\tau\right)}\mathds{1}_2+\left(\cos^2(\Tilde{g}\tau)\sin(\tau)+i\sin^2(\Tilde{g}\tau)\cos(\tau)\right)\boldsymbol{\sigma}_\text{x}\biggr)+\mathcal{O}(\Tilde{g}).\nonumber
			\end{align}}
			\item \textbf{Tackling $\boldsymbol{B}_\textrm{f}^{(2)}(\tau)$}. We conclude that we do not need to compute the expansion of this matrix since it will contribute to higher order, which we can safely ignore.
		\end{itemize}
		Thus, we don't need to compute $\boldsymbol{B}_\textrm{f}^{(2)}(\tau)$ explicitly, as the lowest non-trivial order contribution of $\boldsymbol{B}_\textrm{f}^\dag(\tau)\boldsymbol{B}_\textrm{f}(\tau)$ is given by:
		\begin{align}
			\boldsymbol{B}_\textrm{f}^\dag(\tau)\boldsymbol{B}_\textrm{f}(\tau)&=\boldsymbol{C}_2^\dag(\tau)\boldsymbol{C}_2(\tau)+\mathcal{O}(\Tilde{g}^3),
		\end{align}
		where
		\begin{align}
			\boldsymbol{C}_2(\tau)=&2i\sin\left(\frac{1}{2}\Tilde{g}^2\tau\right)\boldsymbol{\alpha}_0^\dag\boldsymbol{\beta}_0-\frac{i}{2}\Tilde{g}\sin(2\Tilde{g}\tau)\left(\boldsymbol{\alpha}_0^\dag\boldsymbol{\alpha}_0^* e^{i\left(2\tau-\frac{1}{2}\Tilde{g}^2\tau\right)}+\boldsymbol{\beta}_0^\textrm{Tp}\boldsymbol{\beta}_0e^{-i\left(2\tau-\frac{1}{2}\Tilde{g}^2\tau\right)}\right)\nonumber\\
			&+\frac{\tilde{g}}{2}\Bigl((1 -\cos(2\tilde{g}\tau)e^{2i\tau})\boldsymbol{\alpha}_0^\dag\boldsymbol{\sigma}_\text{x}\boldsymbol{\alpha}_0^*-(1 -\cos(2\tilde{g}\tau)e^{-2i\tau})\boldsymbol{\beta}_0^\textrm{Tp}\boldsymbol{\sigma}_\text{x}\boldsymbol{\beta}_0\Bigl).\nonumber
		\end{align}
		Here one notes $\boldsymbol{C}_2(\tau)=\mathcal{O}(\Tilde{g})$ and therefore $\boldsymbol{C}_2^\dag(\tau)\boldsymbol{C}_2(\tau)=\mathcal{O}(\Tilde{g}^2)$. Employing the Mercator series, we find:
		\begin{align}
			\mathcal{F}_\text{eff}(\tau)=1-\frac{1}{2}\text{Tr}\left\{\boldsymbol{C}_2^\dag(\tau)\boldsymbol{C}_2(\tau)\right\}+\mathcal{O}(\Tilde{g}^3).
		\end{align}

		\subsubsection{Squeezed Gaussian states}
		Now we want to consider the case of an initial single-mode squeezed Gaussian state. We simplify the problem and consider squeezed states characterized by one real squeezing parameter $s$. The state $\boldsymbol{\sigma}(0)=\boldsymbol{s}_0\boldsymbol{s}^\dag_0$ is constructed using the following Bogoliubov coefficients: $\boldsymbol{\alpha}_0=\cosh s \mathds{1}_2$ and $\boldsymbol{\beta}_0=\sinh s \mathds{1}_2$. Note that we also assume that the first moments vanish.
		
		This allows us to write:
		\begin{align}
			\boldsymbol{B}_\textrm{f}^{(0)}(\tau)&=\frac{\sinh(2s)}{2}\left(\boldsymbol{f}_0(\tau)-\boldsymbol{f}_0^\dag(\tau)\right),\nonumber\\
			\boldsymbol{B}_\textrm{f}^{(1)}(\tau)&=i\left(\cosh^2(s)\boldsymbol{f}_1(\tau)+\sinh^2(s)\boldsymbol{f}_1^\dag(\tau)\right),\nonumber\\
			\boldsymbol{B}_\textrm{f}^{(2)}(\tau)&=i\left(\frac{\sinh(2s)}{4}\left(\boldsymbol{f}_2(\tau)+\boldsymbol{f}_2^\dag(\tau)\right)-\left(\cosh^2(s)\boldsymbol{f}_2(\tau)+\sinh^2(s)\boldsymbol{f}_2^\dag(\tau)\right)\right).\nonumber
		\end{align}
		We can use these to compute the perturbative expansion of $\boldsymbol{B}_\textrm{f}^\dag(\tau)\boldsymbol{B}_\textrm{f}(\tau)$ to all orders in $\Tilde{g}\ll 1$. The general expression reads
		\begin{align}
			\boldsymbol{B}_\textrm{f}^\dag(\tau)\boldsymbol{B}_\textrm{f}(\tau)=\boldsymbol{O}_0(\tau)+\boldsymbol{O}_1(\tau)\,\Tilde{g}+\boldsymbol{O}_2(\tau)\,\tilde{g}^2+\mathcal{O}(\Tilde{g}^3).\nonumber
		\end{align}
		The zeroth order term is given by:
		\begin{align}
			\boldsymbol{O}_0(\tau)&=\left(\boldsymbol{B}_\textrm{f}^{(0)}(\tau)\right)^\dag\boldsymbol{B}_\textrm{f}^{(0)}(\tau)=\frac{\sinh^2(2s)}{2}\left(\mathds{1}_2-\Re\left\{\boldsymbol{f}_0^2(\tau)\right\}\right)\nonumber,
		\end{align}
		as $\boldsymbol{f}_0(\tau)$ is unitary and symmetric. Some more algebra confirms $\boldsymbol{h}^2(\tau)=\boldsymbol{h}(2\tau)$ and $\Tilde{\boldsymbol{S}}_\textrm{RWA}^2(\tau)=\Tilde{\boldsymbol{S}}_\textrm{RWA}(2\tau)$ and as all matrices we are working with commute, we can conclude $\boldsymbol{f}_0^2(\tau)=\boldsymbol{f}_0(2\tau)$. Thus, we find:
		\begin{align}
			\boldsymbol{O}_0(\tau)=\frac{\sinh^2(2s)}{2}\left(\mathds{1}_2-\Re\left\{\boldsymbol{f}_0(2\tau)\right\}\right)=\frac{\sinh^2(2s)}{2}\left(x^{(0)}(\tau)\mathds{1}_2+y^{(0)}(\tau)\boldsymbol{\sigma}_\text{x}\right),\nonumber
		\end{align}
		where we need to compute an explicit expression{
		\begin{align}
			\boldsymbol{f}_0(2\tau)&=\Tilde{\boldsymbol{S}}_\text{RWA}^\dag(\tau)\boldsymbol{h}(\tau)\nonumber\\
            &=e^{i\tau}\left(\cos(2\Tilde{g}\tau)h_+(2\tau)+i\sin(2\Tilde{g}\tau)h_-(\tau)\right)\mathds{1}_2+e^{i\tau}\left(\cos(2\Tilde{g}\tau)h_-(2\tau)+i\sin(2\Tilde{g}\tau)h_+(\tau)\right)\boldsymbol{\sigma}_\text{x}.\nonumber
		\end{align}}
		Hence, compute
		\begin{align}
			 h_+(2\tau)\cos(2\Tilde{g}\tau)+ih_-(2\tau)\sin(2\Tilde{g}\tau)&=\left(\cos(2\Tilde{g}\tau)\cos((\Tilde{\kappa}_+-\Tilde{\kappa}_-)\tau)+\sin(2\Tilde{g}\tau)\sin((\Tilde{\kappa}_+-\Tilde{\kappa}_-)\tau)\right)e^{-i(\Tilde{\kappa}_++\Tilde{\kappa}_-)\tau},\nonumber\\
			 h_-(2\tau)\cos(2\Tilde{g}\tau)+ih_+(2\tau)\sin(2\Tilde{g}\tau)&=-i\left(\cos(2\Tilde{g}\tau)\sin((\Tilde{\kappa}_+-\Tilde{\kappa}_-)\tau)-\sin(2\Tilde{g}\tau)\cos((\Tilde{\kappa}_+-\Tilde{\kappa}_-)\tau)\right)e^{-i(\Tilde{\kappa}_++\Tilde{\kappa}_-)\tau}.\nonumber
		\end{align}
		One consequently finds:{
		\begin{align}
			 x^{(0)}(\tau)&=1-\left(\cos(2\Tilde{g}\tau)\cos((\Tilde{\kappa}_+-\Tilde{\kappa}_-)\tau)+\sin(2\Tilde{g}\tau)\sin((\Tilde{\kappa}_+-\Tilde{\kappa}_-)\tau)\right)\cos((2-(\Tilde{\kappa}_++\Tilde{\kappa}_-))\tau),\nonumber\\
			 y^{(0)}(\tau)&=-\left(\cos(2\Tilde{g}\tau)\sin((\Tilde{\kappa}_+-\Tilde{\kappa}_-)\tau)-\sin(2\Tilde{g}\tau)\cos((\Tilde{\kappa}_+-\Tilde{\kappa}_-)\tau)\right)\sin((2-(\Tilde{\kappa}_++\Tilde{\kappa}_-))\tau).\nonumber
		\end{align}}
		The first-order term is given by:
		{\small
			\begin{align}
				 \boldsymbol{O}_1(\tau)&=\left(\boldsymbol{B}_\textrm{f}^{(0)}(\tau)\right)^\dag\boldsymbol{B}_\textrm{f}^{(1)}(\tau)+\operatorname{h.c.}\nonumber\\
				&=i\frac{\sinh(2s)}{2}\Bigl(\left(\boldsymbol{f}_0^\dagger(\tau)-\boldsymbol{f}_0(\tau)\right)\left(\cosh^2(s)\boldsymbol{f}_1(\tau)+\sinh^2(s)\boldsymbol{f}_1^\dagger(\tau)\right)-\left(\cosh^2(s)\boldsymbol{f}_1^\dagger(\tau)+\sinh^2(s)\boldsymbol{f}_1(\tau)\right)\left(\boldsymbol{f}_0(\tau)-\boldsymbol{f}^\dagger_0(\tau)\right)\Bigr)\nonumber\\
                &=\frac{\sinh(4s)}{2}\Im\left\{\boldsymbol{h}(\tau)\boldsymbol{p}_1(\tau)\left(\mathds{1}_2+\Tilde{\boldsymbol{S}}_\text{RWA}^\dag(2\tau)\right)\right\}\nonumber\\
				&=:\frac{\sinh(4s)}{2}\left(x^{(1)}(\tau)\mathds{1}_2+y^{(1)}(\tau)\boldsymbol{\sigma}_\text{x}\right).\nonumber
		\end{align}}
		Here one needs to compute $\boldsymbol{h}(\tau)\boldsymbol{p}_1(\tau)=:\psi_1(\tau)\mathds{1}_2+\psi_2(\tau)\boldsymbol{\sigma}_\text{x}$, where we introduced $\psi_1(\tau):=h_+(\tau)p_-(\tau)+h_-(\tau)p_+(\tau)$ and $\psi_2(\tau):= h_+(\tau)p_+(\tau)+h_-(\tau)p_-(\tau)$. These read:{
		\begin{align}
			\psi_1(\tau)&=-\frac{1}{4}\Bigl(\left(e^{-i\Tilde{\kappa}_+\tau}+e^{-i\Tilde{\kappa}_-\tau}\right)\left(\sin(\Tilde{\kappa}_+\tau)-\sin(\Tilde{\kappa}_-\tau)\right)+\left(e^{-i\Tilde{\kappa}_+\tau}-e^{-i\Tilde{\kappa}_-\tau}\right)\left(\sin(\Tilde{\kappa}_+\tau)+\sin(\Tilde{\kappa}_-\tau)\right)\Bigl)\nonumber\\
			&=-\frac{1}{2}\sin((\Tilde{\kappa}_+-\Tilde{\kappa}_-)\tau)e^{-i(\Tilde{\kappa}_++\Tilde{\kappa}_-)\tau},\nonumber\\
			 \psi_2(\tau)&=-\frac{1}{4}\Bigl(\left(e^{-i\Tilde{\kappa}_+\tau}+e^{-i\Tilde{\kappa}_-\tau}\right)\left(\sin(\Tilde{\kappa}_+\tau)+\sin(\Tilde{\kappa}_-\tau)\right)+\left(e^{-i\Tilde{\kappa}_+\tau}-e^{-i\Tilde{\kappa}_-\tau}\right)\left(\sin(\Tilde{\kappa}_+\tau)-\sin(\Tilde{\kappa}_-\tau)\right)\Bigl)\nonumber\\
			&=\frac{i}{2}\left(1-\cos((\Tilde{\kappa}_+-\Tilde{\kappa}_-)\tau)e^{-i(\Tilde{\kappa}_++\Tilde{\kappa}_-)\tau}\right).\nonumber
		\end{align}}
		One finds consequently:
		\begin{align}
			\Im\left\{h_+(\tau)p_-(\tau)+h_-(\tau)p_+(\tau)\right\}&=\frac{1}{2}\sin((\Tilde{\kappa}_+-\Tilde{\kappa}_-)\tau)\sin((\Tilde{\kappa}_++\Tilde{\kappa}_-)\tau),\nonumber\\
			\Im\left\{h_+(\tau)p_+(\tau)+h_+(\tau)p_+(\tau)\right\}&=\frac{1}{2}\left(1-\cos((\Tilde{\kappa}_+-\Tilde{\kappa}_-)\tau)\cos((\Tilde{\kappa}_++\Tilde{
				\kappa}_-)\tau)\right)\nonumber.
		\end{align}
		One needs to compute furthermore the components of $\Im\{\boldsymbol{h}(\tau)\boldsymbol{p}_1(\tau)\Tilde{\boldsymbol{S}}_\text{RWA}^\dag(2\tau)\}$:
		\begin{align}
			\Im\left\{e^{2i\tau}\psi_1(\tau)\cos(2\Tilde{g}\tau)\right\}&=-\frac{1}{2}\cos(2\Tilde{g}\tau)\sin((\Tilde{\kappa}_+-\Tilde{\kappa}_-)\tau)\sin((2-(\Tilde{\kappa}_++\Tilde{\kappa}_-))\tau),\nonumber\\
			\Im\left\{ie^{2i\tau}\sin(2\Tilde{g}\tau)\psi_2(\tau)\right\}&=\frac{1}{2}\sin(2\Tilde{g}\tau)\left(\cos((\Tilde{\kappa}_+-\Tilde{\kappa}_-)\tau)\sin((2-(\Tilde{\kappa}_++\Tilde{\kappa}_-))\tau)-\sin(2\tau)\right),\nonumber\\
			\Im\left\{ie^{2i\tau}\psi_1(\tau)\sin(2\Tilde{g}\tau)\right\}&=-\frac{1}{2}\sin(2\Tilde{g}\tau)\sin((\Tilde{\kappa}_+-\Tilde{\kappa}_-)\tau)\cos((2-(\Tilde{\kappa}_++\Tilde{\kappa}_-))\tau),\nonumber\\
			\Im\left\{e^{2i\tau}\psi_2(\tau)\cos(2\Tilde{g}\tau)\right\}&=\frac{1}{2}\cos(2\Tilde{g}\tau)\left(\cos(2\tau)-\cos((\Tilde{\kappa}_+-\Tilde{\kappa}_-)\tau)\cos((2-(\Tilde{\kappa}_++\Tilde{\kappa}_-))\tau)\right)\nonumber.
		\end{align}
		We found therefore the expressions:
		\begin{align}
			 x^{(1)}(\tau)&=\frac{1}{2}\sin((\Tilde{\kappa}_+-\Tilde{\kappa}_-)\tau)\sin((\Tilde{\kappa}_++\Tilde{\kappa}_-)\tau)-\frac{1}{2}\cos(2\Tilde{g}\tau)\sin((\Tilde{\kappa}_+-\Tilde{\kappa}_-)\tau)\sin((2-(\Tilde{\kappa}_++\Tilde{\kappa}_-))\tau)\nonumber\\
			&\quad+\frac{1}{2}\sin(2\Tilde{g}\tau)\left(\cos((\Tilde{\kappa}_+-\Tilde{\kappa}_-)\tau)\sin((2-(\Tilde{\kappa}_++\Tilde{\kappa}_-))\tau)-\sin(2\tau)\right),\nonumber\\
			 y^{(1)}(\tau)&=\frac{1}{2}\left(1-\cos((\Tilde{\kappa}_+-\Tilde{\kappa}_-)\tau)\cos((\Tilde{\kappa}_++\Tilde{\kappa}_-)\tau)\right)-\frac{1}{2}\sin(2\Tilde{g}\tau)\sin((\Tilde{\kappa}_+-\Tilde{\kappa}_-)\tau)\cos((2-(\Tilde{\kappa}_++\Tilde{\kappa}_-))\tau)\nonumber\\
			&\quad+\frac{1}{2}\cos(2\Tilde{g}\tau)\left(\cos(2\tau)-\cos((\Tilde{\kappa}_+-\Tilde{\kappa}_-)\tau)\cos((2-(\Tilde{\kappa}_++\Tilde{\kappa}_-))\tau)\right)\nonumber.
		\end{align}
		At last, we compute the second-order term of $\boldsymbol{B}_\textrm{f}^\dag(\tau)\boldsymbol{B}_\textrm{f}(\tau)$. To do this  we need to compute:
		\begin{align}
			\boldsymbol{O}_2(\tau)&=\left(\boldsymbol{B}_\textrm{f}^{(0)}(\tau)\right)^\dag\boldsymbol{B}_\textrm{f}^{(2)}(\tau)+\left(\boldsymbol{B}_\textrm{f}^{(1)}(\tau)\right)^\dag\boldsymbol{B}_\textrm{f}^{(1)}(\tau)+\left(\boldsymbol{B}_\textrm{f}^{(2)}(\tau)\right)^\dag\boldsymbol{B}_\textrm{f}^{(0)}(\tau).\nonumber
		\end{align}
		This can be divided into three separate calculations of the terms $\boldsymbol{z}_1(\tau)$, $\boldsymbol{z}_2(\tau)$, and $\boldsymbol{z}_3(\tau)$ defined by:{\small
		\begin{align}
			\left(\boldsymbol{B}_\textrm{f}^{(1)}(\tau)\right)^\dag\boldsymbol{B}_\textrm{f}^{(1)}(\tau)&=\left(\cosh^2(s)\boldsymbol{f}_1^\dag(\tau)+\sinh^2(s)\boldsymbol{f}_1(\tau)\right)\left(\cosh^2(s)\boldsymbol{f}_1(\tau)+\sinh^2(s)\boldsymbol{f}_1^\dag(\tau)\right)\nonumber\\
            &=:\boldsymbol{z}_1(\tau),\nonumber
		\end{align}
		and
		\begin{align}
			&\quad\left(\boldsymbol{B}_\textrm{f}^{(0)}(\tau)\right)^\dag\boldsymbol{B}_\textrm{f}^{(2)}(\tau)+\text{h.c.}\nonumber\\
			&=i\frac{\sinh(2s)}{2}\Biggl(\left(\boldsymbol{f}_0^\dag(\tau)-\boldsymbol{f}_0(\tau)\right)\frac{\sinh(2s)}{4}\left(\boldsymbol{f}_2(\tau)+\boldsymbol{f}_2^\dag(\tau)\right)-\left(\boldsymbol{f}_0^\dag(\tau)-\boldsymbol{f}_0(\tau)\right)\left(\cosh^2(s)\boldsymbol{f}_2(\tau)+\sinh^2(s)\boldsymbol{f}_2^\dag(\tau)\right)-\text{h.c.}\Biggr)\nonumber\\
			&=\underbrace{i\frac{\sinh^2(2s)}{8}\left(\left(\boldsymbol{f}_0^\dag(\tau)-\boldsymbol{f}_0(\tau)\right)\left(\boldsymbol{f}_2(\tau)+\boldsymbol{f}_2^\dag(\tau)\right)-\text{h.c}\right)}_{=:{\boldsymbol{z}_2(\tau)}}\nonumber\\
			&\quad-\underbrace{i\frac{\sinh(2s)}{2}\left(\left(\boldsymbol{f}_0^\dag(\tau)-\boldsymbol{f}_0(\tau)\right)\left(\cosh^2(s)\boldsymbol{f}_2(\tau)+\sinh^2(s)\boldsymbol{f}_2^\dag(\tau)\right)-\text{h.c.}\right)}_{=:-\boldsymbol{z}_3(\tau)}\nonumber,
		\end{align}}
		such that $\boldsymbol{O}_2(\tau)=\boldsymbol{z}_1(\tau)+\boldsymbol{z}_2(\tau)+\boldsymbol{z}_3(\tau)$.
		The terms that need to be evaluated therefore are{
		\begin{align}
			\boldsymbol{z}_1(\tau)&:=\left(\cosh^2(s)\boldsymbol{f}_1^\dagger(\tau)+\sinh^2(s)\boldsymbol{f}_1(\tau)\right)\left(\cosh^2(s)\boldsymbol{f}_1(\tau)+\sinh^2(s)\boldsymbol{f}_1^\dagger(\tau)\right)\nonumber\\
			&=\left(\cosh^4(s)+\sinh^4(s)\right)\boldsymbol{p}_1^2(\tau)+\frac{\sinh^2(2s)}{2}\Re\left\{\boldsymbol{p}_1^2(\tau)\Tilde{\boldsymbol{S}}_\textrm{RWA}(2\tau)\right\},\nonumber\\
			\boldsymbol{z}_2(\tau)&:=i\frac{\sinh^2(2s)}{8}\left(\left(\boldsymbol{f}_0^\dag(\tau)-\boldsymbol{f}_0(\tau)\right)\left(\boldsymbol{f}_2(\tau)+\boldsymbol{f}_2^\dag(\tau)\right)-\left(\boldsymbol{f}_2^\dag(\tau)+\boldsymbol{f}_2(\tau)\right)\left(\boldsymbol{f}_0(\tau)-\boldsymbol{f}_0^\dag(\tau)\right)\right)\nonumber\\
			&=i\frac{\sinh^2(2s)}{4}\left(\boldsymbol{f}_0^\dag(\tau)-\boldsymbol{f}_0(\tau)\right)\left(\boldsymbol{f}_2(\tau)+\boldsymbol{f}_2^\dag(\tau)\right),\nonumber\\
			\boldsymbol{z}_3(\tau)&:=-i\frac{\sinh(2s)}{2}\Bigl(\left(\boldsymbol{f}_0^\dag(\tau)-\boldsymbol{f}_0(\tau)\right)\left(\cosh^2(s)\boldsymbol{f}_2(\tau)+\sinh^2(s)\boldsymbol{f}_2^\dag(\tau)\right)\nonumber\\
			&\quad-\left(\cosh^2(s)\boldsymbol{f}_2^\dag(\tau)+\sinh^2(s)\boldsymbol{f}_2(\tau)\right)\left(\boldsymbol{f}_0(\tau)-\boldsymbol{f}_0^\dag(\tau)\right)\Bigl)\nonumber\\
			&=-i\frac{\sinh(4s)}{4}\left(\boldsymbol{f}_0^\dag(\tau)-\boldsymbol{f}_0(\tau)\right)\left(\boldsymbol{f}_2(\tau)+\boldsymbol{f}_2^\dag(\tau)\right).\nonumber
		\end{align}}
		For later convenience we introduce $x_1^{(2)}(\tau,s)$, $y_1^{(2)}(\tau,s)$, $x^{(2)}_2(\tau,s)$, $y^{(2)}_2(\tau,s)$, $x^{(2)}(\tau,s)$, and $y^{(2)}(\tau,s)$ via the equations
		\begin{align}
			 x_1^{(2)}(\tau,s)\mathds{1}_2+y_1^{(2)}(\tau,s)\boldsymbol{\sigma}_\textrm{x}&:=\left(\cosh^4(s)+\sinh^4(s)\right)\boldsymbol{p}_1^2(\tau)+\frac{\sinh^2(2s)}{2}\Re\left\{\boldsymbol{p}_1^2(\tau)\Tilde{\boldsymbol{S}}_\textrm{RWA}(2\tau)\right\},\nonumber\\
			 x^{(2)}_2(\tau,s)\mathds{1}_2+y^{(2)}_2(\tau,s)\boldsymbol{\sigma}_\text{x}&:=\boldsymbol{z}_2(\tau)+\boldsymbol{z}_3(\tau)=\left(\sinh^2(2s)-\sinh(4s)\right)\Im\left\{\boldsymbol{f}_0(\tau)\right\}\Re\left\{\boldsymbol{f}_2(\tau)\right\},\nonumber\\
			 x^{(2)}(\tau,s)\mathds{1}_2+y^{(2)}(\tau,s)\boldsymbol{\sigma}_\text{x}&:=\left(x^{(2)}_1(\tau,s)+x^{(2)}_2(\tau,s)\right)\mathds{1}_2+\left(y^{(2)}_1(\tau,s)+y^{(2)}_2(\tau,s)\right)\boldsymbol{\sigma}_\text{x}\nonumber.
		\end{align}
		Here we need to compute the components of $\boldsymbol{p}_1^2(\tau)$:
		\begin{align}
			p_+^2(\tau)+p_-^2(\tau)&=\frac{1}{4}\left(\left(\sin(\Tilde{\kappa}_+\tau)+\sin(\Tilde{\kappa}_-\tau)\right)^2+\left(\sin(\Tilde{\kappa}_+\tau)-\sin(\Tilde{\kappa}_-\tau)\right)^2\right)=\frac{1}{2}\left(\sin^2(\Tilde{\kappa}_+\tau)+\sin^2(\Tilde{\kappa}_-\tau)\right),\nonumber\\
			2p_+(\tau)p_-(\tau)&=\frac{1}{2}\left(\sin^2(\Tilde{\kappa}_+\tau)-\sin^2(\Tilde{\kappa}_-\tau)\right)=\frac{1}{2}\sin((\Tilde{\kappa}_++\Tilde{\kappa}_-)\tau)\sin((\Tilde{\kappa}_+-\Tilde{\kappa}_-)\tau),\nonumber
		\end{align}
		as well as the components of $\Re\{\boldsymbol{p}_1^2(\tau)\Tilde{\boldsymbol{S}}_\text{RWA}^\dag(\tau)\}$:
		\begin{align}
			&\Re\left\{e^{2i\tau}\left(\left(p_+^2(\tau)+p_-^2(\tau)\right)\cos(2\Tilde{g}\tau)+i2p_+(\tau)p_-(\tau)\sin(2\Tilde{g}\tau)\right)\right\}\nonumber\\
			\quad&=\frac{1}{2}\cos(2\tau)\cos(2\Tilde{g}\tau)\left(\sin^2(\Tilde{\kappa}_+\tau)+\sin^2(\Tilde{\kappa}_-\tau)\right)-\frac{1}{2}\sin(2\tau)\sin(2\Tilde{g}\tau)\sin((\Tilde{\kappa}_++\Tilde{\kappa}_-)\tau)\sin((\Tilde{\kappa}_+-\Tilde{\kappa}_-)\tau),\nonumber\\
			&\Re\left\{e^{2i\tau}\left(i\left(p_+^2(\tau)+p_-^2(\tau)\right)\sin(2\Tilde{g}\tau)+2p_+(\tau)p_-(\tau)\cos(2\Tilde{g}\tau)\right)\right\}\nonumber\\
			\quad&=-\frac{1}{2}\sin(2\tau)\sin(2\Tilde{g}\tau)\left(\sin^2(\Tilde{\kappa}_+\tau)+\sin^2(\Tilde{\kappa}_-\tau)\right)+\frac{1}{2}\cos(2\tau)\cos(2\Tilde{g}\tau)\sin((\Tilde{\kappa}_++\Tilde{\kappa}_-)\tau)\sin((\Tilde{\kappa}_+-\Tilde{\kappa}_-)\tau).\nonumber
		\end{align}
		This implies:
		{\small
			\begin{align}
				x^{(2)}_1(\tau,s)&=\frac{1}{2}\left(\cosh^4(s)+\sinh^4(s)\right)\left(\sin^2(\Tilde{\kappa}_+\tau)+\sin^2(\Tilde{\kappa}_-\tau)\right)+\frac{\sinh^2(2s)}{4}\Bigl(\cos(2\tau)\cos(2\Tilde{g}\tau)\left(\sin^2(\Tilde{\kappa}_+\tau)+\sin^2(\Tilde{\kappa}_-\tau)\right)\nonumber\\
				&\quad-\sin(2\tau)\sin(2\Tilde{g}\tau)\sin((\Tilde{\kappa}_++\Tilde{\kappa}_-)\tau)\sin((\Tilde{\kappa}_+-\Tilde{\kappa}_-)\tau)\Bigl),\nonumber\\
				y^{(2)}_1(\tau,s)&=\frac{1}{2}\left(\cosh^4(s)+\sinh^4(s)\right)\sin((\Tilde{\kappa}_++\Tilde{\kappa}_-)\tau)\sin((\Tilde{\kappa}_+-\Tilde{\kappa}_-)\tau)-\frac{\sinh^2(2s)}{4}\Bigl(\sin(2\tau)\sin(2\Tilde{g}\tau)\left(\sin^2(\Tilde{\kappa}_+\tau)+\sin^2(\Tilde{\kappa}_-\tau)\right)\nonumber\\
				&\quad-\cos(2\tau)\cos(2\Tilde{g}\tau)\sin((\Tilde{\kappa}_++\Tilde{\kappa}_-)\tau)\sin((\Tilde{\kappa}_+-\Tilde{\kappa}_-)\tau)\Bigl).\nonumber
			\end{align}
		}
		At last, we need to compute the components of $\Im\{\boldsymbol{f}_0(\tau)\}$:
		{\small
			\begin{align}
				\Im\left\{e^{i\tau}\left(\cos(\Tilde{g}\tau)h_+(\tau)+i\sin(\Tilde{g}\tau)h_-(\tau)\right)\right\}&=\left(\cos(\Tilde{g}\tau)\cos\left(\frac{\Tilde{\kappa}_+-\Tilde{\kappa}_-}{2}\tau\right)+\sin(\Tilde{g}\tau)\sin\left(\frac{\Tilde{\kappa}_+-\Tilde{\kappa}_-}{2}\tau\right)\right)\sin\left(\tau-\frac{\Tilde{\kappa}_++\Tilde{\kappa}_-}{2}\tau\right),\nonumber\\
				\Im\left\{e^{i\tau}\left(\cos(\Tilde{g}\tau)h_-(\tau)+i\sin(\Tilde{g}\tau)h_+(\tau)\right)\right\}&=\left(\sin(\Tilde{g}\tau)\cos\left(\frac{\Tilde{\kappa}_+-\Tilde{\kappa}_-}{2}\tau\right)-\cos(\Tilde{g}\tau)\sin\left(\frac{\Tilde{\kappa}_+-\Tilde{\kappa}_-}{2}\tau\right)\right)\cos\left(\tau-\frac{\Tilde{\kappa}_++\Tilde{\kappa}_-}{2}\tau\right)\nonumber.
			\end{align}
		}
		And also the components of $\Re\{\boldsymbol{f}_2(\tau)\}$:
		\begin{align}
			\Re\left\{e^{i\tau}\left(\cos(\Tilde{g}\tau)p_+(\tau)+i\sin(\Tilde{g}\tau)p_-(\tau)\right)\right\}&=\cos(\tau)\cos(\Tilde{g}\tau)p_+(\tau)-\sin(\tau)\sin(\Tilde{g}\tau)p_-(\tau),\nonumber\\
			\Re\left\{e^{i\tau}\left(\cos(\Tilde{g}\tau)p_-(\tau)+i\sin(\Tilde{g}\tau)p_+(\tau)\right)\right\}&=\cos(\tau)\cos(\Tilde{g}\tau)p_-(\tau)-\sin(\tau)\sin(\Tilde{g}\tau)p_+(\tau).\nonumber
		\end{align}
		This implies:{\small
		\begin{align}
			 x^{(2)}_2(\tau,s)&=\left(\sinh^2(2s)-\sinh(4s)\right)\Biggl(\left(\cos(\Tilde{g}\tau)\cos\left(\frac{\Tilde{\kappa}_+-\Tilde{\kappa}_-}{2}\tau\right)+\sin(\Tilde{g}\tau)\sin\left(\frac{\Tilde{\kappa}_+-\Tilde{\kappa}_-}{2}\tau\right)\right)\sin\left(\tau-\frac{\Tilde{\kappa}_++\Tilde{\kappa}_-}{2}\tau\right)\nonumber\\
			&\quad\times\left(\cos(\tau)\cos(\Tilde{g}\tau)p_+(\tau)-\sin(\tau)\sin(\Tilde{g}\tau)p_-(\tau)\right)\nonumber\\
			&\quad-\left(\cos(\Tilde{g}\tau)\sin\left(\frac{\Tilde{\kappa}_+-\Tilde{\kappa}_-}{2}\tau\right)-\sin(\Tilde{g}\tau)\cos\left(\frac{\Tilde{\kappa}_+-\Tilde{\kappa}_-}{2}\tau\right)\right)\cos\left(\tau-\frac{\Tilde{\kappa}_++\Tilde{\kappa}_-}{2}\tau\right)\nonumber\\
			&\quad\times\left(\cos(\tau)\cos(\Tilde{g}\tau)p_-(\tau)-\sin(\tau)\sin(\Tilde{g}\tau)p_+(\tau)\right)\Biggl),\nonumber\\
			 y^{(2)}_2(\tau,s)&=\left(\sinh^2(2s)-\sinh(4s)\right)\Biggl(\left(\cos(\Tilde{g}\tau)\cos\left(\frac{\Tilde{\kappa}_+-\Tilde{\kappa}_-}{2}\tau\right)+\sin(\Tilde{g}\tau)\sin\left(\frac{\Tilde{\kappa}_+-\Tilde{\kappa}_-}{2}\tau\right)\right)\sin\left(\tau-\frac{\Tilde{\kappa}_++\Tilde{\kappa}_-}{2}\tau\right)\nonumber\\
			&\quad\times\left(\cos(\tau)\cos(\Tilde{g}\tau)p_-(\tau)-\sin(\tau)\sin(\Tilde{g}\tau)p_+(\tau)\right)\nonumber\\
			&\quad-\left(\cos(\Tilde{g}\tau)\sin\left(\frac{\Tilde{\kappa}_+-\Tilde{\kappa}_-}{2}\tau\right)-\sin(\Tilde{g}\tau)\cos\left(\frac{\Tilde{\kappa}_+-\Tilde{\kappa}_-}{2}\tau\right)\right)\cos\left(\tau-\frac{\Tilde{\kappa}_++\Tilde{\kappa}_-}{2}\tau\right)\nonumber\\
			&\quad\times\left(\cos(\tau)\cos(\Tilde{g}\tau)p_+(\tau)-\sin(\tau)\sin(\Tilde{g}\tau)p_-(\tau)\right)\nonumber\Biggl).
		\end{align}}
		Thus, one finds:
		\begin{align}
			\mathcal{F}_\text{eff}^{-2}(\tau)&=1+c_{0}(\tau,s)+\Tilde{g}c_1(\tau,s)+\Tilde{g}^2c_2(\tau,s)+\mathcal{O}(\Tilde{g}^3),\nonumber
		\end{align}
		where we have introduced the following coefficients for the sake of simplicity:
		\begin{align}
			 c_0(\tau,s)&=\sinh^2(2s)x^{(0)}(\tau)+\frac{\sinh^4(2s)}{4}\left((x^{(0)}(\tau))^2-(y^{(0)}(\tau))^2\right),\nonumber\\
			 c_1(\tau,s)&=\sinh(4s)\left(x^{(1)}(\tau)\left(1+\frac{\sinh^2(2s)}{2}x^{(0)}(\tau)\right)-y^{(1)}(\tau)\frac{\sinh^2(2s)}{4}y^{(0)}(\tau)\right),\nonumber\\
			 c_2(\tau,s)&=\frac{\sinh^2(4s)}{4}\left((x^{(1)}(\tau))^2-(y^{(1)}(\tau))^2\right)+2\left(x^{(2)}(\tau,s)\left(1+\frac{\sinh^2(2s)}{2}x^{(0)}(\tau)\right)-y^{(2)}(\tau)\frac{\sinh^2(2s)}{2}y^{(0)}(\tau,s)\right).\nonumber
		\end{align}
		Setting $s=0$ yields the same result for the vacuum state as obtained in Equation~\eqref{Eqn:Eff:Fidelity:Vacuum:State:result:Appendix}, as expected.
		
		We are now equipped with the means to recover the RWA. We only need to assume the limit $\Tilde{g}\ll 1$ and set $\Tilde{g}\tau\equiv\operatorname{const.}$, such that $\Tilde{g}^2\tau\ll 1$. This implies $\Tilde{g}^2\tau=\mathcal{O}(\Tilde{g})$ and we also need to require that the squeezing parameter $s$ is constant and small enough. In particular, $c_n(\tau,s)\tilde{g}^n\ll1$. This provides a constraint on $s$. 
		
		Expanding $\Tilde{\kappa}_\pm=1\pm\Tilde{g}-\Tilde{g}^2/2\pm\Tilde{g}^3/2-5\Tilde{g}^4/8+\mathcal{O}(\Tilde{g}^5)$ yields: $\Tilde{\kappa}_+-\Tilde{\kappa}_-=2\Tilde{g}+\Tilde{g}^3+\mathcal{O}(\Tilde{g}^5)$ and $\Tilde{\kappa}_++\Tilde{\kappa}_-=2-\Tilde{g}^2+\mathcal{O}(\Tilde{g}^4)$. 
		
		We proceed with obtaining the coefficients of the Fidelity.
		\begin{itemize}
			\item \textbf{Computing} $\boldsymbol{c_0(\tau,s)}$: We start with obtaining
			{
				\begin{align}
					 x^{(0)}&=1-\left(\cos(2\Tilde{g}\tau)\cos\left(2\Tilde{g}\tau+\Tilde{g}^3\tau\right)+\sin(2\Tilde{g}\tau)\sin(2\Tilde{g}\tau+\Tilde{g}^3\tau)\right)\cos(2\tau-(2\tau-\Tilde{g}^2\tau))+\mathcal{O}(\Tilde{g}^4\tau)\nonumber\\
					 &=\frac{1}{2}\Tilde{g}^4\tau^2+\mathcal{O}(\Tilde{g}^6\tau^2)\approx\frac{1}{2}\Tilde{g}^4\tau^2+\mathcal{O}(\Tilde{g}^3),\nonumber\\
					 y^{(0)}&=-\left(\cos(2\Tilde{g}\tau)\sin(2\Tilde{g}\tau+\Tilde{g}^3\tau)-\sin(2\Tilde{g}\tau)\cos(2\Tilde{g}\tau+\Tilde{g}^3\tau)\right)\sin(\Tilde{g}^2\tau)+\mathcal{O}(\Tilde{g}^4\tau)\nonumber\\
					 &=-\sin(\Tilde{g}^3\tau)\sin(\Tilde{g}^2\tau)+\mathcal{O}(\Tilde{g}^4\tau)\approx\mathcal{O}(\Tilde{g}^3),\nonumber
				\end{align}
			}
			which implies
			\begin{align}
				c_0(\tau,s)&=\frac{\sinh^2(2s)}{2}\Tilde{g}^4\tau^2+\mathcal{O}(\Tilde{g}^3).\nonumber
			\end{align}
			\item \textbf{Computing} $\boldsymbol{c_1(\tau,s)}$: We continue with the next step, that is, to compute $x^{(1)}(\tau)$, $y^{(1)}(\tau)$, and $x^{(2)}(\tau,s)$. We proceed and find
			{\small
				\begin{align}
					x^{(1)}(\tau)&=\frac{1}{2}\sin(2\Tilde{g}\tau)\sin(2\tau-\Tilde{g}^2\tau)-\frac{1}{2}\cos(2\Tilde{g}\tau)\sin(2\Tilde{g}\tau)\sin(\Tilde{g}^2\tau)\nonumber\\
                    &\qquad+\frac{1}{2}\sin(2\Tilde{g}\tau)\left(\cos(2\Tilde{g}\tau)\sin(\Tilde{g}^2\tau)-\sin(2\tau)\right)+\mathcal{O}(\Tilde{g}^3\tau)\nonumber\\
					&=-\frac{1}{2}\Tilde{g}^2\tau\sin(2\Tilde{g}\tau)\cos(2\tau)+\mathcal{O}(\Tilde{g}^2)=\mathcal{O}(\Tilde{g}),\nonumber\\
					y^{(1)}(\tau)&=\frac{1}{2}\Bigl(1-\cos(2\Tilde{g}\tau)\cos(2\tau-\Tilde{g}^2\tau)-\sin(2\Tilde{g}\tau)\sin(2\Tilde{g}\tau)\cos(\Tilde{g}^2\tau)\nonumber\\
                    &\qquad+\cos(2\Tilde{g}\tau)\left(\cos(2\tau)-\cos(2\Tilde{g}\tau)\cos(\Tilde{g}^2\tau)\right)\Bigl)+\mathcal{O}(\Tilde{g}^3\tau)\nonumber\\
					&=-\frac{1}{2}\Tilde{g}^2\tau\sin(2\tau)\cos(2\Tilde{g}\tau)+\mathcal{O}(\Tilde{g}^2)=\mathcal{O}(\Tilde{g}).\nonumber
				\end{align}
			}
			This implies
			\begin{align}
				c_1(\tau,s)&=-\frac{1}{2}\sinh(4s)\cos(2\tau)\sin(2\Tilde{g}\tau)\Tilde{g}^2\tau+\mathcal{O}(\Tilde{g}^2).\nonumber
			\end{align}
			\item \textbf{Computing} $\boldsymbol{c_2(\tau,s)}$:
			At last, we need to compute $x^{(2)}(\tau,s)$. This calculation can be reduced by noticing $x^{(2)}_2(\tau,s)=\mathcal{O}(\Tilde{g})$. Thus, we have
			{
				\begin{align}
					 x^{(2)}(\tau,s)&=\frac{1}{2}\left(\cosh^4(s)+\sinh^4(s)\right)\left(1-\cos(2\tau)\cos(2\Tilde{g}\tau)\right)\nonumber\\
                    &\qquad+\frac{1}{8}\sinh^2(2s)\left(2\cos(2\tau)\cos(2\Tilde{g}\tau)-\cos(4\tau)\cos(4\Tilde{g}\tau)-1\right)+\mathcal{O}(\Tilde{g}),\nonumber
				\end{align}
			}
			which in turn implies
			\begin{align}
				  c_2(\tau,s)&=\left(\cosh^4(s)+\sinh^4(s)\right)\left(1-\cos(2\tau)\cos(2\Tilde{g}\tau)\right)\nonumber\\
                &\qquad+\frac{1}{4}\sinh^2(2s)\left(2\cos(2\tau)\cos(2\Tilde{g}\tau)-\cos(4\tau)\cos(4\Tilde{g}\tau)-1\right)+\mathcal{O}(\Tilde{g})\nonumber.
			\end{align}
		\end{itemize}
		Putting all together, we finally have the following perturbative expression for the fidelity:
		\begin{align}
			\mathcal{F}_\textrm{eff}^{-2}(\tau)&=1+ C_2(\tau,s)\Tilde{g}^2+\mathcal{O}(\Tilde{g}^3),
		\end{align}
		where
		\begin{align}
			 C_2(\tau,s)&:=\frac{\sinh^2(2s)}{2}\Tilde{g}^2\tau^2-\frac{\sinh(4s)}{2}\cos(2\tau)\sin(2\Tilde{g}\tau)\Tilde{g}\tau+\left(\cosh^4(s)+\sinh^4(s)\right)\left(1-\cos(2\tau)\cos(2\Tilde{g}\tau)\right)\nonumber\\
			 &\quad+\frac{1}{4}\sinh^2(2s)\left(2\cos(2\tau)\cos(2\Tilde{g}\tau)-\cos(4\tau)\cos(4\Tilde{g}\tau)-1\right)\label{Eqn:Coefficient:Squeezed:States:Appendix}.
		\end{align}
		
		\section{Bounding the fidelity}\label{sec:bounding:fidelity:appendix}
		Here, we aim to obtain an upper and lower bound for the fidelity \eqref{distance:fidelity:final:general} from section \ref{Sec:Distinguishing:two:Gaussian:States}.
		Thus, it is useful to realize that the matrix $\boldsymbol{B}_\text{f}^\dagger(t)\boldsymbol{B}_\text{f}(t)$ is Hermitian and positive semidefinite. The eigenvalues of $\boldsymbol{B}_\text{f}^\dagger(t)\boldsymbol{B}_\text{f}(t)$ can therefore be written as $\text{Tr}(\boldsymbol{B}_\text{f}^\dagger(t)\boldsymbol{B}_\text{f}(t))\cos^2(\phi)$ and $\text{Tr}(\boldsymbol{B}_\text{f}^\dagger(t)\boldsymbol{B}_\text{f}(t))\sin^2(\phi)$, where $\phi$ is some real number. This allows us to compute $\operatorname{det}(\boldsymbol{B}_\text{f}^\dagger(t)\boldsymbol{B}_\text{f}(t))=1+\text{Tr}(\boldsymbol{B}_\text{f}^\dagger(t)\boldsymbol{B}_\text{f}(t))+\text{Tr}^2(\boldsymbol{B}_\text{f}^\dagger(t)\boldsymbol{B}_\text{f}(t))\sin^2(2\theta)/4$, which can simply be bounded by:{\small
			\begin{align}
				1+\text{Tr}\left(\boldsymbol{B}_\text{f}^\dagger(t)\boldsymbol{B}_\text{f}(t)\right)\leq\det(\mathds{1}_2+\boldsymbol{B}_\text{f}^\dagger(t)\boldsymbol{B}_\text{f}(t))&\leq 1+\text{Tr}\left(\boldsymbol{B}_\text{f}^\dagger(t)\boldsymbol{B}_\text{f}(t)\right)+\frac{1}{4}\text{Tr}^2\left(\boldsymbol{B}_\text{f}^\dagger(t)\boldsymbol{B}_\text{f}(t)\right)\nonumber\\
                &=\left(1+\frac{1}{2}\text{Tr}\left(\boldsymbol{B}_\text{f}^\dagger(t)\boldsymbol{B}_\text{f}(t)\right)\right)^2\nonumber.
		\end{align}}
		The fidelity \eqref{distance:fidelity:final:general} is thus generally bounded as follows:
		\begin{align}
			\frac{1}{1+\frac{1}{2}\text{Tr}\bigl(\boldsymbol{B}_\text{f}^\dagger(t)\boldsymbol{B}_\text{f}(t)\bigr)}\leq\mathcal{F}_\text{eff}(t)\leq\frac{1}{\sqrt{1+\text{Tr}\bigl(\boldsymbol{B}_\text{f}^\dagger(t)\boldsymbol{B}_\text{f}(t)\bigr)}}.\label{Eqn:Bounds:For:Fidelity:General}
		\end{align}
		We can also aim, to find an upper bound for the fidelity \eqref{distance:fidelity:final:general} depending on the excitation numbers $N_\text{ab}(0)$ and $N_\text{ab}(t)$. In order to do this, we will again only consider pure centered Gaussian states and assume that the reference evolution operator $\hat{U}_\text{R}(t)$ only induces a passive transformation.
		
		For a pure two-mode centered Gaussian state, i.e. a two-mode state satisfying $\Tr(\hat{\rho}_\text{G})=1$ with vanishing first moments, one finds generally:
		\begin{align}
			 N_\text{ab}(t)&=\text{Tr}\left(\left(\hat{a}^\dagger(t)\hat{a}(t)+\hat{b}^\dagger(t)\hat{b}(t)\right)\hat{\rho}_\text{G}\right)\nonumber\\
			 &=\frac{1}{4}\left(\text{Tr}\left(\left\{\hat{a}(t),\hat{a}^\dagger(t)\right\}+\left\{\hat{a}^\dagger(t),\hat{a}(t)\right\}+\left\{\hat{b}(t),\hat{b}^\dagger(t)\right\}+\left\{\hat{b}^\dagger(t),\hat{b}(t)\right\}-4\right)\hat{\rho}_\text{G}\right)\nonumber\\
			 &=\frac{1}{4}\left(\Tr(\boldsymbol{\sigma}_\text{eff}(t))-4\right),
		\end{align}
		where $\boldsymbol{\sigma}_\text{eff}(t)=\boldsymbol{S}_\text{eff}(t)\boldsymbol{\sigma}_0\boldsymbol{S}_\text{eff}^\dagger(t)$. We can now write the initial covariance matrix $\boldsymbol{\sigma}_0$ again as $\boldsymbol{\sigma}_0=\boldsymbol{s}_0\boldsymbol{s}_0^\dagger$ and do the following manipulation
		\begin{align}
			\text{Tr}\left(\boldsymbol{S}_\text{eff}(t)\boldsymbol{\sigma}_0\boldsymbol{S}_\text{eff}^\dagger(t)\right)=\text{Tr}\left(\boldsymbol{s}_0\boldsymbol{s}_0^{-1}\boldsymbol{S}_\text{eff}(t)\boldsymbol{s}_0\boldsymbol{s}_0^\dagger\boldsymbol{S}_\text{eff}^\dagger(t)\left(\boldsymbol{s}_0^{-1}\right)^\dagger\boldsymbol{s}_0^\dagger\right)=\text{Tr}\left(\boldsymbol{s}_0\boldsymbol{S}_\text{f}(t)\boldsymbol{S}_\text{f}^\dagger(t)\boldsymbol{s}_0^\dagger\right)\nonumber,
		\end{align}
		where we used the definition of the final symplectic matrix $\boldsymbol{S}_\text{f}(t):=\boldsymbol{s}_0^{-1}\boldsymbol{S}_\text{eff}(t)\boldsymbol{s}_0$ from section~\ref{Sec:Distinguishing:two:Gaussian:States}. We can now this result for a establishing a rough upper bound, one only has to recall that for Hermitian positive definite matrices $\boldsymbol{A}$ and $\boldsymbol{B}$ the inequality $0\leq\text{Tr}(\boldsymbol{A}\boldsymbol{B})\leq\text{Tr}(\boldsymbol{A})\text{Tr}(\boldsymbol{B})$ holds \cite{Shebraw:Albadwi:2013}. The matrices $\boldsymbol{S}_\text{f}(t)\boldsymbol{S}_\text{f}^\dagger(t)$ and $\boldsymbol{s}_0^\dagger\boldsymbol{s}_0$ satisfy these conditions neccassrily, thus, we find:
		\begin{align}
			N_\text{ab}(t)+1&=\frac{1}{4}\text{Tr}\left(\boldsymbol{S}_\text{f}(t)\boldsymbol{S}_\text{f}^\dagger(t)\boldsymbol{s}_0^\dagger\boldsymbol{s}_0\right)\leq\frac{1}{4}\text{Tr}\left(\boldsymbol{S}_\text{f}(t)\boldsymbol{S}_\text{f}^\dagger(t)\right)\text{Tr}(\boldsymbol{\sigma}_0)\nonumber\\
            &=4\left(1+\text{Tr}\left(\boldsymbol{B}_\text{f}^\dagger(t)\boldsymbol{B}_\text{f}(t)\right)\right)(N_\text{ab}(0)+1)\nonumber.
		\end{align} 
		The previously established bounds \eqref{Eqn:Bounds:For:Fidelity:General} for the fidelity yield consequently:
		\begin{align}
			\mathcal{F}_\text{eff}(t)\leq2\sqrt{\frac{N_\text{ab}(0)+1}{N_\text{ab}(t)+1}}.\label{Eqn:First:Bound:Using:Ex:Number}
		\end{align}
		This bound isn't as strict, as one could hope for: If one assumes the vacuum state as the initial state, i.e. $\boldsymbol{\sigma}_0=\mathds{1}_4$, one finds:
		\begin{align}
			N_\text{ab}(t)+1=\frac{1}{4}\text{Tr}\left(\boldsymbol{S}_\text{f}(t)\boldsymbol{S}_\text{f}^\dagger(t)\right)=1+\text{Tr}\left(\boldsymbol{B}_\text{f}(t)\boldsymbol{B}_\text{f}^\dagger(t)\right)\leq \mathcal{F}^{-2}_\text{eff}(t)\nonumber.
		\end{align}
		This enhances the previous bound \eqref{Eqn:First:Bound:Using:Ex:Number} by a factor of two:
		\begin{align}
			\mathcal{F}_\text{eff}(t)\leq\sqrt{\frac{1}{N_\text{ab}(t)+1}}=\sqrt{\frac{N_\text{ab}(0)+1}{N_\text{ab}(t)+1}}\nonumber.
		\end{align}
		We aim now, to find a stricter bound. For this, we start by computing $\text{Tr}(\boldsymbol{S}_\text{f}(t)\boldsymbol{S}_\text{f}^\dagger(t)\boldsymbol{s}_0^\dagger\boldsymbol{s}_0)$ explicitly:{
			\begin{align}
				\text{Tr}\left(\boldsymbol{S}_\text{f}(t)\boldsymbol{S}_\text{f}^\dagger(t)\boldsymbol{s}_0^\dagger\boldsymbol{s}_0\right)
                =\text{Tr}\left(\begin{pmatrix}
					\mathds{1}_2+2\boldsymbol{B}_\text{f}(t)\boldsymbol{B}_\text{f}^\dagger(t)&2\boldsymbol{A}_\text{f}(t)\boldsymbol{B}_\text{f}^\text{Tp}(t)\\
					2\boldsymbol{A}_\text{f}^*(t)\boldsymbol{B}_\text{f}^\dagger(t)&\mathds{1}_2+2\boldsymbol{B}_\text{f}^*(t)\boldsymbol{B}_\text{f}^\text{Tp}(t)
				\end{pmatrix}\begin{pmatrix}
					\mathds{1}_2+2\boldsymbol{\beta}_0^\text{Tp}\boldsymbol{\beta}_0^*&2\boldsymbol{\alpha}_0^\dagger\boldsymbol{\beta}_0\\
					2\boldsymbol{\alpha}_0^\text{Tp}\boldsymbol{\beta}_0^*&\mathds{1}_2+2\boldsymbol{\beta}_0^\dagger\boldsymbol{\beta}_0
				\end{pmatrix}\right)\nonumber\\
				=2\Re{\text{Tr}\left(\mathds{1}_2+2\boldsymbol{B}_\text{f}(t)\boldsymbol{B}_\text{f}^\dagger(t)+2\boldsymbol{\beta}_0^\text{Tp}\boldsymbol{\beta}_0^*+4\boldsymbol{B}_\text{f}(t)\boldsymbol{B}_\text{f}^\dagger(t)\boldsymbol{\beta}_0^\text{Tp}\boldsymbol{\beta}_0^*+4\boldsymbol{A}_\text{f}(t)\boldsymbol{B}_\text{f}^\text{Tp}(t)\boldsymbol{\alpha}_0^\text{Tp}\boldsymbol{\beta}_0^*\right)},
		\end{align}}
		where we used the Bogoliubov identities to simplify the result. Now we need to find an upper bound for {\small$\text{Re}\{\text{Tr}(\boldsymbol{A}_\text{f}(t)\boldsymbol{B}_\text{f}^\text{Tp}(t)\boldsymbol{\alpha}_0^\text{Tp}\boldsymbol{\beta}_0^*)\}$}, thus, we recall the von Neumann's trace inequality \cite{Mirsky:1975}, which implies that for any matrix $\boldsymbol{A}$ one has $|\text{Tr}(\boldsymbol{A})|\leq \sum_j\sigma_j$, where $\sigma_j$ are the singular values of the matrix. Next, we use that the singular values $\sigma_j$ of the matrix $\boldsymbol{A}$ are the eigenvalues of the Hermitian and positive definite matrix $(\boldsymbol{A}^\dagger\boldsymbol{A})^{1/2}$ \cite{Shebraw:Albadwi:2013}. This implies $|\text{Tr}(\boldsymbol{A})|\leq \text{Tr}((\boldsymbol{A}^\dagger\boldsymbol{A})^{1/2})$ and consequently:{\small
			\begin{align}
				 \Re{\text{Tr}\left(\boldsymbol{A}_\text{f}(t)\boldsymbol{B}_\text{f}^\text{Tp}(t)\boldsymbol{\alpha}_0^\text{Tp}\boldsymbol{\beta}_0^*\right)}&\leq\left|\text{Tr}\left(\boldsymbol{A}_\text{f}(t)\boldsymbol{B}_\text{f}^\text{Tp}(t)\boldsymbol{\alpha}_0^\text{Tp}\boldsymbol{\beta}_0^*\right)\right|\nonumber\\
                &\leq\frac{1}{2}\left(\Tr\left(\boldsymbol{B}_\text{f}^*(t)\boldsymbol{A}_\text{f}^\dagger(t)\boldsymbol{A}_\text{f}(t)\boldsymbol{B}_\text{f}^\text{Tp}(t)\right)+\text{Tr}\left(\boldsymbol{\beta}_0^\text{Tp}\boldsymbol{\alpha}_0^*\boldsymbol{\alpha}_0^\text{Tp}\boldsymbol{\beta}_0^*\right)\right),\nonumber
		\end{align}}
		where we used theorem 2.5 from \cite{Shebraw:Albadwi:2013}. Now we can simply use the Bogoliubov identities and the usual inequalities to find:
		\begin{align}
			\Re{\text{Tr}\left(\boldsymbol{A}_\text{f}(t)\boldsymbol{B}_\text{f}^\text{Tp}(t)\boldsymbol{\alpha}_0^\text{Tp}\boldsymbol{\beta}_0^*\right)}&\leq \frac{1}{2}\text{Tr}\left(\boldsymbol{B}_\text{f}^*(t)\left(\mathds{1}_2+\boldsymbol{B}_\text{f}^\text{Tp}(t)\boldsymbol{B}_\text{f}^*(t)\right)\boldsymbol{B}_\text{f}^\text{Tp}(t)\right)+\frac{1}{2}\text{Tr}\left(\boldsymbol{\beta}_0^\text{Tp}\left(\mathds{1}_2+\boldsymbol{\beta}_0^*\boldsymbol{\beta}_0^\text{Tp}\right)\boldsymbol{\beta}_0^*\right)\nonumber\\
			&\leq \frac{1}{2}\text{Tr}\left(\boldsymbol{B}_\text{f}(t)\boldsymbol{B}_\text{f}^\dagger(t)\right)\left(1+\text{Tr}\left(\boldsymbol{B}_\text{f}(t)\boldsymbol{B}_\text{f}^\dagger(t)\right)\right)+\frac{1}{2}\text{Tr}\left(\boldsymbol{\beta}_0\boldsymbol{\beta}_0^\dagger\right)\left(1+\text{Tr}\left(\boldsymbol{\beta}_0\boldsymbol{\beta}_0^\dagger\right)\right)\nonumber.
		\end{align}
		We can now recall that $N_\text{ab}(0)=\text{Tr}(\boldsymbol{\beta}_0\boldsymbol{\beta}_0^\dagger)$ and conclude:
		\begin{align}
			 N_\text{ab}(t)+1&\leq 1+\text{Tr}\left(\boldsymbol{B}_\text{f}(t)\boldsymbol{B}_\text{f}^\dagger(t)\right)+N_\text{ab}(0)+2\text{Tr}\left(\boldsymbol{B}_\text{f}(t)\boldsymbol{B}_\text{f}^\dagger(t)\right)N_\text{ab}(0)\nonumber\\
			&\quad+\text{Tr}\left(\boldsymbol{B}_\text{f}(t)\boldsymbol{B}_\text{f}^\dagger(t)\right)\left(1+\text{Tr}\left(\boldsymbol{B}_\text{f}(t)\boldsymbol{B}_\text{f}^\dagger(t)\right)\right)+N_\text{ab}(0)(N_\text{ab}(0)+1)\nonumber\\
			&=\left(1+\text{Tr}\left(\boldsymbol{B}_\text{f}(t)\boldsymbol{B}_\text{f}^\dagger(t)\right)\right)\left(1+\text{Tr}\left(\boldsymbol{B}_\text{f}(t)\boldsymbol{B}_\text{f}^\dagger(t)\right)+2N_\text{ab}(0)\right)+N_\text{ab}^2(0)\nonumber\\
			&\leq\mathcal{F}_\text{eff}^{-2}(t)\left(\mathcal{F}_\text{eff}^{-2}(t)+2N_\text{ab}(0)\right)+N_\text{ab}^2(0)\nonumber.
		\end{align}
		This implies, as long as $N_\text{ab}(t)+1\geq N_\text{ab}^2(0)$:
		\begin{align}
			\mathcal{F}^2(t)\leq \frac{\mathcal{F}^{-2}_\text{eff}(t)+2N_\text{ab}(0)}{N_\text{ab}(t)+1-N_\text{ab}^2(0)}\nonumber,
		\end{align}
		which in turn implies:
		\begin{align}
			\mathcal{F}^4(t)\leq\frac{1+2\mathcal{F}^2_\text{eff}(t)N_\text{ab}(0)}{N_\text{ab}(t)+1-N_\text{ab}^2(0)}\leq\frac{1+2N_\text{ab}(0)}{N_\text{ab}(t)+1-N_\text{ab}^2(0)}\nonumber.
		\end{align}
		Thus, the final result reads:
		\begin{align}
			\mathcal{F}_\text{eff}(t)\leq\min\left\{2\sqrt{\frac{N_\text{ab}(0)+1}{N_\text{ab}(t)+1}},\sqrt[4]{\frac{1+2N_\text{ab}(0)}{N_\text{ab}(t)+1-N_\text{ab}^2(0)}}\right\},
		\end{align}
		where the former is more strict for a large initial excitation number and the latter for a small amount of initial excitations. The latter one is also only valid as long as the condition $N_\text{ab}(t)+1\geq N_\text{ab}^2(0)$ is satisfied.

\end{document}